\documentclass[aps,twocolumn,pra,notitlepage,superscriptaddress]{revtex4-2}
\usepackage{amsmath,amssymb}
\usepackage{dsfont}
\usepackage{algorithm}
\usepackage{algpseudocode}
\usepackage{graphicx}
\usepackage{physics}
\usepackage{mathtools}
\usepackage{bbold}
\usepackage{bm}
\usepackage{cancel}
\usepackage{bm}
\usepackage[dvipsnames]{xcolor}
\usepackage[T1]{fontenc}
\usepackage[colorlinks,urlcolor=blue,linkcolor=blue,citecolor=blue]{hyperref}
\usepackage[normalem]{ulem}

\newcommand{\Var}{\operatorname{Var}}
\newcommand{\Cov}{\operatorname{Cov}}

\newcommand{\myvec}[1]{\vb{#1}}

\begin{document}

\title{Evidence of Quantum Machine Learning Advantage with Tens of Noisy Qubits}
\author{Onur Danaci}
\thanks{Equal contribution}
\affiliation{Applied Quantum Algorithms $\langle aQa^L \rangle$, Leiden University, The Netherlands}
\affiliation{QuTech and Kavli Institute of Nanoscience, Delft University of Technology, 2628 CJ, Delft, The Netherlands}
\author{Yash J. Patel}
\thanks{Equal contribution}
\affiliation{Applied Quantum Algorithms $\langle aQa^L \rangle$, Leiden University, The Netherlands}
\author{Riccardo Molteni}
\affiliation{Applied Quantum Algorithms $\langle aQa^L \rangle$, Leiden University, The Netherlands}
\author{Evert van Nieuwenburg}
\affiliation{Applied Quantum Algorithms $\langle aQa^L \rangle$, Leiden University, The Netherlands}
\author{Vedran Dunjko}
\affiliation{Applied Quantum Algorithms $\langle aQa^L \rangle$, Leiden University, The Netherlands}
\author{Jan A. Krzywda}\email{j.a.krzywda@liacs.leidenuniv.nl}
\affiliation{Applied Quantum Algorithms $\langle aQa^L \rangle$, Leiden University, The Netherlands}

% \affiliation{Applied Quantum Algorithms $\langle aQa^L \rangle$, Leiden University, The Netherlands}
\date{\today}

\begin{abstract}

Learning problems involving quantum data are natural candidates for demonstrating an advantage in quantum machine learning. Recent results indicate that, for certain tasks and under noiseless conditions, coherent processing of quantum data outperforms fixed-measurement schemes followed by classical processing. It remained uncertain whether this performance gap persists at a finite scale, and in the presence of noise that is unavoidable with current quantum devices. In this work, we present simulations and analysis of the performance of existing hardware on a learning problem known to exhibit asymptotic advantage, now subjected to noisy quantum data. Comparing coherent quantum processing directly against fixed-measurement schemes, our results demonstrate a clear performance separation at a scale of just 30 to 40 noisy qubits. Already at this scale, the fundamental bottleneck is no longer classical computation but data acquisition; matching the noisy coherent protocol with measure-first strategies would still require months or even years of measurements. By systematically evaluating hardware constraints such as state preparation, gate errors, readout errors, connectivity, and coherence times, we provide evidence that a demonstration of such a strong learning advantage is accessible on near-term devices.

\end{abstract}

\maketitle

% ==================================================================
% SECTION I: INTRODUCTION
% ==================================================================
\section{Introduction}
Quantum algorithms have been theoretically shown to outperform classical methods on specific computational tasks. However, most practical applications require quantum devices that exceed the capabilities of current hardware.
This challenge has motivated the search for near-term scenarios in which meaningful quantum advantages can still be achieved, with quantum machine learning (QML) emerging as a particularly promising candidate. 

Although several learning tasks with classical data admit provable quantum advantages \cite{liu2021rigorous,gyurik2023exponential,molteni2026exponential,anschuetz2026arbitrary}, learning from \emph{quantum data} is expected to be a natural setting for such an advantage, since both the inputs and the processing are inherently quantum. In this setting, one is given labeled quantum states as training data, and the objective is to predict labels of previously unseen states \cite{huang2020predicting}. For such tasks, coherent processing emerges as a proven asset; related work has already established its advantage in extracting physical properties directly from quantum state \cite{chen2022exponential,huang2022quantum}. This has significant potential for scenarios where quantum data is generated by an analog physical simulation and processed digitally using a quantum circuit \cite{daley2022practical}.

To compare classical and quantum strategies in a quantum machine learning setting, Ref.~\cite{gyurik2023limitations} introduced two learning protocols. The \emph{measure-first} (MF) protocol relies on fixed measurements, such as classical shadows \cite{huang2020predicting}, followed by purely classical processing. Conversely, the \emph{fully quantum} (FQ) protocol employs task-specific coherent operations. While FQ exhibits an asymptotic exponential advantage in sample complexity for specific relational tasks \cite{aaronson2023qubit,gyurik2023limitations}, the central question is whether this advantage survives at finite scales under realistic data and hardware noise.

Recent experiments have begun to probe theoretically predicted quantum advantages over fixed measurement strategies for computing properties of quantum states under realistic hardware constraints.
For example, Kretschmer et al. demonstrated a quantum-classical separation in memory using a distributed version of linear cross-entropy benchmarking~\cite{kretschmer2025demonstrating}, Kumar et al. observed a separation in transmitted information using linear optics~\cite{kumar2019experimental}, Benedetti et al. proposed \cite{benedetti2026provable} and experimentally demonstrated \cite{benedetti2025unconditional} an unconditional quantum-classical separation using a complement sampling game on up to 55 trapped-ion qubits.
While these results represent significant milestones, here we focus on a concrete learning task with noisy quantum data, which, in general, is difficult to extract from a quantum device. 
More generally, how different noise mechanisms reshape the relative performance of concrete MF and FQ protocols at experimentally relevant scales has remained largely unexplored.

\begin{figure*}[htb!]
    \centering
    \includegraphics[width=1\linewidth]{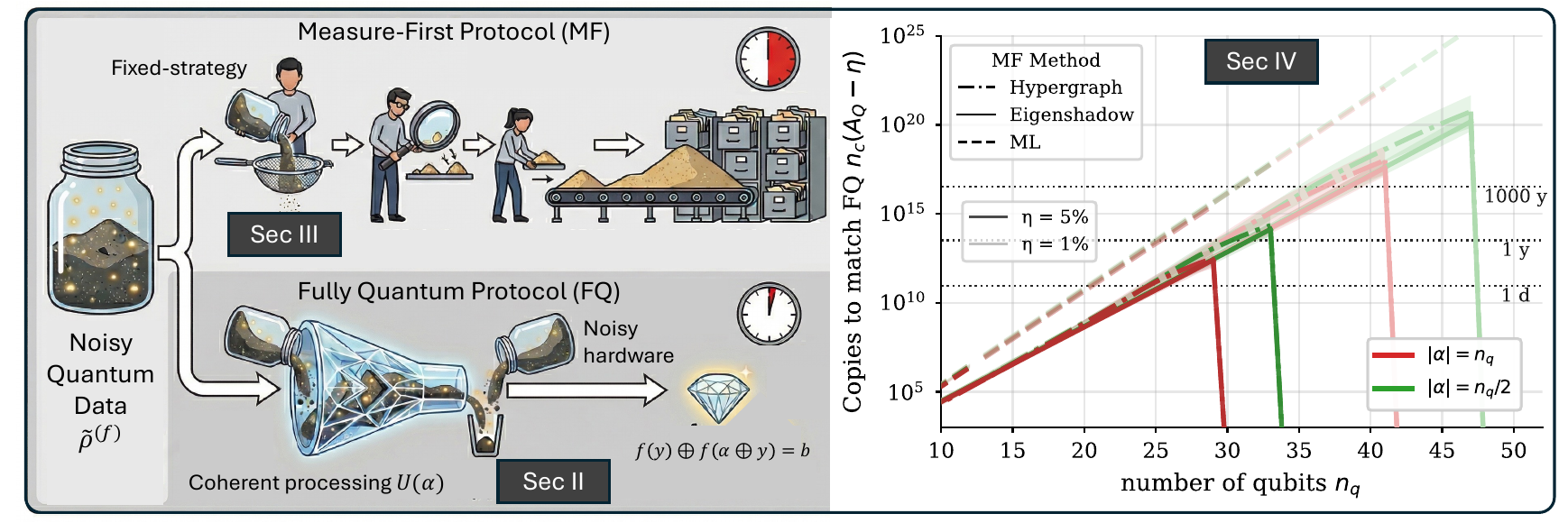}
    \caption{\textbf{Framework for evaluating quantum advantage in learning from noisy quantum data.} \textbf{Left:} Schematic of the two competing paradigms for the learning task described in the main text. A noisy quantum state \(\tilde\rho^{(f)}\) is prepared subject to a local error channel. 
    The fully quantum protocol (Sec.~\ref{sec:quantum_protocol}) coherently processes the state through a potentially noisy entangling measurement circuit to extract the parity bit \(b\). 
    In contrast, measure-first protocols (Sec.~\ref{sec:mf_protocols}) immediately measure \(n_c\) copies of the state to build a classical representation for subsequent statistical inference. \textbf{Right:} Representative scaling of the number of state copies \(n_c\) required by classical measure-first methods to match the fully quantum accuracy within a margin \(\eta \in \{1\%, 5\%\}\). 
    The example shown corresponds to the quantum data degraded by a thermal-relaxation channel with amplitude \(\epsilon_p = 0.1\). See Sec.~\ref {sec:results} and Fig.~\ref{fig:scaling} for other parameters and channels.
    Colors distinguish concepts $|\alpha|$, while the line styles denote the specific measure-first method. 
    The secondary right-hand axis converts \(n_c\) into physical runtime assuming a \(1\,\mu\mathrm{s}\) experimental cycle. Sharp vertical drop-offs indicate the system size \(n_q\) at which the fully quantum protocol loses its coherent advantage because of measurement circuit noise.}
    \label{fig1}
\end{figure*} 

Establishing whether quantum advantages persist at a finite scale is non-trivial because practical comparisons cannot rely solely on asymptotic complexity; they require evaluating the impact of physical noise while simultaneously optimizing over classical measure-first strategies to ensure a relevant baseline. Idealized assumptions are significantly challenged by hardware limitations, where data may be corrupted during preparation, for instance, due to imperfect transduction \cite{kurizki2015quantum, zhang2018quantum} or analog control \cite{kokail2021entanglement,rao2025stability}. The coherent processing is further hindered by gate errors \cite{proctor2019direct}, limited connectivity, finite coherence times \cite{hu2024overcoming}, and readout imperfections \cite{bravyi2021mitigating}. 

These limitations are fundamentally platform-specific: superconducting architectures face $T_1$ relaxation and connectivity constraints \cite{Krantz2019, tuokkola2025methods}, while atomic platforms leverage all-to-all connectivity \cite{ransford2025helios98qubittrappedionquantum, kaushal2020shuttling, moses2023race} and extended coherence \cite{Henriet2020, manovitz2022trapped} but suffer from slower operation times \cite{Bruzewicz2019, wintersperger2023neutral}. In photonic architectures, probabilistic two-qubit gates naturally favor measurement-based sampling \cite{Slussarenko2019, bromley2020applications}, whereas semiconductor spin qubits offer speed \cite{George2024} and connectivity \cite{De_Smet_2025} but remain limited by low-frequency noise \cite{burkard2023semiconductor}. Consequently, modeling noise through a single effective error rate is insufficient; the specific noise type is decisive for learning performance.

To address these challenges, we develop a framework to determine when fully quantum learning advantages persist for this specific learning task \cite{gyurik2023limitations} at a finite scale and under realistic noise. By analyzing the interplay between imperfect state preparation, hardware limitations, and optimized measure-first protocols for the considered task, we establish three main results:
\begin{itemize}
    \item Empirical advantage: Already at the 30-40 qubit scale, optimized measure-first (MF) protocols require years of data acquisition to match fully quantum (FQ) performance. 
    \item Robust near-term signal: Fully quantum (FQ) processing on current hardware extracts a finite learning signal at a relevant scale of $>30$ qubits, effectively overcoming the overhead of noisy measurement circuits.
    \item Noise-dependent advantage: This performance gap actively widens under specific physical noise in the input data; for instance, thermal relaxation severely degrades efficient MF protocols while the FQ method remains comparatively robust.
\end{itemize}

These results are supported by several technical components: a scalable simulation pipeline for fully quantum protocols under realistic hardware constraints (App.~\ref{app:quantum_sim}), a surrogate model for large-scale shadow tomography at high qubit counts (App.~\ref{app:shadow_surrogate}), an analytical characterization of measure-first sample complexity across various physical channels (App.~\ref{app:measure_first_analytics}), and a task-specific measure-first protocol that exploits the algebraic structure of random-phase states (App.~\ref{app:hypergraph_method}).

\paragraph*{Structure of the paper.}
Figure~\ref{fig1} provides an overview of the framework and the main result. 
Section~\ref{sec:quantum_protocol} introduces the learning problem and examines the effects of state-preparation noise, hardware imperfections, and readout on the FQ protocol. 
Section~\ref{sec:mf_protocols} presents several measure-first protocols and analyzes their sample complexity under state-preparation noise. 
Section~\ref{sec:results} details the main results regarding the number of state copies required for MF protocols to achieve the fully quantum accuracy under realistic hardware constraints, followed by a discussion in Section~\ref{sec:discussion}.

% ==================================================================
% SECTION II: THE FULLY QUANTUM PROTOCOL
% ==================================================================
\section{Fully Quantum Protocol with Noisy Data}
\label{sec:quantum_protocol}
\subsection{The Learning Task}
Learning with quantum data involves extracting information from quantum states produced by physical processes or quantum circuits. 
In particular, quantum data can be defined as pairs of quantum states and classical labels, coming from optimized measurements. The objective is to accurately predict the labels for previously unseen input states. The advantage of a quantum protocol is the ability to learn the optimal measurement during training, whereas an MF protocol is limited to measuring each input state using a fixed-measurement strategy. 
While such advantages are expected to arise in many learning tasks of this kind, in this work, we focus on a contrived problem introduced in Ref.~\cite{gyurik2023limitations}, for which a provable advantage was established using results from communication complexity~\cite{bar2004exponential,aaronson2023qubit}. We later discuss possible generalizations of the problem in Sec.~\ref{sec:discussion}.

Specifically, the task is to learn the output of a quantum measurement applied to a set of input states. We consider $n_q$-qubit binary phase-states \cite{BostanciHPS2025} of the following form:
\begin{equation}
\label{eq:state_def}
\ket{\psi_f} = \frac{1}{\sqrt{2^{n_q}}} \sum_{y\in\{0,1\}^{n_q}} (-1)^{f(y)}\ket{y}
\end{equation}
Here, each state encodes a distinct hidden Boolean function $f:\{0,1\}^{n_q}\to\{0,1\}$ in its phases. The learner's goal is to predict the outcome of a target measurement, labeled by a bitstring $\alpha \in \{0,1\}^{n_q}$\footnote{We note a change in notation with respect to Ref.~\cite{gyurik2023limitations} where the measurements were labeled by $x\in\{0,1\}^{n_q}$.}. Specifically, for a new input state $\ket{\psi_f}$, the learner must output a pair $y$ and $b$ that satisfies the relation:
\begin{equation}
b = f(y)\oplus f(y\oplus\alpha)
\end{equation}
where $y = (y', 0)$ and $y' \in \{0,1\}^{n_q-1}$ is a bitstring of $n_q-1$ bits. In the absence of noise, there exists a unitary $U(\alpha)$ (shown in Fig.~\ref{fig:circuit}) such that, when applied to $\ket{\psi_f}$, a subsequent measurement in the computational basis yields an outcome $z=(y',b)$ from which the correct pair $y$ and $b$ is obtained \cite{aaronson2023qubit}. This circuit uses $|\alpha|-1$ CNOT gates, where $|\alpha|$ denotes the Hamming weight of $\alpha$, followed by a final Hadamard gate. Each CNOT is controlled on the last of the $n_q$ qubits (where $\alpha_{n_q} = 1$) and targets a qubit $i$ if $\alpha_i=1$. We define these targeted qubits as \emph{active} ($\alpha_i=1$), while the remaining qubits that do not participate in the CNOT operations are termed \emph{passive} ($\alpha_i=0$). As detailed in Appendix~\ref{app:quantum_analytics}, this circuit leverages constructive interference to amplify the amplitudes of the correct outcomes $y$ and $b$.

To execute the task, the learner is provided with a training set $T_\alpha = \{(\ket{\psi_f}, (\alpha, y, b))\}$. In principle, when information about the measurement setup $\alpha$ is not included in the labels, it must be learned from the quantum data. That can be done, for example, by slightly modifying the distribution of the input states to include a $1/\mathrm{poly}(n_q)$ probability of outputting states in the computational basis $\ket{0\ldots 1_j\ldots0}$ for which the $U(\alpha)$ measurement circuit outputs $b = \alpha_j$. 

\subsection{Fully Quantum vs. Measure-First Protocols}
\label{sec:sq_mf_training}
Access to a training set is relevant for the boundary between the fully quantum protocol (FQ) and any measure-first (MF) one. While the FQ protocol is allowed to learn the correct measurement circuit $U(\alpha)$ from the training data $T_
\alpha$, any MF protocol must commit to a fixed-measurement scheme before observing the labels, and only use the information from the labels after the measurement data is collected.

In principle, one might consider a brute-force strategy that stands in between FQ and MF, where one commits to measuring the same state across all $2^{n_q-1}$ possible $U(\alpha_k)$ circuits and post-selects valid outcomes based on $\alpha$ learned from the labeled data. While this approach uses a fixed-measurement scheme, it should rather be seen as the untrained version of the FQ protocol, as clearly its sample complexity can be reduced from exponential to a single-shot by first learning $\alpha$ from the training data. Without training, this approach would simultaneously suffer from the exponential sample complexity of the brute-force strategy, the noise sensitivity of the FQ protocol, and the difficulty of implementing exponentially many highly entangling measurement circuits $U(\alpha_k)$, making it an impractical baseline for comparison.

\subsection{Noise Modeling in the Fully Quantum Protocol}
Focusing on the trained FQ protocol, its perfect performance naturally degrades under realistic conditions due to noise within both the input quantum data and the measurement circuit. For a fixed $\alpha$, the probability of outputting the correct bit $b$ is defined as the \emph{accuracy} of the protocol,
\begin{equation}
    A_Q(\alpha,y) = \frac{1+V_Q(\alpha, y)}{2},
\end{equation}
where $V_Q(\alpha,y) = |\Re(\bra{y}\tilde\rho^{(f)}\ket{y\oplus \alpha})| \leq 1$ represents the \emph{visibility} of the interference pattern. 
We decompose the $V_Q$ into three components: the visibility reduction $V_p$ due to noisy input states \(\ket{\psi_f}\), the reduction $V_m$ from noise in the measurement circuit \(U(\alpha)\), and the readout noise $V_r$. 
For our chosen noise models, we derive simple analytical expressions for the contributions $V_p$ and $V_r$ below. Because $V_m$ lacks a simple analytical form, we evaluate it by simulating the noisy measurement circuit $U(\alpha)$ with the scalable quantum simulator detailed in Appendix~\ref{app:quantum_sim}. Finally, in Appendix~\ref{app:quantum_numerics}, we empirically verify that the total visibility factorizes as $V_Q \approx V_p V_m V_r$ for the parameters of interest, and we show that $V_m$ is driven primarily by the circuit structure $\alpha$ rather than the total number of qubits.

\emph{Noise on the input data.}
First, we examine the decrease in accuracy of the fully quantum protocol caused solely by the noisy quantum state $V_p$, assuming a noiseless measurement circuit and readout. In our model, the ideal pure state \(\rho^{(f)} = \ketbra{\psi_f}\) is replaced by a mixed state, assumed to take the following form:
\begin{equation}
    \tilde \rho^{(f)}
    =
    \left(\bigotimes_{i=0}^{n_q-1}\mathcal{E}_{i,\epsilon_p}\right)
    \!\left(\ketbra{\psi_f}{\psi_f}\right),
\end{equation}
where \(\mathcal{E}_{i,\epsilon_p}\) denotes a single-qubit channel applied to each qubit with probability \(\epsilon_p\). 
We consider three physically motivated noise channels that affect the quantum data before it enters the measurement circuit $U(\alpha)$. The first, the dephasing channel, models the frequency fluctuations that typically dominate during the analog simulation of an Ising Hamiltonian \cite{shaffer2021practical, daley2022practical}. The second is a zero-temperature thermal-relaxation (amplitude damping) channel, which acts as a primary decoherence mechanism in superconducting charge qubits, such as transmons \cite{Krantz2019}. The third, the depolarizing channel, models unbiased stochastic noise and is often used as a proxy for twirled error channels \cite{Wallman_2016}. This selection spans both unital channels (dephasing and depolarizing) and a non-unital channel (thermal relaxation).

\begin{table}[h]
\caption{Effective attenuation factors \(\bar{\gamma}\) for different preparation-noise channels acting on active (\(\alpha_i=1\)) and passive (\(\alpha_i=0\)) qubits.}
\label{tab:gamma_factors}
\centering
\renewcommand{\arraystretch}{1.4}
\begin{tabular}{lcc}
\hline\hline
Channel & Active \(\bigl(\bar{\gamma}_{\mathrm{act}}\bigr)\) & Passive \(\bigl(\bar{\gamma}_{\mathrm{pass}}\bigr)\) \\
\hline
Dephasing    & \(1-2\epsilon_p\)                   & \(1\) \\
Relaxation   & \(\sqrt{1-\epsilon_p}\)             & \(1-\epsilon_p/2\) \\
Depolarizing & \(1-\frac{4\epsilon_p}{3}\)         & \(1-\frac{2\epsilon_p}{3}\) \\
\hline\hline
\end{tabular}
\end{table}

The presence of noise in the quantum data can significantly reduce the contrast and, consequently, the accuracy of a learning task, even in the absence of imperfections in the measurement circuit $U(\alpha)$.
The extent of degradation depends on the interplay between the channel type, its amplitude, and the target measurement $\alpha$.
Following the derivation in Appendix~\ref{app:quantum_analytics}, the expected visibility of quantum data $V_p(n_q,\alpha)$, averaged over random function inputs \(f\) and observed \(y\) bitstrings, is given by:

\begin{equation}
    V_p(n_q,\alpha) =   \left(\bar{\gamma}_{\mathrm{act}}\right)^{|\alpha|} \left(\bar{\gamma}_{\mathrm{pass}}\right)^{n_q-|\alpha|}.
\end{equation}
The exact expressions for the effective attenuation factors for active ($\bar{\gamma}_{\mathrm{act}}$) and passive qubits ($\bar{\gamma}_{\mathrm{pass}}$) are derived in Appendix~\ref{app:prep_noise} and are summarized in Table~\ref{tab:gamma_factors}. 

\begin{figure}[htb!]
    \centering
    \includegraphics[width=1\linewidth]{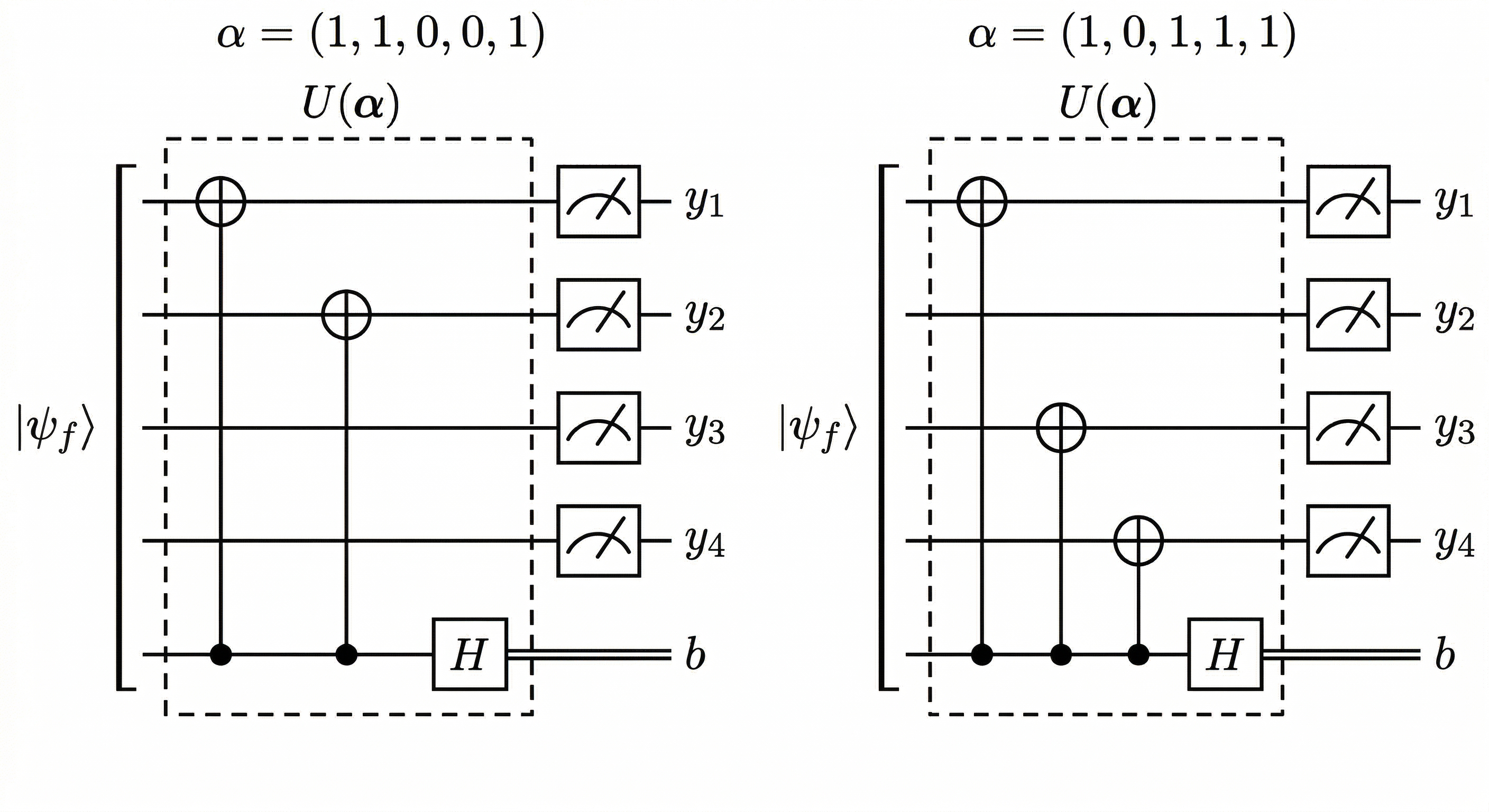}
    \caption{Measurement circuit for two different concepts $\alpha$.}
    \label{fig:circuit}
\end{figure}

\emph{Measurement-circuit noise.}
In addition to the noise on the input states, the fully quantum protocol is also affected by imperfections during the measurement circuit \(U(\alpha)\), as shown in Fig.~\ref{fig:circuit}. 
To estimate performance on near-term devices, we model gate errors, decoherence, and readout faults using the parameter combinations specified in Table~\ref{tab:qc_comparison_final}, chosen to span a contrastive yet realistic range of near-term hardware characteristics.

Specifically, we investigate the trade-off between connectivity, coherence time, and gate fidelity by comparing three distinct device models. 
Device A features all-to-all connectivity; to account for the physical complexity of non-local operations, we assume a lower two-qubit gate fidelity, but zero routing overhead and negligible idling errors.

Devices B and C, in contrast, feature square-lattice connectivity and incur additional SWAP overhead when implementing $U(\alpha)$. 
To capture the impact of routing-induced idle errors, device C is assigned a coherence time that is an order of magnitude shorter than that of device B. 
Additionally, to align device C with widely accessible cloud-based hardware, its two-qubit gate fidelity is set to a conservative 99\%. 
A detailed mapping of these effective parameters to realistic hardware platforms is provided in Appendix~\ref{app:real_hardware}.

\emph{Readout noise.}
Finally, the protocol's sensitivity to readout errors is modeled as independent bit flips with probability $\epsilon_r$. While a flip on the final qubit directly corrupts the inferred parity, the flips on the remaining $n_q-1$ qubits yield a wrong parity with a 50\% probability, resulting in a compound error:
\begin{equation}
V_r(n_q) = (1-2\epsilon_r)(1-\epsilon_r)^{n_q-1}.
\end{equation}
Unless stated otherwise, we assume $\epsilon_r=0.1$ for devices A and B, and $\epsilon_r=1\%$ for device C.

\begin{table}[h]
\caption{Hardware parameters used to simulate the noisy measurement circuit \(U(\alpha)\).
The ratio \(T_{\mathrm{idle}}/T_{2q}\) compares the relevant idle-noise timescale to the two-qubit gate duration.
See Appendix~\ref{app:real_hardware} for a discussion of how these effective parameters relate to different hardware platforms.}
\label{tab:qc_comparison_final}
\centering
\renewcommand{\arraystretch}{1.4}
\begin{tabular}{lccc}
\hline\hline
Setup & A & B & C \\
\hline
1-Q gate \(F_{1q}\)                  & \(99.99\%\)   & \(99.99\%\)   & \(99.99\%\) \\
2-Q gate \(F_{2q}\)                  & \(99\%\)    & \(99.9\%\)    & \(99\%\) \\
Dominant idle noise                  & \(T_2\)       & \(T_1\)       & \(T_2\) \\
Quality \(T_{\mathrm{idle}}/T_{2q}\) & \(10^{6}\)    & \(2\times10^{3}\) & \(2\times10^{2}\) \\
Readout \(1-\epsilon_r\)             & \(99.9\%\)    & \(99.9\%\)    & \(99\%\) \\
Connectivity                         & All-to-all    & Square        & Square \\
\hline\hline
\end{tabular}
\end{table}

\begin{figure}[htb!]
    \centering
    \includegraphics[width=1\linewidth]{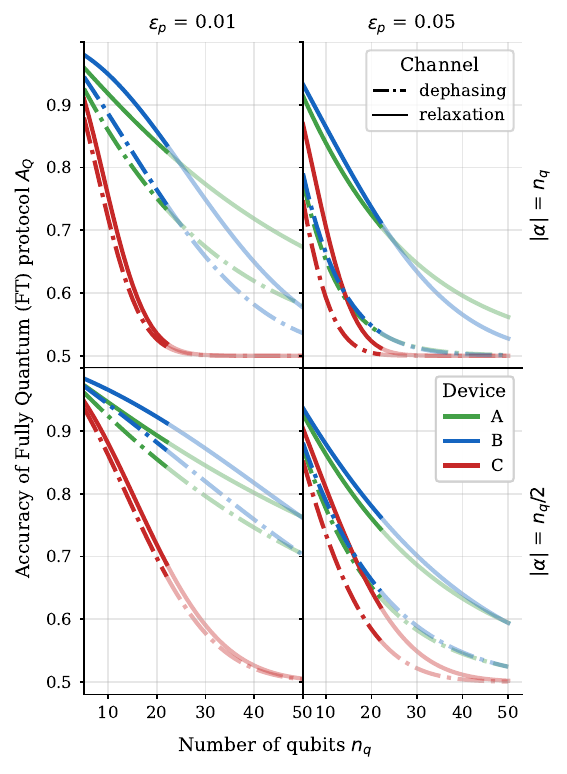}
    \caption{\textbf{Accuracy of the fully quantum protocol under realistic hardware constraints.} 
    The expected accuracy \(A_Q\) is shown as a function of qubit number \(n_q\) for device models A, B, and C (distinguished by color) and for three state-preparation noise channels (dephasing, depolarizing, and relaxation, distinguished by line style).
    The columns compare high (\(\epsilon_p=0.05\), left) and low (\(\epsilon_p=0.01\), right) preparation-noise rates, while the top and bottom rows correspond to concept sizes \(|\alpha|=n_q\) and \(|\alpha|=n_q/2\), respectively. These results are obtained by extrapolating simulations of $V_m$ at $n_q \leq 22$ (darker lines) to larger system sizes using the factorized error model $V_Q = V_p V_m V_r$. See Appendix~\ref{app:quantum_numerics} for extrapolation details. }
    \label{fig:fully_quantum}
\end{figure}

\subsection{Hardware Performance and Scaling}
Figure~\ref{fig:fully_quantum} presents the accuracy of the fully quantum protocol as a function of qubit number for all device models. Extrapolating from simulations up to $|\alpha| = n_q \leq 22$ (Appendix~\ref{app:quantum_numerics}), we observe a clear error transition in the nearest-neighbor devices (B and C): they shift from being gate-fidelity dominated at low $|\alpha|$ to idling-error dominated at $|\alpha| > 10$. In contrast, the all-to-all device (A) remains gate-dominated throughout due to the absence of additional swap gates.

Reducing the target measurement weight from $|\alpha| = n_q$ to $|\alpha| = n_q/2$ yields substantial improvements, particularly for devices B and C, where routing overhead increases significantly with circuit depth. This improvement arises from both a shorter measurement circuit and, for the dephasing and depolarizing channels, a reduced influence of noise on the quantum data. Consequently, under low preparation noise ($\epsilon_p = 0.01$), the performance crossover between devices A and B shifts from $n_q \approx 20$ (at full weight) to $n_q \approx 40$ (at half weight). While circuit errors dominate and obscure the distinctions between noise channels in this low-noise regime, the specific signatures of the preparation channels remain distinct at higher preparation noise ($\epsilon_p = 0.05$).

Ultimately, two key features emerge for this learning task. First, the expected error increases significantly with the Hamming weight $|\alpha|$ due to both an extended measurement circuit and the amplified impact of noise on the quantum data. Second, this data degradation is least severe under the thermal-relaxation channel. Measure-first protocols modify both of these relations.

% ==================================================================
% ==================================================================
\section{Measure-First Protocols}
\label{sec:mf_protocols}
To establish classical baselines, we analyze the sample complexity in terms of \(n_c\), the number of measured state copies required to achieve a target accuracy in correctly outputting $y$ and $b$.
We analyze this for several MF strategies under realistic noise in quantum data, in the absence of which the MF protocol requires exponentially many copies to match the quantum result \cite{gyurik2023limitations}. 
We investigate whether this separation persists when the FQ protocol's performance is degraded by noise.

To isolate the information-acquisition aspect of the problem, we allow unrestricted classical computational power, as any limitation here would strictly increase the quantum advantage. For the quantum measurements, we restrict the MF protocol to single copies of $\ket{\psi_f}$. This aligns with theoretical results showing that even multiple noiseless copies cannot enable the MF protocol to efficiently solve the task \cite{gyurik2023limitations}, and avoids the increased resource demands of multi-copy entangled measurements, ensuring our setup remains a genuinely near-term application. The design space of MF protocols is broad and highly task-dependent. 
Following extensive empirical studies, we focus on the most relevant MF approaches, ranging from generic shadow-based methods to task-specific strategies that exploit structural priors.

\subsection{Shadow-based methods}
Classical shadows allow the estimation of observables from randomized single-qubit measurements \cite{huang2020predicting}. 
The absence of entangling gates, combined with standard single-qubit operations (Clifford gates), makes this approach hardware-agnostic and experimentally feasible. 
From the resulting $n_c$ randomized measurement snapshots, one can form an estimator of the full density matrix $\tilde \rho^{(f)}(n_c)$ by averaging the appropriately inverted snapshots. 
As demonstrated empirically in Appendix~\ref{app:shadow_surrogate}, elements of these shadow density matrices can be efficiently sampled \emph{in silico} using our surrogate model. 
This capability enables the design and testing of strategies prior to running inference on physical quantum hardware. By leveraging the central limit theorem, the surrogate simulates entries of the reconstructed density matrix at arbitrary qubit numbers and shot counts, making large-scale evaluations of shadow-based protocols tractable. We then utilize this framework in Appendix~\ref{app:measure_first_analytics} to analytically derive the scaling laws for all shadow-based methods. 

\emph{Local methods.} Consistent with the assumption of unrestricted classical computational resources, we now analyze how this estimate of $\tilde\rho$ can be used to solve the learning task. As shown in Appendix~\ref{app:quantum_analytics}, the learning task can be solved by returning the sign of a relevant element of the density matrix \(\tilde \rho_{y,y\oplus\alpha} \propto (-1)^{f(y) \oplus f(y\oplus \alpha)}\). However, the corresponding sample complexity scales exponentially: $ {
n_c \sim 4^{n_q}
\bigl(1.5/\bar{\gamma}_{\mathrm{act}}^{2}\bigr)^{|\alpha|}
\bigl(1/\bar{\gamma}_{\mathrm{pass}}^{2}\bigr)^{n_q-|\alpha|}}$, with $\overline{\gamma}$ defined previously in Tab.~\ref{tab:gamma_factors}. 
An attempt to mimic the quantum interference by estimating the sign of $\sum_k  \tilde \rho^{(f)}_{y,k}  \tilde \rho^{(f)}_{k,y\oplus\alpha}$ yields a small improvement,
\[
n_c \sim 4^{n_q}
\bigl(0.87/\bar{\gamma}_{\mathrm{act}}\bigr)^{|\alpha|}
\bigl(1.80/(1+\bar{\gamma}_{\mathrm{pass}}^{2})\bigr)^{n_q-|\alpha|},
\]
but even in the noiseless limit, this approach still suffers from an exponential cost of \(n_c\sim 3.6^{n_q}\) and exhibits only a weak dependence on \(|\alpha|\). 
This scaling remains dominant in the presence of weak noise in quantum data, where the dependence on \(|\alpha|\) and channel remains as a multiplicative factor that is only weakly dependent on the total qubit number.

\emph{Eigenshadow method.} Among the shadow-based approaches, computing the principal eigenvector of $\tilde \rho^{(f,n_c)}$ yields the best performance. 
As shown in Fig.~\ref{fig:eigenshadow_accuracies}, we observe an empirical scaling of $n_c\sim 2.8^{n_q}$ for relaxation noise and $n_c\sim 3.2^{n_q}$ for dephasing noise at $\epsilon_p=0.1$ and $|\alpha|=n_q/2$, with only negligible dependence on $|\alpha|$. 
These numerical results are consistent with the derived scaling law (See Fig.~\ref{fig:eigenshadow_scaling} for validation):
\[
n_c \sim \bigl(2.5/\bar{\gamma}_\text{eff}^{2}\bigr)^{n_q},
\qquad
\bar{\gamma}_\text{eff} = (\bar{\gamma}_{\mathrm{act}}+\bar{\gamma}_{\mathrm{pass}})/2.
\]
Thus, while this method shares the FQ protocol's robustness against thermal relaxation, it differs in a crucial aspect: the sample complexity depends almost entirely on system size and is minimally sensitive to the Hamming weight $|\alpha|$.

\emph{Machine-learning methods.} While the preceding shadow-based methods rely on fixed, heuristic strategies, such as extracting the principal eigenvector, an alternative is to make the classical processing learnable directly from the measurement data. Such a data-driven approach is highly motivated by computational efficiency, as it bypasses the need to reconstruct the full underlying function $f(y)$ or the complete density matrix. However, while raw snapshots of classical shadows can be used as direct inputs to standard machine learning models, we empirically observed that training unstructured deep neural networks becomes increasingly intractable beyond the few-qubit regime. 
This scaling bottleneck arises from the exponential growth in the dimensionality and variance of the raw shadow representation.

To overcome this scaling bottleneck, we replace direct high-dimensional classification with physics-informed feature extraction. Specifically, we compress the estimated density matrix into an $\mathcal{O}(n_q)$-dimensional summary by grouping elements of $\tilde{\rho}^{(f,n_c)}$ into ``shells'' based on their Hamming distance from the target off-diagonal element. This low-dimensional representation, spanning either $n_q+2$ or $2n_q+1$ features depending on the chosen map, exploits a key physical property of the ideal phase state: products of the form $\tilde{\rho}_{y\oplus\alpha,i}\tilde{\rho}_{i,y}$ probe the target coherence through multiple intermediate pathways. Under realistic noise, deviations in these shell summaries act as highly sensitive indicators of coherence loss and purity contraction (see Appendix~\ref{app:ml_method} for details).

To obtain a noise-adaptive classifier from these features, we train a regularized logistic regression model directly on the shell sums. Despite its simplicity, this linear model is exceptionally well-suited to the physics-informed compression: it remains stable across train-validation splits, successfully extracts the signal from the noise-dominated raw shadows, and yields coefficients that provide an interpretable, shell-by-shell breakdown of the learning process. Full construction and training details are provided in Appendix~\ref{app:ml_method}.

\subsection{Hypergraph-based method}
Finally, we explore a structural approach that explicitly leverages the mathematical properties of the input quantum data. Since each binary phase state corresponds to a distinct hypergraph \cite{Rossi_2013, Qu_2013}, general randomized Pauli measurements can be replaced with a fixed scheme optimized in the graph basis. Building upon hypergraph certification techniques \cite{zhu2019efficient} and methods for learning random-phase states \cite{arunachalam2023optimalalgorithmslearningquantum}, our approach reconstructs the algebraic normal form of the underlying Boolean function $f(y)$. By leveraging this structure, we can directly target phase reconstruction even in the presence of noise in the quantum data. Further details are provided in Appendix~\ref{app:hypergraph_method}, while comprehensive theoretical derivations and numerical tests can be found in Ref.~\cite{danaci_in_prep}.

\begin{figure}[htb!]
    \centering
    \includegraphics[width=1\linewidth]{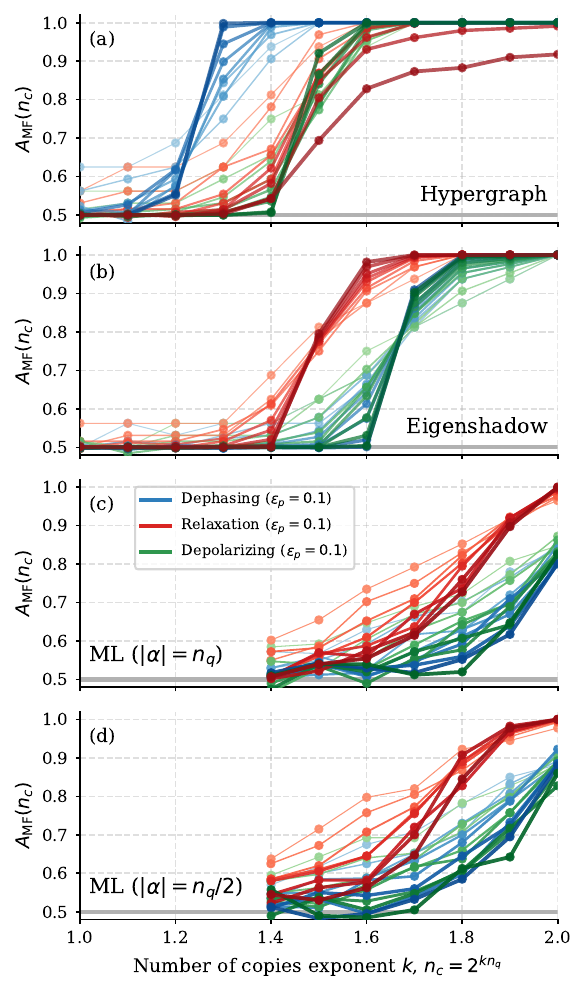}
    \caption{\textbf{Sample complexity of measure-first protocols.} The expected accuracy is plotted against the scaled copy-complexity exponent \(k\), defined by \(n_c=2^{k n_q}\), at preparation-noise rate \(\epsilon_p=0.1\). 
    The panels show (a) the Hypergraph method, (b) the Eigenshadow method, and the ML method for (c) \(|\alpha|=n_q\) and (d) \(|\alpha|=n_q/2\). 
    Noise channels are distinguished by color (blue: dephasing, red: relaxation, green: depolarizing), and increasing system size (\(n_q=5\) to \(15\)) is indicated by increasing color intensity and line thickness. Results for smaller $
    \epsilon_p$ are shown in Fig.~\ref{fig:all_mf}}
    \label{fig:eigenshadow_accuracies}.
\end{figure}

In the noiseless limit, the hypergraph method efficiently samples hyperedges by employing $n_q$ distinct measurement configurations. In each configuration, a single target qubit is measured in the $X$-basis while the remaining spectator qubits are measured in the $Z$-basis. Ideally, this requires $O(2^{n_q})$ samples per configuration to perfectly learn the underlying hypergraph. Under realistic conditions, however, the sampled data is corrupted by bit flips and phase flips, producing invalid bitstrings. To recover the true state, we repeat the experiment and filter the correct equations from the erroneous ones by sorting based on the empirical frequency of the bitstrings. The success of this sorting procedure is highly sensitive to the nature of the noise. Dephasing proves to be the most sample-efficient; because $Z$-errors commute with the spectator measurements, only the parity outcome is affected, allowing the sample complexity to retain its $2^{n_q}$ scaling subject only to a mild multiplicative factor. This ceases to be true for the depolarizing channel, where spectator (parity) qubits experience bit (phase) flips with a probability of $\frac{2}{3}\epsilon_p$. In the limit of large $n_q$, this drives the average-case sample complexity to scale as $O(2^{n_q}/[1-\frac{4}{3}\epsilon_p]^{2n_q})$. Similarly, for a randomly selected function $f$, both spectator and $X$-qubits have a $\frac{1}{2}\epsilon_p$ probability to be flipped under thermal relaxation, and hence the scaling diverges as $O(2^{n_q}/[1-\epsilon_p]^{2n_q})$. While these scaling bounds might suggest that the depolarizing channel performs worse than relaxation, the latter additionally suffers from the bias towards low-parity bits oversamples low-degree hyperedges, causing the frequency-based sorting method to collapse entirely. Consequently, the protocol's performance is strictly limited at higher values of $\epsilon_p$ and $n_q$. This limitation is empirically evident in Fig.~\ref{fig:eigenshadow_accuracies}, where even in the limit of large $n_c$, the protocol fails to resolve the task at $n_q = 15$ and $\epsilon_p = 0.1$ with the thermal relaxation channel. 

\subsection{Sample Complexity of MF Protocols}
We finally compare the resources required by the relevant measure-first methods in Fig.~\ref{fig:eigenshadow_accuracies}. 
To isolate the role of noisy quantum data, we compare the expected accuracy as a function of the number of copies \(n_c\) across all preparation-noise channels (distinguished by colors) at a relatively high noise rate \(\epsilon_p=0.1\). 
Given that sample complexity scales approximately exponentially, accuracies are plotted as a function of the exponent \(k\in[1,2]\), defined by \(n_c=2^{k n_q}\). This approach enables results for different qubit numbers to be presented on a unified axis, with color intensity representing system size. The influence of the concept Hamming weight \(|\alpha|\) is shown explicitly only for the ML method, as the other approaches are considerably less sensitive to this parameter, as previously discussed.

Several distinct trends emerge. First, the Eigenshadow method (Fig.~\ref{fig:eigenshadow_accuracies}(b)) exhibits a sharp, nearly system-size-independent transition in \(k\)-space. 
The close overlap of the curves validates the exponential scaling law \(n_c\sim 2^{k n_q}\), confirming that the threshold exponent \(k\) remains essentially constant across different qubit counts. 
Second, we observe a reversal in noise-channel robustness: while the Eigenshadow and ML methods require significantly fewer copies under relaxation than under dephasing or depolarizing noise, the Hypergraph method exhibits the opposite trend. In particular, we see a separation in performance between the dephasing channel and the remaining two channels. While dephasing requires the fewest samples, the relaxation and depolarizing channels share a similar sample complexity up to a finite system size, above which the measurement bias under thermal relaxation fails to approach unity, leading to a complete collapse of the accuracy. This collapse is clearly evident for the largest simulated system ($n_q=15$, darkest red line).

As a result, the optimal MF strategy at finite target accuracies depends on the dominant physical noise. Under thermal relaxation, while the Hypergraph and Eigenshadow methods may perform similarly at small qubit counts, their behavior diverges sharply at scale. By $n_q=15$, the Hypergraph method fails entirely due to measurement bias. In stark contrast, the Eigenshadow method successfully resolves the signal with a transition near $k \approx 1.5$ (corresponding to $n_c \approx 6 \times 10^6$). For higher error rates ($\epsilon_p > 0.1$), this gap will only widen, cementing Eigenshadow as the strictly more scalable approach for relaxation noise.

Conversely, for the dephasing and depolarizing channels, the Hypergraph method remains substantially more efficient across a wide range of $\epsilon_p$. At a system size of $n_q=15$, the Hypergraph transition occurs near $k=1.25$ ($n_c \approx 5 \times 10^5$) under dephasing and $k=1.5$ ($n_c \approx 6 \times 10^6$) under depolarizing noise. By comparison, the Eigenshadow method requires a much higher $k=1.65$ ($n_c \approx 3 \times 10^7$) to resolve either of these channels.

In contrast to the sharp transitions observed in the algebraic and spectral methods, the ML method displays a more gradual increase in accuracy. 
This characteristic enhances its relative competitiveness at moderate target accuracies, such as around \(0.6\), where noisy quantum protocols may realistically operate.

\begin{figure}[htb!]
    \centering
    \includegraphics[width=1.0\linewidth]{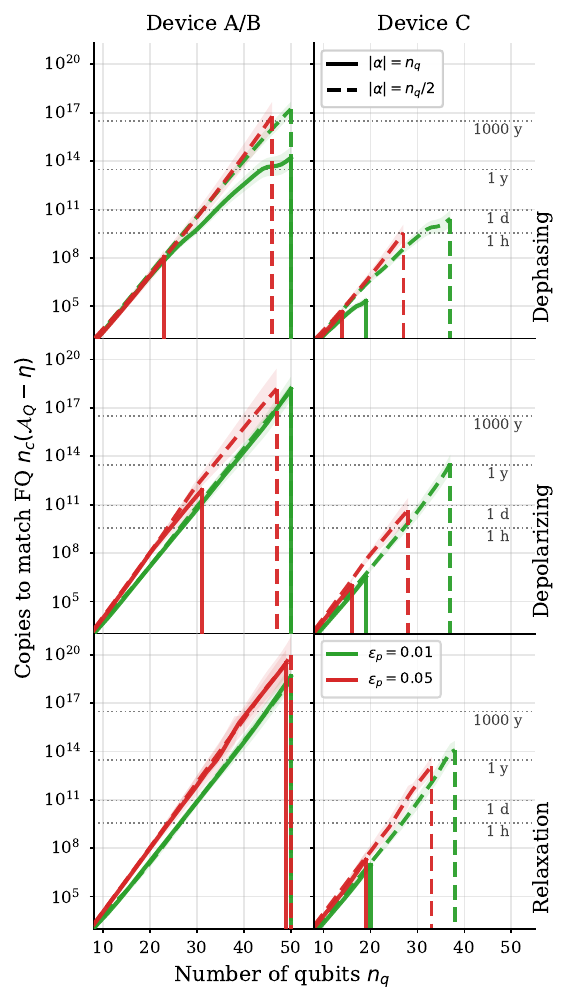}
    \caption{\textbf{Sample complexity required to match the fully quantum protocol.} The required number of copies \(n_c\) for measure-first protocols is plotted as a function of qubit number \(n_q\). Columns correspond to devices B and C, while rows show different state-preparation noise channels (dephasing, depolarizing and relaxation). Device A yields a similar result to B. Line styles indicate the concept size, \(|\alpha|=n_q\) (solid) and \(|\alpha|=n_q/2\) (dashed). Colors distinguish the preparation-noise amplitude, \(\epsilon_p=0.01\) (light) and \(\epsilon_p=0.05\) (dark). Horizontal dotted lines indicate the corresponding experimental acquisition time assuming a \(1\,\mu\mathrm{s}\) cycle time. Curves terminate when the fully quantum accuracy falls below the target threshold \(A_{\mathrm{target}} = A_Q-\eta\), so the matching condition can no longer be satisfied. The results are plotted for $\eta = 1\%$. For similar plot for $\eta = 5\%$ and including $\epsilon_p = 0.1$ see Fig.~\ref{fig:eta5}.}
    \label{fig:scaling}
\end{figure}

% ==================================================================
% SECTION V: RESULTS
% ==================================================================
\section{Evidencing of a quantum advantage}
\label{sec:results}

We now compare the number of measurements required by the MF protocols to achieve the accuracy of FQ at a common noise level in the quantum data. For simplicity we count only the measurement cost incurred at the inference stage (For a training of the FQ protocol see Sec.~\ref{sec:sq_mf_training}).

Because the accuracy of MF protocols exhibits a sharp threshold (Fig.~\ref{fig:eigenshadow_accuracies}), distinguishing it from the accuracy of the quantum protocol in the regime close to random guessing ($A_Q, A_{MF}\sim 0.5$) would require exponentially many shots. To achieve a resolvable quantum advantage, we therefore introduce a finite significance threshold $\eta$ and require:
\begin{equation}\label{eq: advantage_condition}
    A_Q(n_q)-A_{\mathrm{MF}}(n_q,n_c)\geq \eta.
\end{equation}
We then benchmark each MF method by finding the number of copies $n_c$ needed to reach this target accuracy, $A_{\mathrm{target}}(n_q) = A_Q(n_q) - \eta$. At the baseline where $A_{\mathrm{target}}(n_q) = 0.5$, random guessing suffices and the required copy budget naturally collapses to $n_c = 1$.

We extract \(n_c(A_{\mathrm{target}}, n_q)\) from the numerical MF accuracy curves by fitting the finite-size scaling law \(k_x(T, n_q) = C_{\star}(T) + \beta_{\star}(T)/n_q\), where \(\log_2n_c = n_q \, k_x\).
Uncertainty bands are attached via validation across different \(n_q\), ensuring that reported extrapolations are supported by empirical data (see Appendix~\ref{app:extrapolation_details} for full details). 
For \(A_Q\), we use the extrapolated values from Sec.~\ref{sec:quantum_protocol} for devices B and C. 
Device A behaves similarly to device B, and is shown separately in Fig.~\ref{fig1}, where we also compare the $\eta = 0.01$ against the $\eta = 0.05$ case. 
From now on, unless stated otherwise, we set \(\eta = 0.01\); Appendix~\ref{app:extrapolation_details} discusses the corresponding confidence level and the case \(\eta = 0.05\). 
Beyond the observed range, the bands quantify uncertainty under the assumed \(\log(n_c) \sim 1/n_q\) scaling.

%but not model-form validity, which could in principle be tested by running the MF protocols on a real quantum device at larger system sizes.

The results are shown in Fig.~\ref{fig:scaling}, where we plot the required number of copies $n_c$ for locally the most sample-efficient MF protocol as a function of qubit number $n_q$. 
We evaluate concept sizes $|\alpha|=n_q$ (solid lines) and $|\alpha|=n_q/2$ (dashed lines) under preparation-noise amplitudes $\epsilon_p=0.05$ (red) and $\epsilon_p=0.01$ (green). 
The panels are divided by noise channel, dephasing (top row), depolarizing (middle row) and thermal-relaxation (bottom row), and by hardware profile, contrasting Device B (left column) with the lower-quality Device C (right column). In the considered regime, Device A offers advantages similar to those of Device B. 

In most cases, the required MF copy budget grows exponentially with $n_q$. 
These curves terminate sharply when the fully quantum (FQ) accuracy itself drops below the target $A_{\mathrm{target}} = A_Q - \eta$, at which point the condition for quantum advantage in Eq.~(\ref{eq: advantage_condition}) can no longer be satisfied. 
To provide physical context, the horizontal dotted lines convert the required MF copy budget into experimental runtime. 
This assumes an illustrative cycle time of $1\,\mu\mathrm{s}$ per state preparation and measurement \cite{google2025quantum}, which is ultimately lower-bounded by hundreds of nanosecond readout time in superconducting architectures \cite{swiadek2024enhancing}. 
Under this conversion, the required MF runtime approaches or exceeds the one-day threshold in most of the considered regimes.

The magnitude of the quantum advantage varies gradually depending on the noise channel affecting the quantum data. The weakest quantum advantage is observed under the dephasing channel, where Hypergraph methods remain relatively efficient. The advantage becomes moderate under depolarizing noise, and is most pronounced under thermal relaxation, where measure-first (MF) methods perform poorly while the full quantum (FQ) protocol remains robust. Despite dephasing yielding the weakest relative advantage, the gap remains significant at low preparation noise ($\epsilon_p=0.01$, green lines). Across both Devices A/B and Device C, MF protocols require anywhere from hours to thousands of years of data acquisition to match the FQ accuracy, depending on the system size ($n_q$) and the concept size ($|\alpha|$). For the lower-quality Device C, achieving any advantage requires restricting to the smaller concept size of $|\alpha|=n_q/2$. Even under these constraints at $\epsilon_p=0.01$, MF methods still require roughly a few hours to match FQ performance at $n_q \approx 30$ across all noise channels. At stronger preparation noise ($\epsilon_p=0.05$, red lines), the accessible advantage is generally reduced. For dephasing, the advantage is largely restricted to the smaller concept size ($|\alpha|=n_q/2$) even on the higher-quality Device A/B. Here, matching the FQ protocol requires days of MF data acquisition at $n_q \approx 30$, and years at $n_q \approx 40$, consistently across both noise levels. 

Interestingly, at a fixed system size, the relative quantum advantage actually grows with $\epsilon_p$ when the quantum data is degraded by depolarizing or thermal relaxation noise. Combined with the quantum protocol's inherent robustness against thermal relaxation, we observe that at $n_q = 33$, the MF protocol requires a full year to match quantum performance on Device C, albeit only for $|\alpha| = n_q/2$. This trend is even more striking for the larger $\epsilon_p=0.1$ illustrated in Fig.~\ref{fig1}, where the one-year mark drops to $n_q = 30$ on Devices A and B for any $|\alpha|$. Ultimately, the numerical study confirms two competing mechanisms: while increased noise in the quantum data degrades the absolute performance of the quantum protocol, it severely penalizes classical data acquisition, often amplifying the relative quantum advantage under certain noise channels.
    
\section{Discussion}
\label{sec:discussion}
    
Our results show that a significant quantum advantage for a certain learning task with quantum data may be experimentally observed in noisy regimes relevant to current hardware. Specifically, we consider the task introduced in Ref.~\cite{gyurik2023limitations}, which is formalized within the rigorous probably approximately correct (PAC) framework, where an asymptotic advantage has been demonstrated in an idealized setting. Here, we consider realistic conditions at a finite scale by modeling both state-preparation noise and hardware imperfections for the task. We compare the fully quantum protocol against several purposely developed measure-first methods. We identify regimes in which coherent quantum processing retains a substantial advantage, observable with as few as 30-40 qubits. 

To support these findings, we derived and numerically verified analytical models for both approaches in the presence of noise. This required developing a specialized simulation pipeline capable of handling up to 22 noisy qubits, enabling us to extrapolate the quantum protocol's performance to larger system sizes. 
Furthermore, to ensure a rigorous baseline, we introduced a selection of theoretically and numerically motivated measure-first strategies. The innovations include: 1) A surrogate model for shadow tomography designed to streamline data-driven training. 2) A novel hypergraph-based method that efficiently exploits the underlying problem structure. 3) A lightweight, noise-adaptive machine learning model that performs robustly across diverse noise channels.
    
\subsection{Hardware and Algorithmic Trade-offs}
Our study allows us to identify certain relationships and trade-offs between noise in quantum data, its amplitude, hardware specifications, and their combined effect on the size of the quantum advantage.
    
Starting with the noise degrading quantum data, our results highlight the distinct role of the noise \emph{type}, not only its magnitude. 
In the fully quantum protocol, the effect of degrading quantum data significantly depends on $|\alpha|$ only for the dephasing channel, while under the thermal-relaxation channel, the quantum protocol is comparatively robust and $|\alpha|$ independent. In particular, this same noise type severely degrades the most sample-efficient measure-first baselines, which leads to a counterintuitive dynamic: when quantum data is degraded by the thermal-relaxation channel (and to a lesser extend, depolarizing noise), increasing the noise level can initially \emph{widen} the quantum advantage, up to the point where the Eigenshadow method becomes the optimal classical strategy. We observe this empirically for the relaxation channel around an error probability of $\epsilon_p \approx 0.1$. Consequently, for the relaxation channel at $\epsilon_p = 0.1$, matching the fully quantum protocol would require at least a year of continuous MF data acquisition at \(n_q=30\), assuming a state-of-the-art $1\mu$s cycle-time. We highlight that for the other channels degrading quantum data, hitting the one-year mark would require $n_q = 40$ qubits and $\epsilon_p \leq 0.01$ for all $|\alpha|$, and slightly more relaxed $\epsilon_p \leq 0.05$ if restricted to $|\alpha| \leq n_q/2$. 
    
We validate that this result can be achieved by deploying the measurement circuit on a subset of currently existing quantum hardware; however, the range of parameters that yield quantum advantage widens with connectivity beyond nearest-neighbor devices, as otherwise routing overhead exposes the qubits to decoherence during idling ($T_1, T_2$ times). Such a penalty for limited connectivity becomes more severe as the system size grows; in particular, in the relevant range of $n_q > 15$ qubits, it dominates other noise sources, including error caused by gate errors (e.g., $F_{2q}>99\%$). We confirm that all of the above advantages can be achieved on connectivity-restricted devices if the coherence times are on the order of thousands of two-qubit gate times. Below this threshold, and for the lower-quality Device C, we observe that an advantage on the timescale of hours can be achieved only in specific settings, including a smaller concept size $|\alpha| = n_q/2$ and higher fidelity of quantum data $\epsilon_p \ll 0.1$. 
    
In generally, a smaller $|\alpha|$ is favorable for the quantum protocol, as the measurement circuit scales linearly with $|\alpha|$ for all-to-all connected devices, and shows an empirically observed scaling of $\mathcal{O}(
|\alpha|^{1.5})$ for nearest-neighbor devices, using a standard transpiler. Additionally, at a fixed amplitude of noise in the quantum data, a large $|\alpha|$ corresponds to higher-order coherences, which are generally more prone to decoherence. In contrast, the most efficient measure-first methods show little dependence on $|\alpha|$. Therefore, reducing the value of $|\alpha|$ increases the experimental separation between the fully quantum and measure-first protocols, as reflected in the orders-of-magnitude rise in the number of classical samples needed to equal the quantum protocol when comparing the cases $|\alpha| = n_q$ and $|\alpha| = n_q/2$. However, it is important to note that the theoretical quantum advantage is not guaranteed to hold when $|\alpha|$ is restricted.

\subsection{Practical Relevance}
In our benchmark, all measure-first (MF) protocols operate on single copies of the input state. 
In practice, this means that the total time needed to learn many states during inference is simply the single-state learning time (shown in the plots) multiplied by the number of states.
The separation result of Ref.~\cite{gyurik2023limitations} does not inherently restrict measure-first strategies 
to single-copy measurements, and in principle, MF protocols could use collective measurements on multiple copies of the state to suppress preparation errors~\cite{huggins2021virtual, koczor2021exponential}. 
However, these methods require additional inter-copy entangling operations immediately before measurement. On near-term hardware, this would significantly increase the qubit count, two-qubit depth, and routing overhead, thereby introducing an additional noise source. 
Motivated by the search for a near-term advantage, we therefore restricted our attention to single-copy MF schemes and leave a systematic multi-copy benchmark for future work.

    Regarding the experimental demonstrations of this protocol, while our paper remains agnostic to the specific method of random phase state preparation, several practical approaches can be considered. 
    Depending on the platform, preparing a random phase state might require the simulation of an Ising Hamiltonian with precise control over couplings \cite{Shen_2026, BostanciHPS2025}, or a sophisticated state-preparation circuit, potentially including nontrivial multi-qubit gates \cite{Rossi_2013, Qu_2013}. 
    While the preparation of such few-qubit hypergraph states has been demonstrated in small-scale photonic systems \cite{huang2024demonstration}, scaling this approach remains challenging. 
    The more practical route is to sample from sufficiently rich pseudorandom phase-state ensembles generated by polynomial-depth constructions, such as pseudorandom quantum circuits based on Goldreich--Goldwasser--Micali protocol~\cite{goldreich1986construct, zhandry2021construct} or polynomially many layers of Toffoli gates \cite{brakerski2019pseudo}. 
    While exploring such implementations in experimentally realistic settings is an important direction for future work, we already note that a practical demonstration would strongly benefit from using an analog-digital platform (e.g., an Ising simulator) or a platform with native multi-qubit gates (e.g., Rydberg atoms \cite{bai2026multipartitecontrollednotgatesusing} or spin qubits \cite{nguyen2025single}).

\subsection{Broader Context}
Finally, we reflect on the practical relevance of the task studied in this work. In this paper, we experimentally studied the contrived learning problem formalized in~\cite{gyurik2023exponential}, chosen precisely because a quantum advantage for it had been formally proven. Although we later argue that a broader range of more natural learning tasks with quantum inputs may also exhibit such advantages, we first show that the specific task considered in this work already has a direct practical application for several reasons.  
Firstly, our protocol for the exact task considered here can serve as a benchmark for random-phase state preparation and coherent processing, as it allows verifying the circuit's output against a desired target with a single shot (albeit for a random $y$) \cite{huang2025certifying}. Secondly, the task itself can be viewed as a constrained structured-output prediction problem \cite{jiang2022constraint, tsochantaridis2005large}, in which the valid outputs $(y, b)$ must satisfy the relational constraint. From a physical perspective, it provides a method for observing many-body symmetries and correlations in random-phase states, with possible applications in frustrated magnetism, spin glasses, and many-body localization \cite{zhou2024efficient, anschuetz2025efficient}. In these contexts, our algorithm can be used to measure the response of a quantum system under an $\alpha$ perturbation, augmenting learning problems with coherently processed quantum data \cite{zhang2025quantumcomputationmoleculargeometry}.

Beyond the specific task considered in this work, we expect advantages of quantum protocols over MF strategies to appear in many other types of learning problems with quantum data as input. In particular, the ability to process quantum input states directly on a quantum device is likely to be beneficial even in learning tasks where the labels are not only samples drawn from the quantum states, but instead correspond, for example, to deterministic quantities computed from them although more research in this direction is needed.

In summary, by focusing on a specific learning problem, we have provided evidence for near-term quantum advantage in learning from noisy quantum data. We identified the noise regimes and hardware trade-offs where this advantage is most pronounced, demonstrating that the relevant comparison depends critically on the physical character of the noise rather than on a single abstract error rate. Most notably, our results indicate that even with substantial noise in the quantum data, the fully quantum protocol maintains a significant practical advantage over measure-first methods. This advantage persists when the data is degraded by the relaxation channel (observable above 30 qubits) as well as other noise channels (above 40 qubits). These findings offer concrete evidence for quantum advantage in learning tasks and pave the way for its demonstration on currently available hardware.

\section*{acknowledgments}
The authors acknowledge insightful discussions with Owidiusz Makuta and Marcello Benedetti. This work was supported by the Dutch National Growth Fund (NGF) as part of the Quantum Delta NL programme, which provided specific support for J.A.K. and E.V.N. O.D. acknowledges support from the MAZeLTof-Q project (Machine Assisted Zero-Knowledge Tune-up of Superconducting Qubits, Grant No. NGF.1582.22.031). This publication is part of the NWA research program “Research on Routes by Consortia (ORC),” financed by the Dutch Research Council (NWO), through the projects "Divide \& Quantum" (Project No. 1389.20.241) and "Quantum Inspire – the Dutch Quantum Computer in the Cloud" (Project No. NWA.1292.19.194). Additionally, this work was supported by the European Union’s Horizon Europe program through the ERC CoG BeMAIQuantum (Grant No. 101124342).

\section*{\label{sec:contrib}Data and code availability}
The data and code used to generate the figures can be found on GitHub at \href{https://github.com/jaq-lab/noisy-learning-advantage}{https://github.com/jaq-lab/noisy-learning-advantage}.
\onecolumngrid

\bibliography{references}

\appendix
\newpage

\section{Fully quantum protocol}
\label{app:quantum_analytics}

In this appendix, we derive the measurement statistics of the fully quantum (FQ) protocol, first in the ideal noiseless setting and then in a form suitable for the noise analysis used later in the text. We consider a register of \(n_q\) qubits labeled \(1, \dots, n_q\), with qubit \(n_q\) serving as the control qubit and the remaining \(n_q - 1\) qubits forming the target register. For a target-register bitstring \(y' \in \{0,1\}^{n_q-1}\), we define the associated full \(n_q\)-bit string \(y := (y', 0)\). Likewise, we write \(\alpha = (\alpha', 1)\), where \(\alpha' \in \{0,1\}^{n_q-1}\) specifies which target qubits are coupled to the control qubit.

\subsection{Input state}

The ideal input is the random phase state
\begin{equation}
\ket{\psi_f} = \frac{1}{\sqrt{2^{n_q}}} \sum_{k \in \{0,1\}^{n_q}} (-1)^{f(k)} \ket{k},
\end{equation}
where \(f:\{0,1\}^{n_q} \to \{0,1\}\) is a Boolean function. It is convenient to introduce the phase coefficients \(c_k := (-1)^{f(k)}\). In terms of the target register \(y' \in \{0,1\}^{n_q-1}\) and control bit \(b \in \{0,1\}\), the state can be decomposed as
\begin{equation}
\ket{\psi_f} = \frac{1}{\sqrt{2^{n_q}}} \sum_{y' \in \{0,1\}^{n_q-1}} \left( c_{(y',0)}\ket{y',0} + c_{(y',1)}\ket{y',1} \right).
\end{equation}

\subsection{Measurement circuit}

The measurement circuit \(U(\alpha)\) consists of a layer of CNOT gates, all controlled on the final qubit \(n_q\), followed by a Hadamard gate on that same qubit (See Fig.~\ref{fig:circuit}). Let \(S_T := \{ j \in \{1, \dots, n_q-1\} : \alpha'_j = 1 \}\) denote the set of active target qubits. The unitary operation is defined as
\begin{equation}
U(\alpha) = H_{n_q} \prod_{j \in S_T} CX_{n_q \to j},
\end{equation}
where \(CX_{n_q \to j}\) denotes a CNOT with control \(n_q\) and target \(j\). The CNOT layer leaves the component \(\ket{y', 0}\) unchanged and maps \(\ket{y', 1} \mapsto \ket{y' \oplus \alpha', 1}\). Applying this transformation to \(\ket{\psi_f}\) yields
\begin{equation}
\left( \prod_{j \in S_T} CX_{n_q \to j} \right) \ket{\psi_f} = \frac{1}{\sqrt{2^{n_q}}} \sum_{y' \in \{0,1\}^{n_q-1}} \left( c_{(y',0)}\ket{y',0} + c_{(y' \oplus \alpha', 1)}\ket{y',1} \right).
\end{equation}
Using the identities \(y = (y', 0)\) and \(y \oplus \alpha = (y' \oplus \alpha', 1)\), this expression simplifies to
\begin{equation}
\frac{1}{\sqrt{2^{n_q}}} \sum_{y' \in \{0,1\}^{n_q-1}} \left( c_y\ket{y',0} + c_{y\oplus\alpha}\ket{y',1} \right).
\end{equation}
Applying the final Hadamard gate to the control qubit results in the pre-measurement state:
\begin{equation}
U(\alpha)\ket{\psi_f} = \frac{1}{\sqrt{2^{n_q+1}}} \sum_{y' \in \{0,1\}^{n_q-1}} \Big[ \bigl(c_y + c_{y\oplus\alpha}\bigr)\ket{y',0} + \bigl(c_y - c_{y\oplus\alpha}\bigr)\ket{y',1} \Big].
\end{equation}

\subsection{Measurement probabilities in the noiseless setting}

The probability of observing the computational-basis outcome \(\ket{y', b}\), with \(b \in \{0,1\}\), is
\begin{equation}
\label{eq:prob_yb_app}
P(y', b \mid f) = \left| \bra{y',b} U(\alpha) \ket{\psi_f} \right|^2 = \frac{\left| c_y + (-1)^b c_{y\oplus\alpha} \right|^2}{2^{n_q+1}}.
\end{equation}
Expanding the modulus gives
\begin{equation}
P(y', b \mid f) = \frac{ |c_y|^2 + |c_{y\oplus\alpha}|^2 + (-1)^b 2 \Re[c_y^* c_{y\oplus\alpha}] }{2^{n_q+1}}.
\end{equation}
In the ideal setting, \(c_k \in \{\pm 1\}\), so \(|c_y|^2 = |c_{y\oplus\alpha}|^2 = 1\). The interference term becomes
\begin{equation}
c_y^* c_{y\oplus\alpha} = (-1)^{f(y)} (-1)^{f(y\oplus\alpha)} = (-1)^{f(y) \oplus f(y\oplus\alpha)}.
\end{equation}
It follows that
\begin{equation}
P(y', b \mid f) = \begin{cases} 2^{-(n_q-1)}, & b = f(y) \oplus f(y\oplus\alpha), \\ 0, & \text{otherwise}. \end{cases}
\end{equation}
Thus, in the noiseless protocol, the output bit \(b\) always matches the desired parity, while the target-register outcome \(y'\) is uniformly distributed.

\subsection{Relation to accuracy}

The accuracy \(A_Q(\alpha \mid f)\) of the fully quantum protocol is the total probability of obtaining any \(y'\) together with its corresponding correct output bit \(b_{\text{target}}(y) := f(y) \oplus f(y\oplus\alpha)\):
\begin{equation}
A_Q(\alpha \mid f) = \sum_{y' \in \{0,1\}^{n_q-1}} P\bigl(y', b_{\text{target}}(y) \mid f\bigr) = \sum_{y' \in \{0,1\}^{n_q-1}} 2^{-(n_q-1)} = 1.
\end{equation}
Hence, the ideal protocol succeeds deterministically.

\subsection{Extension to noisy states}

To analyze the impact of noise, let \(\tilde{\rho}^{(f)}\) denote the noisy input state immediately before the measurement circuit. The measurement probability generalizes to
\begin{equation}
P(y', b \mid f) = \bra{y',b} U(\alpha) \tilde{\rho}^{(f)} U(\alpha)^\dagger \ket{y',b} = \frac{ \tilde{\rho}^{(f)}_{y,y} + \tilde{\rho}^{(f)}_{y\oplus\alpha, y\oplus\alpha} + (-1)^b 2 \Re[\tilde{\rho}^{(f)}_{y, y\oplus\alpha}] }{2}.
\end{equation}
Summing the diagonal elements over all $y'$ yields exactly half the trace of the density matrix, $\frac{1}{2}\mathrm{Tr}(\tilde{\rho}^{(f)}) = \frac{1}{2}$. The corresponding accuracy therefore evaluates to
\begin{equation}
A_Q(\alpha \mid f) = \frac{1}{2} + \frac{1}{2} V_Q(\alpha \mid f),
\end{equation}
where the visibility \(V_Q\) is defined as
\begin{equation}
V_Q(\alpha \mid f) := 2 \sum_{y' \in \{0,1\}^{n_q-1}} (-1)^{b_{\text{target}}(y)} \Re[\tilde{\rho}^{(f)}_{y, y\oplus\alpha}].
\end{equation}
For the preparation-noise models analyzed in this work, the noise channels suppress the off-diagonal elements by strictly positive, real attenuation factors without inducing phase rotations. Consequently, the coherences preserve the sign structure of the noiseless target parity, allowing us to absorb the alternating sign and write the visibility as a sum of absolute values:
\begin{equation}
V_Q(\alpha \mid f) = 2 \sum_{y' \in {0,1}^{n_q-1}} \left| \Re[\tilde{\rho}^{(f)}_{y, y\oplus\alpha}] \right|.
\end{equation}
In Appendix~\ref{app:prep_noise}, we analyze how this visibility is modified by different noise channels during random phase-state preparation. Averaging this visibility over all possible Boolean functions $f$ directly yields the expected preparation visibility $V_p(n_q, \alpha)$ introduced in the main text.

\newpage

\section{Efficient quantum protocol simulator}
\label{app:quantum_sim}

This appendix describes the simulation framework used to estimate the accuracy of the fully quantum (FQ) protocol in the presence of state-preparation noise and hardware-level execution noise. The framework combines a quantum-trajectory treatment of the preparation channel with a hardware-aware circuit simulation of the subsequent measurement circuit. This avoids the \(4^{n_q}\) memory overhead of full density-matrix propagation, while retaining the exact state-vector evolution within each sampled trajectory.

A key optimization is \emph{trajectory decimation}. Under stochastic unraveling of the preparation channel, each sampled trajectory is assigned a discrete jump code that records the Kraus index selected for each qubit. 
For a fixed input state, channel, and qubit order, trajectories with the same jump code yield the same normalized post-noise state. 
We therefore group trajectories by jump code and simulate the downstream noisy circuit only once per group. The final protocol accuracy is then recovered by weighting each group contribution by its multiplicity in the original trajectory ensemble.

\subsection{Simulation pipeline}

Our goal is to estimate the average accuracy of the fully quantum protocol for a preparation-noise channel \(\mathcal{E}\) and a hardware noise model $M$ (parameterized by connectivity, $T_1$, $T_2$, and gate fidelities $F_{1q}, F_{2q}$ as summarized in Table~\ref{tab:qc_comparison_final}). 
For each pair \((\mathcal{E}, M)\), we sample \(N_f\) random Boolean functions \(f\), and for each function we estimate the protocol accuracy by Monte Carlo simulation.

For a fixed function \(f_i\), the simulation proceeds as follows.

\begin{enumerate}
    \item \textbf{State preparation.} Construct the ideal input state \(\ket{\psi_{f_i}}\) associated with the Boolean function \(f_i\).

    \item \textbf{Trajectory sampling.} Sample \(N_{\mathrm{mcs}}\) independent pure-state trajectories of the preparation channel \(\mathcal{E}\). We assume that \(\mathcal{E}\) factorizes across qubits,
    \begin{equation}
        \mathcal{E} = \bigotimes_{k=1}^{n_q} \mathcal{E}_k,
    \end{equation}
    where each single-qubit channel \(\mathcal{E}_k\) is represented by Kraus operators \(\{K^{(k)}_j\}\). Along each trajectory, one Kraus operator is sampled on each qubit according to the usual trajectory probabilities, yielding a normalized output state.

    \item \textbf{Decimation.} Each sampled trajectory is assigned a discrete jump code that records the realized Kraus indices. Trajectories with identical jump codes are grouped together, producing \(N_u\) unique post-noise states \(\{\ket{\psi_{i,u}}\}_{u=1}^{N_u}\) with multiplicities \(\{m_u\}_{u=1}^{N_u}\), where \(\sum_{u=1}^{N_u} m_u = N_{\mathrm{mcs}}\).

    \item \textbf{Hardware-aware circuit execution.} For each unique state \(\ket{\psi_{i,u}}\), we simulate the fully quantum measurement circuit, including transpilation to the target device connectivity and the application of gate and idle-noise channels specified by \(M\). The noisy circuit is sampled \(N_{\mathrm{shots}}\) times.

    \item \textbf{Decoding.} For each shot, the simulator returns an outcome \((y'_{i,u,s}, b_{i,u,s})\), where \(y'_{i,u,s} \in \{0,1\}^{n_q-1}\) is the measured target-register string and \(b_{i,u,s} \in \{0,1\}\) is the measured control bit. We define the corresponding full string \(y_{i,u,s} := (y'_{i,u,s}, 0)\), and evaluate whether the relational condition is satisfied. The empirical accuracy for unique state \(u\) is
    \begin{equation}
        a_{i,u}
        =
        \frac{1}{N_{\mathrm{shots}}}
        \sum_{s=1}^{N_{\mathrm{shots}}}
        \mathds{1}\!\left[
            f_i(y_{i,u,s}) \oplus f_i(y_{i,u,s}\oplus\alpha)
            =
            b_{i,u,s}
        \right].
    \end{equation}

    \item \textbf{Recombination.} The protocol accuracy for function \(f_i\) is estimated by the multiplicity-weighted average
    \begin{equation}
        \hat{A}_{Q,i}(\alpha \mid \mathcal{E}, M)
        =
        \frac{1}{N_{\mathrm{mcs}}}
        \sum_{u=1}^{N_u} m_u\, a_{i,u}.
    \end{equation}

    \item \textbf{Averaging over functions.} The reported accuracy is obtained by averaging over the sampled Boolean functions:
    \begin{equation}
        \hat{A}_Q(\alpha \mid \mathcal{E}, M)
        =
        \frac{1}{N_f}
        \sum_{i=1}^{N_f}
        \hat{A}_{Q,i}(\alpha \mid \mathcal{E}, M).
    \end{equation}
\end{enumerate}

In practice, the number of unique states \(N_u\) is typically much smaller than \(N_{\mathrm{mcs}}\), so the decimation step substantially reduces the number of expensive noisy-circuit simulations.

The full procedure is summarized in Algorithm~\ref{alg:qmc_fq}.

\begin{algorithm}[H]
\caption{Quantum-trajectory simulation of the fully quantum protocol}
\label{alg:qmc_fq}
\begin{algorithmic}[1]

\Statex \textbf{Input:} Boolean functions \(\{f_1,\dots,f_{N_f}\}\), single-qubit Kraus operators \(\{K^{(k)}_j\}\), hardware noise model \(M\), number of trajectories \(N_{\mathrm{mcs}}\), number of circuit shots \(N_{\mathrm{shots}}\)
\Statex \textbf{Output:} Estimated protocol accuracy \(\hat{A}_Q\)

\For{$i = 1,\dots,N_f$}
    \For{$s = 1,\dots,N_{\mathrm{mcs}}$}
        \State \(\ket{\psi} \gets \ket{\psi_{f_i}}\)
        \State Initialize jump code \(C_s \gets []\)
        \For{$k = 1,\dots,n_q$}
            \State Compute \(\ket{\phi_j} = K^{(k)}_j \ket{\psi}\)
            \State Compute \(p_j = \langle \phi_j \mid \phi_j \rangle\)
            \State Sample Kraus index \(\hat{j}\) with probability \(p_j\)
            \State Update \(\ket{\psi} \gets \ket{\phi_{\hat{j}}}/\sqrt{p_{\hat{j}}}\)
            \State Append \(\hat{j}\) to \(C_s\)
        \EndFor
        \State Store final state \(\ket{\psi_{i,s}} \gets \ket{\psi}\)
    \EndFor

    \State Group the sampled trajectories by jump code \(C_s\)
    \Statex For the local product channels considered here, each jump code determines a unique
    \Statex normalized post-noise state. Let \(\{\ket{\psi_{i,u}}\}_{u=1}^{N_u}\) denote the resulting
    \Statex distinct groups, with multiplicities \(\{m_u\}_{u=1}^{N_u}\).

    \For{$u = 1,\dots,N_u$}
        \State Simulate the transpiled noisy measurement circuit on input \(\ket{\psi_{i,u}}\) under hardware model \(M\)
        \State Sample \(N_{\mathrm{shots}}\) measurement outcomes \((y'_{i,u,s}, b_{i,u,s})\)
        \State Decode the outcomes and compute \(a_{i,u}\)
    \EndFor

    \State \(\hat{A}_{Q,i} \gets \frac{1}{N_{\mathrm{mcs}}}\sum_{u=1}^{N_u} m_u a_{i,u}\)
\EndFor

\State \(\hat{A}_Q \gets \frac{1}{N_f}\sum_{i=1}^{N_f} \hat{A}_{Q,i}\)
\State \Return \(\hat{A}_Q\)

\end{algorithmic}
\end{algorithm}

\subsection{Hardware-aware transpilation}

We now describe how the measurement circuit is compiled to a realistic device model. The transpilation procedure is designed to minimize two-qubit gate overhead while respecting the target hardware's connectivity constraints. Throughout this work, we consider square-lattice connectivity, which is representative of superconducting-qubit architectures, although the same strategy applies more generally.

The transpilation consists of three steps.

\begin{enumerate}
    \item \textbf{Stochastic routing.} We transpile the circuit using Qiskit's SABRE layout and routing passes, repeated over multiple random seeds and trials (e.g., 200). The trial with the lowest two-qubit gate count is retained to minimize cumulative two-qubit gate error.

    \item \textbf{Schedule pruning.} Qiskit's ASAP scheduling inserts explicit \texttt{delay} instructions to synchronize qubits. We remove delays that do not contribute to physically relevant idling, such as terminal delays immediately preceding the measurement, to avoid artificially inflating decoherence.

    \item \textbf{Noise attachment.} The remaining delays are retained as explicit idle intervals. In the final simulator, these intervals are mapped to identity-equivalent gate placeholders, allowing calibrated idle-noise channels (e.g., relaxation channel) to be attached during execution, effectively simulating continuous $T_1/T_2$ decay with the discrete gate framework.
\end{enumerate}

\begin{figure}[h]
    \centering
    \includegraphics[width=\textwidth]{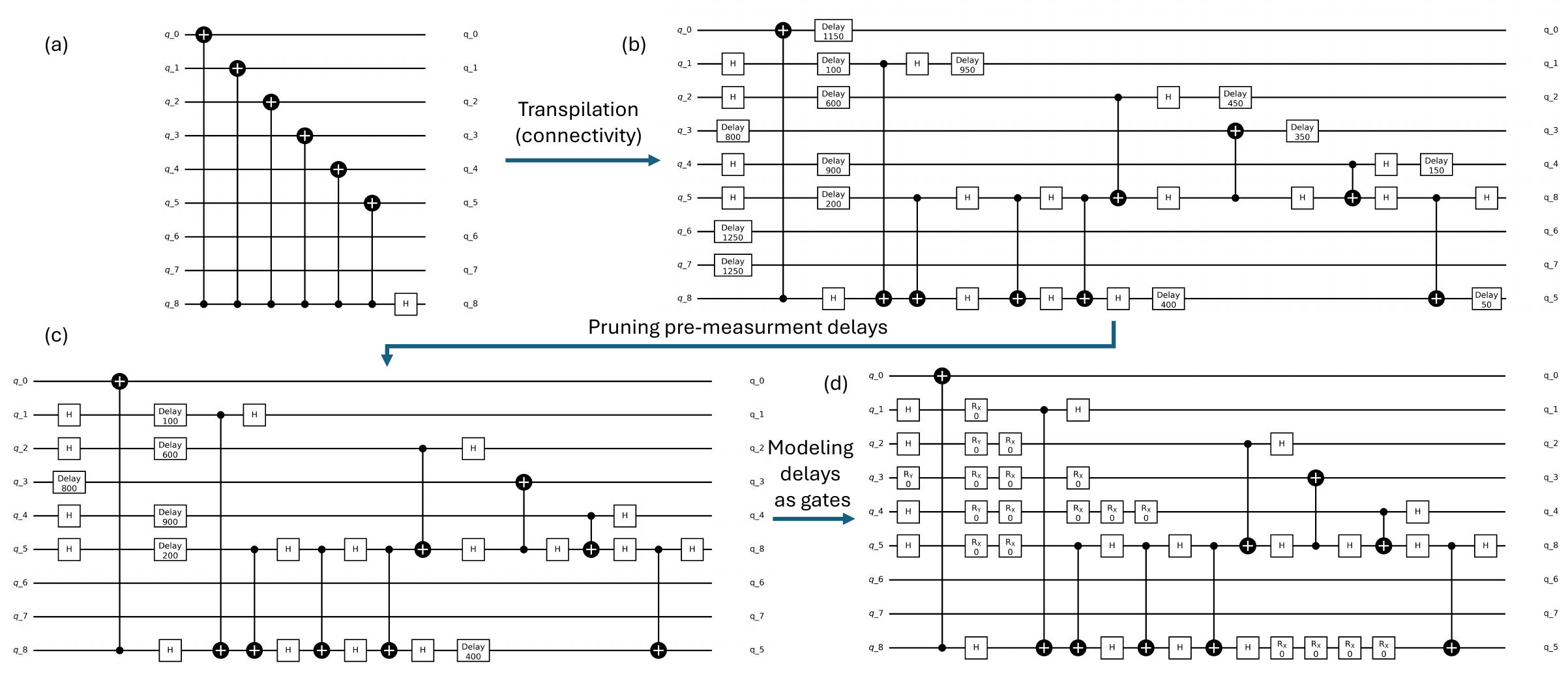}
    \caption{\textbf{Hardware-aware transpilation of the fully quantum measurement circuit.} The original circuit is routed to the target connectivity using stochastic transpilation, explicit idle periods are introduced by scheduling, and non-physical delays are pruned before calibrated noise channels are attached.}
    \label{fig:transpilation}
\end{figure}

The procedure is illustrated in Fig.~\ref{fig:transpilation}, which shows the original circuit, the routed circuit with inserted delays, the pruned schedule, and the final circuit used for noisy simulation.
The resulting circuit is simulated in TensorCircuit, which allows custom noise channels to be attached at the gate level.

\subsection{Transpilation statistics}
\begin{figure}[htb!]
    \centering
    \includegraphics[width=0.99\linewidth]{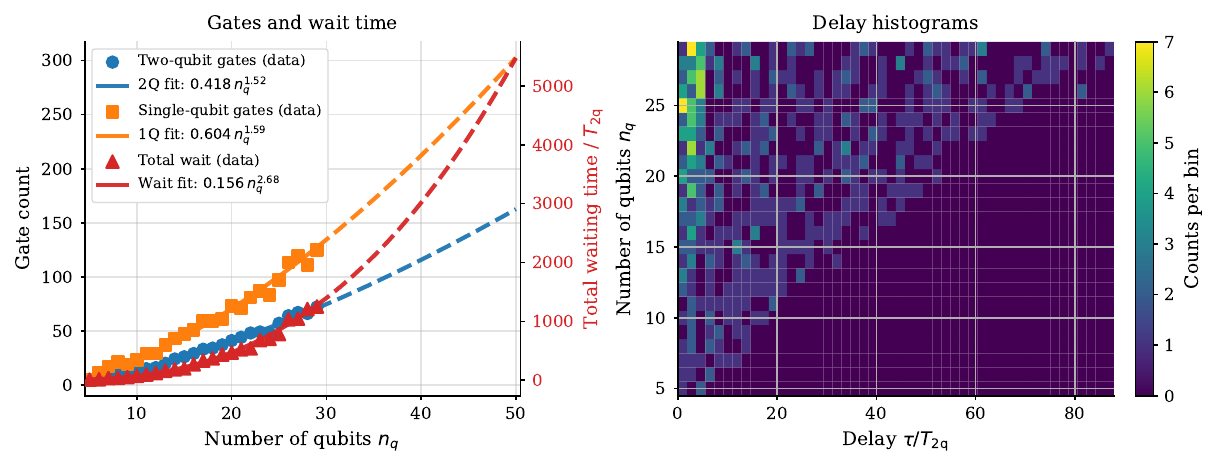}
    \caption{\textbf{Scaling of gate counts and idle times for a square-connectivity device.} \textbf{(Left)} The number of single-qubit gates (orange squares) and two-qubit gates (blue circles) plotted on the left axis, alongside the total waiting time normalized by the two-qubit gate duration $T_{2q}$ (red triangles) on the right axis, as a function of the number of qubits $n_q$. The dashed lines represent power-law fits to the data, demonstrating that while the gate counts scale approximately as $\mathcal{O}(n_q^{1.5})$, the total wait time incurred by routing overhead scales more steeply at $\mathcal{O}(n_q^{2.7})$. \textbf{(Right)} A 2D histogram illustrating the distribution of individual delay times $\tau/T_{2q}$ across varying system sizes $n_q$. The colorbar indicates the occurrence count per bin, showing that while short idle periods remain the most frequent, the maximum delay and overall spread of the distribution increase significantly as the system scales.}
    \label{fig:transpilation_stat}
\end{figure}

In Fig.~\ref{fig:transpilation_stat} we illustrate the scaling of the number of two-qubit gates and the resulting statistics of idle times for the square-connectivity device.

\subsection{Modeling idle decoherence}

Idle periods are modeled by attaching relaxation and dephasing channels to the explicit delay intervals produced by transpilation. Because TensorCircuit is a discrete-gate simulator, each delay of duration \(t\) is represented by a short sequence of identity-equivalent placeholder operations (modeled as zero-angle rotations $R_X(0), R_Y(0), R_Z(0)$), allowing the corresponding noise channels to be applied within the gate-based framework. 
These placeholder operations are used only to carry idle decoherence and are excluded from the gate-error model.
Very small delays (e.g., $\frac{t}{ T_{\text{idle}}} < 10^{-4}$ where $T_{\text{idle}}$ is the characteristic idle time constant of the device) are discarded when their duration is negligible relative to the relevant coherence times.

For an idle interval of duration \(t\), energy relaxation is described by an amplitude-damping channel with probability
\begin{equation}
    p_{1}(t) = 1-\exp(-t/T_1).
\end{equation}
Pure dephasing during the same interval is described by a phase-damping channel with probability
\begin{equation}
    p_{\phi}(t) = 1-\exp(-t/T_{\phi}),
\end{equation}
where \(T_{\phi}\) is the effective pure-dephasing time. If device calibration data are reported in terms of \(T_1\) and \(T_2\), we use the standard relation
\begin{equation}
    \frac{1}{T_{\phi}}
    =
    \max\!\left\{0,\frac{1}{T_2}-\frac{1}{2T_1}\right\}.
\end{equation}
The idle channel is then implemented as the composition of amplitude damping and phase damping over the same delay interval. While this approach does not model the low-frequency noise components of decoherence, it can be considered
equivalent to the effect of a Hahn echo sequence with perfect pulses, which is a common practical approximation for
simulating $T_1/T_2$ effects in discrete-gate simulators.

\subsection{Gate-error model}

In addition to idle decoherence, we attach depolarizing noise channels to the gates in the circuit. The depolarizing channel is parameterized as
\begin{equation}
    \mathcal{E}_{\mathrm{depol}}(\rho)
    =
    (1-\epsilon_{\mathrm{gate}})\rho
    +
    \epsilon_{\mathrm{gate}}\frac{I}{d},
\end{equation}
where \(d\) is the Hilbert-space dimension of the gate and \(\epsilon_{\mathrm{gate}}\) is the backend-specific depolarizing parameter.

Within TensorCircuit's \texttt{isotropicdepolarizingchannel} convention, the channel is parameterized such that the input probability relates to the average gate fidelity via the scaling factor $\epsilon_n = \frac{2^n+1}{2^n}(1-F)$. Consequently, we set the gate error parameters to
\begin{equation}
    \epsilon_{1q} = \frac{3}{2}(1-F_{1q}), \qquad \epsilon_{2q} = \frac{5}{4}(1-F_{2q}),
\end{equation}
for one- and two-qubit gates, respectively.

\subsection{Computational implementation}

The simulator is implemented in JAX and TensorCircuit and executed on GPU hardware. To accelerate the noisy-circuit evaluations, unique post-noise states are batched and propagated in parallel using vectorized execution. Because the state dimension still grows as \(2^{n_q}\), the implementation uses adaptive chunking: both the number of trajectories processed simultaneously and the number of unique states simulated in parallel are reduced as \(n_q\) increases. This allows intermediate-scale instances to be simulated without exceeding device memory.

For large shot counts, measurement sampling is performed iteratively in fixed-size chunks rather than by fully unrolling the computation graph. Together with batched evaluation of unique states, this keeps memory usage controlled while preserving efficient hardware utilization. Additional implementation details are available in the official GitHub repository at \href{https://github.com/jaq-lab/noisy-learning-advantage}{https://github.com/jaq-lab/noisy-learning-advantage}.

\newpage
\section{Numerical validation of the fully quantum protocol}
\label{app:quantum_numerics}

In this section, we perform numerical experiments to validate our method of simulating measurement circuits on realistic, noisy quantum hardware. We also verify our model for combining this circuit noise with state-preparation noise and readout errors, specifically validating the approximation:
\begin{equation}
V_Q(n_q,\alpha) \approx V_p(n_q,\alpha)V_m(|\alpha|)V_r(n_q)
\end{equation}

\subsection{Readout error}
First, we derive the readout error contribution using the pre-measurement state derived in Appendix~\ref{app:quantum_analytics}. We model the readout error as an independent bit-flip channel with probability $\epsilon_r$ on each qubit. In our relational learning task, the final qubit ($n_q$) encodes the target parity bit $b$, while the remaining $n_q - 1$ qubits encode the measurement string $y'$. 

\begin{figure}[htpb]
    \centering
    \includegraphics[width=0.6\linewidth]{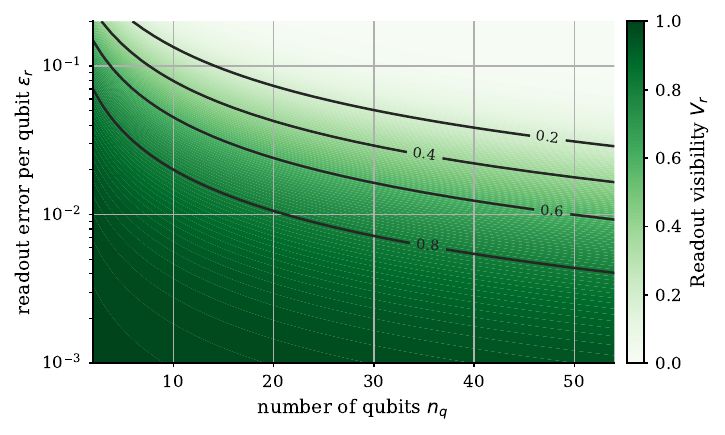}
    \caption{\textbf{Readout visibility} $V_r$ as a function of the number of qubits $n_q$ and the readout error per qubit $\epsilon_r$. The contour lines map the exponential decay of the signal visibility as the system size and per-qubit error rates increase.}
    \label{fig:readout_visibility}
\end{figure}

A readout error on the final qubit directly inverts the inferred parity, immediately reducing the visibility by a factor of $1 - 2\epsilon_r$. Conversely, an error on any of the first $n_q - 1$ qubits results in an incorrect string $y'$. Because the underlying Boolean function $f$ is uniformly random, evaluating the parity for an incorrect string yields a completely uncorrelated random bit. This effectively randomizes the outcome, contributing a survival factor of $1 - \epsilon_r$ for each of these qubits. Combining these effects yields the total readout visibility factor $V_r(n_q) = (1 - 2\epsilon_r)(1 - \epsilon_r)^{n_q-1}$. In Fig.~\ref{fig:readout_visibility}, we plot $V_r$ as a function of the qubit number $n_q$ and error rate $\epsilon_r$.

\subsection{Measurement circuit}
Next, we perform numerical simulations of the measurement circuit described in Appendix~\ref{app:quantum_sim}, assuming perfect initial quantum data ($V_p = 1$). We first compute the signal amplitude $V_m(|\alpha|, n_q)$ as a function of the size $|\alpha|$ and qubit number $n_q$. These simulations are executed for the three distinct hardware classes defined in Appendix~\ref{app:real_hardware}. 

In Fig.~\ref{fig:circuit_visibility}, we plot $V_m(|\alpha|, n_q)$ for the full-weight case $|\alpha| = n_q$ as a function of the system size $n_q$, using circles with error bars derived from averaging over random $f$ states with 1000 shots each. To manage computational complexity, we scale the number of averaged $f$ states based on system size: 1000 states (for $n_q \leq 11$), 50 states (for $n_q \leq 16$), 20 states (for $n_q \leq 20$)and 10 states (for $n_q > 20$). We employ full density-matrix simulations for $n_q \leq 10$ and a Monte Carlo trajectory simulator for larger systems. Each curve is fitted to a stretched-exponential decay $V_m(|\alpha|) = A \exp(-c|\alpha|^\beta)$. Minor deviations from the fit can be attributed to finite-size effects and hardware-specific circuit transpilation overhead.

\begin{table}[h]
\centering
\caption{Fitted parameters for the measurement circuit visibility $V_m(|\alpha|)$ using the stretched-exponential model $A \exp(-c |\alpha|^\beta)$ with $A=1$. Devices A, B, and C correspond to the hardware archetypes defined in Appendix~\ref{app:real_hardware}.}
\label{tab:fit_parameters}
\renewcommand{\arraystretch}{1.3}
\begin{tabular}{lcc}
\hline\hline
\textbf{Device} & \textbf{Coefficient ($c$)} & \textbf{Stretch Exponent ($\beta$)} \\ \hline
A  & $0.00851$ & $1.1477$ \\
B  & $0.00032$ & $2.1760$ \\
C  & $0.00342$ & $2.1803$ \\ \hline\hline
\end{tabular}
\end{table}

\begin{figure}[htpb]
    \centering
    \includegraphics[width=0.9\linewidth]{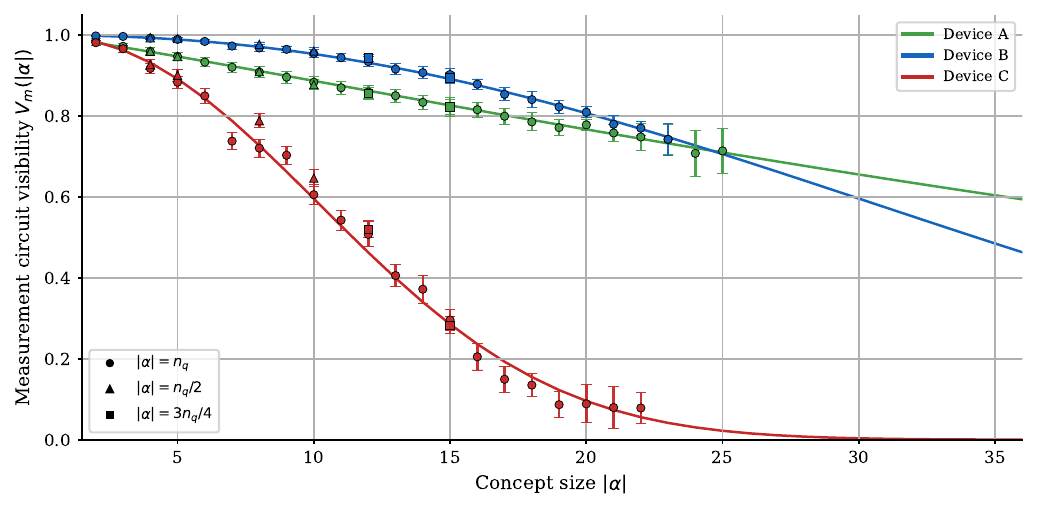}
    \caption{\textbf{Measurement circuit visibility} $V_m(|\alpha|)$ as a function of the size $|\alpha|$ which labels the target concepts for Devices A, B, and C. Solid lines represent the stretched-exponential fits for $|\alpha| = n_q$ (circles). Markers represent different relative sizes: $|\alpha| = n_q$ (circles), $|\alpha| = n_q/2$ (triangles), and $|\alpha| = 3n_q/4$ (squares).}
    \label{fig:circuit_visibility}
\end{figure}

On the same plot, we overlay a selection of points for $V_m(|\alpha| = n_q/2, n_q)$ (triangles) and $V_m(|\alpha| = 3n_q/4, n_q)$ (squares). By observing how closely these points align with the best-fit line, we confirm that the possibility of measuring passive qubits immediately after state preparation makes the signal amplitude only weakly dependent on the total qubit number. Instead, it is heavily dominated by the number of active CNOT gates, $|\alpha|-1$. For Device C, we observe that these points remain slightly above the best-fit line. This can be attributed to the fact that circuit transpilation at a fixed $|\alpha|$ becomes more efficient when distributed across a larger total number of qubits $n_q$. This effect is less pronounced at larger $|\alpha|$, confirming our working assumption that $V_m(|\alpha|, n_q) \approx V_m(|\alpha|)$.

\subsection{Combining the noise sources}

\begin{figure}[htb!]
    \centering
    \includegraphics[width=\linewidth]{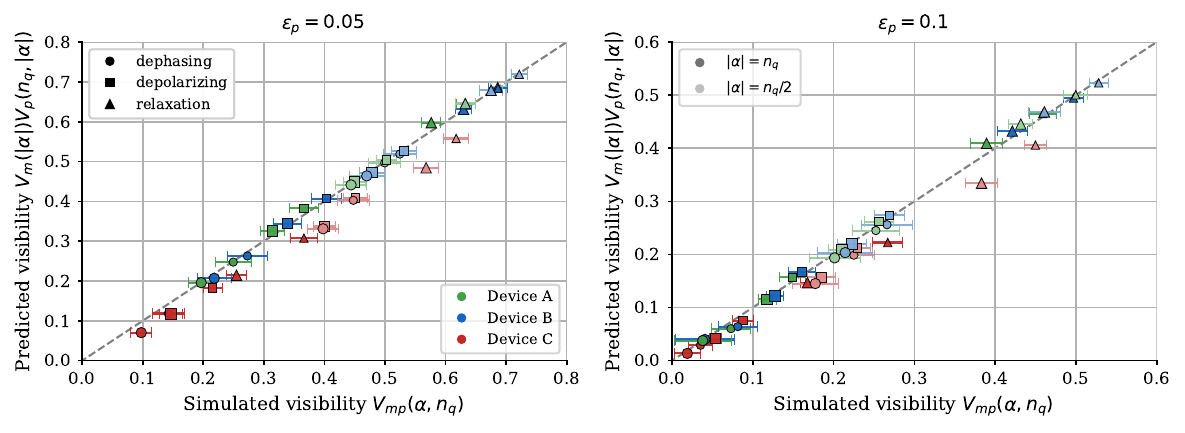}
    \caption{\textbf{Validation of the combined noise factorization.} The scatter plots compare the predicted visibility against the fully simulated visibility $V_{mp}(\alpha, n_q)$ for state preparation noise amplitudes $\epsilon_p = 0.05$ (left) and $\epsilon_p = 0.1$ (right). Different noise channels are distinguished by marker shape (circle: dephasing, square: depolarizing, triangle: relaxation), devices by color (green: A, yellow: B, red: C), and the concept class size by shading (dark: $|\alpha| = n_q$, light: $|\alpha| = n_q/2$). The dashed line indicates perfect agreement.}
    \label{fig:combined_noise}
\end{figure}

Finally, we synthesize the individual noise sources to verify that the factorized approximation $V_Q(n_q,\alpha) \approx V_p(n_q,\alpha)V_m(|\alpha|)V_r(n_q)$ holds. Utilizing the full pipeline from Appendix~\ref{app:quantum_sim}, we compute exact points at $n_q \in \{12, 14\}$ and $|\alpha| \in \{n_q/2, n_q\}$ across a selection of devices and noise channels affecting quantum data. 
In Fig.~\ref{fig:combined_noise}, we compare these full simulations against the approximate formula. The plot distinguishes different noise channels by symbol shape, devices by color, qubit numbers $n_q$ by marker size, and concept sizes ($|\alpha| = n_q$ versus $|\alpha| = n_q/2$) by color intensity. We observe a slight underestimation of the true visibility, which is particularly visible for Device C. This can be traced back to the minor discrepancy between the best-fit line and the empirical points in the $V_m \approx 0.7$ regime. However, this yields an absolute error no larger than 0.1. Crucially, because this approximation systematically underestimates the true visibility, it provides a strictly conservative bound on any resulting quantum advantage.

\newpage
\,
\newpage
\newpage
\section{Preparation noise for random phase states}
\label{app:prep_noise}

In this appendix, we derive the effect of local preparation noise on the coherences that determine the success probability of the fully quantum (FQ) protocol. We consider the random phase states introduced in Appendix~\ref{app:quantum_analytics},
\begin{equation}
    \ket{\psi_f} = \frac{1}{\sqrt{2^{n_q}}} \sum_{k\in\{0,1\}^{n_q}} c_k \ket{k}, \qquad c_k := (-1)^{f(k)} \in \{\pm1\},
\end{equation}
where \(f\) is a uniformly random Boolean function. The corresponding ideal density matrix is
\begin{equation}
    \rho^{(f)} = \frac{1}{2^{n_q}} \sum_{k,k'\in\{0,1\}^{n_q}} c_k c_{k'} \ketbra{k}{k'}.
\end{equation}

Our interest lies in the matrix elements \(\rho^{(f)}_{y,y\oplus\alpha}\), as these are the coherences that enter the visibility \(V_Q(\alpha\mid f)\) and determine the accuracy of the fully quantum protocol. We assume that preparation noise acts independently on each qubit as a local channel:
\begin{equation}
    \mathcal{E} = \bigotimes_{q=1}^{n_q} \mathcal{E}_q,
\end{equation}
so that the damping of the relevant coherence factorizes across the register.

\subsection{Active and passive qubits}

For a fixed concept bitstring \(\alpha \in \{0,1\}^{n_q}\), the coherence \(\rho^{(f)}_{y,y\oplus\alpha}\) involves a bit-flip on qubit \(i\) precisely when \(\alpha_i=1\). This motivates the decomposition of the register into:
\begin{equation}
    \mathcal{A}(\alpha) := \{ i : \alpha_i = 1 \}, \qquad \mathcal{P}(\alpha) := \{ i : \alpha_i = 0 \},
\end{equation}
with \(|\mathcal{A}| = |\alpha|\). We refer to qubits in \(\mathcal{A}\) as \emph{active} and those in \(\mathcal{P}\) as \emph{passive}. For passive qubits, it is useful (especially for amplitude damping) to distinguish whether the corresponding bit value in \(y\) is \(0\) or \(1\):
\begin{equation}
    \mathcal{P}_0(y,\alpha) := \{ i \in \mathcal{P}(\alpha) : y_i=0 \}, \qquad \mathcal{P}_1(y,\alpha) := \{ i \in \mathcal{P}(\alpha) : y_i=1 \}.
\end{equation}
These sets partition the passive qubits such that \(|\mathcal{P}_0(y,\alpha)| + |\mathcal{P}_1(y,\alpha)| = n_q - |\alpha|\). By the convention established in the main text, the final qubit $n_q$ is always active (\(\alpha_{n_q}=1\)).

\subsection{General structure of the damping}

Let the noisy state obtained from \(\rho^{(f)}\) be \(\tilde{\rho}^{(f)} = \mathcal{E}(\rho^{(f)})\). For the coherence of interest, we may write:
\begin{equation}
    \tilde{\rho}^{(f)}_{y,y\oplus\alpha} = \gamma_{y,\alpha}\,\rho^{(f)}_{y,y\oplus\alpha} + \kappa^{(f)}_{y,\alpha},
\end{equation}
where \(\gamma_{y,\alpha} \ge 0\) collects the contributions of the local Kraus components that preserve the index pair \((y,y\oplus\alpha)\), while the "feeding terms" \(\kappa^{(f)}_{y,\alpha}\) arise from Kraus components that map other coherences into the same matrix element.

For random phase states, these feeding terms do not contribute to the ensemble-averaged visibility. Any Kraus term that maps a different coherence \(\rho^{(f)}_{k,k'}\) into \(\tilde{\rho}^{(f)}_{y,y\oplus\alpha}\) contributes a phase factor \(c_k c_{k'}\). Because the \(c_k\) are independent random variables taking values in $\{+1, -1\}$ with equal probability for a uniformly random \(f\), their ensemble average vanishes. Thus, only the index-preserving part of the channel contributes to the average performance. Because the noise channel factorizes across qubits, the damping factor also factorizes:
\begin{equation}
\gamma_{y,\alpha} = \prod_{i\in\mathcal{A}(\alpha)} \gamma^{(i)}_{\mathrm{act}} \prod_{i\in\mathcal{P}_0(y,\alpha)} \gamma^{(i)}_{\mathrm{P0}} \prod_{i\in\mathcal{P}_1(y,\alpha)} \gamma^{(i)}_{\mathrm{P1}}.
\end{equation}

\subsection{Dephasing}

The local dephasing channel is \(\mathcal{D}_{\mathrm{deph}}(\rho) = (1-\epsilon_p)\rho + \epsilon_p Z\rho Z\). On an active qubit, the relevant local operator is \(\ketbra{0}{1}\) or \(\ketbra{1}{0}\), both of which acquire a minus sign under conjugation by \(Z\). Thus, \(\gamma_{\mathrm{act}}^{\mathrm{(deph)}} = 1-2\epsilon_p\). On a passive qubit, the local operator is diagonal (\(\ketbra{0}{0}\) or \(\ketbra{1}{1}\)) and is unchanged by \(Z\). Therefore, \(\gamma_{\mathrm{P0}}^{\mathrm{(deph)}} = \gamma_{\mathrm{P1}}^{\mathrm{(deph)}} = 1\). The total damping factor is:
\begin{equation}
    \gamma^{\mathrm{(deph)}}_{y,\alpha} = (1-2\epsilon_p)^{|\alpha|},
\end{equation}
which is independent of \(y\).

\subsection{Depolarizing noise}

The single-qubit depolarizing channel is \(\mathcal{D}_{\mathrm{depol}}(\rho) = (1-\epsilon_p)\rho + \frac{\epsilon_p}{3}(X\rho X + Y\rho Y + Z\rho Z)\). For an active qubit, the \(X\)- and \(Y\)-terms map \(\ketbra{0}{1}\) to \(\ketbra{1}{0}\) (feeding terms that average to zero), while the \(Z\)-term contributes a minus sign. The surviving factor is:
\begin{equation}
    \gamma_{\mathrm{act}}^{\mathrm{(depol)}} = (1-\epsilon_p) - \frac{\epsilon_p}{3} = 1-\frac{4\epsilon_p}{3}.
\end{equation}
For a passive qubit, the identity and \(Z\)-terms preserve the diagonal, while \(X\) and \(Y\) feed into off-diagonals. Therefore:
\begin{equation}
    \gamma_{\mathrm{P0}}^{\mathrm{(depol)}} = \gamma_{\mathrm{P1}}^{\mathrm{(depol)}} = (1-\epsilon_p) + \frac{\epsilon_p}{3} = 1-\frac{2\epsilon_p}{3}.
\end{equation}
The total damping factor is:
\begin{equation}
    \gamma^{\mathrm{(depol)}}_{y,\alpha} = \left(1-\frac{4\epsilon_p}{3}\right)^{|\alpha|} \left(1-\frac{2\epsilon_p}{3}\right)^{n_q-|\alpha|}.
\end{equation}

\subsection{Thermal relaxation (Amplitude damping)}

The Kraus operators for zero-temperature amplitude damping are \(K_0 = \ketbra{0}{0} + \sqrt{1-\epsilon_p}\ketbra{1}{1}\) and \(K_1 = \sqrt{\epsilon_p}\ketbra{0}{1}\). For an active qubit, the only index-preserving contribution comes from \(K_0\), giving \(K_0 \ketbra{0}{1} K_0^\dagger = \sqrt{1-\epsilon_p}\ketbra{0}{1}\). Hence, \(\gamma_{\mathrm{act}}^{\mathrm{(rel)}} = \sqrt{1-\epsilon_p}\). For a passive qubit in \(\ketbra{0}{0}\), \(\gamma_{\mathrm{P0}}^{\mathrm{(rel)}} = 1\), while for \(\ketbra{1}{1}\), the survival contribution yields \(\gamma_{\mathrm{P1}}^{\mathrm{(rel)}} = 1-\epsilon_p\). Therefore, for fixed \(y\):
\begin{equation}
    \gamma^{\mathrm{(rel)}}_{y,\alpha} = (\sqrt{1-\epsilon_p})^{|\alpha|}(1-\epsilon_p)^{|\mathcal{P}_1(y,\alpha)|}.
\end{equation}
\begin{table}[h]
\caption{Effective single-qubit damping factors governing the coherence \(\tilde{\rho}_{y,y\oplus\alpha}\) under local preparation noise.}
\centering
\renewcommand{\arraystretch}{1.4}
\begin{tabular}{lccc}
\hline\hline
Channel & Active (\(\mathcal{A}\)) & Passive 0 (\(\mathcal{P}_0\)) & Passive 1 (\(\mathcal{P}_1\)) \\ \hline
Dephasing & \(1-2\epsilon_p\) & \(1\) & \(1\) \\
Depolarizing & \(1-\frac{4\epsilon_p}{3}\) & \(1-\frac{2\epsilon_p}{3}\) & \(1-\frac{2\epsilon_p}{3}\) \\
Relaxation & \(\sqrt{1-\epsilon_p}\) & \(1\) & \(1-\epsilon_p\) \\ \hline\hline
\end{tabular}
\label{tab:damping_factors}
\end{table}
\subsection{Expected protocol accuracy}
\label{app:accuracy}

The analytical damping factors are summarized in Table~\ref{tab:damping_factors}. From Appendix~\ref{app:quantum_analytics}, the ensemble-averaged accuracy is:
\begin{equation}
    A_Q(\alpha) := \mathbb{E}_f[A_Q(\alpha\mid f)] = \frac{1}{2}\bigl(1+\bar{\gamma}(\alpha)\bigr),
\end{equation}
where \(\bar{\gamma}(\alpha) = \frac{1}{2^{n_q-1}} \sum_{y'} \gamma_{y,\alpha}\). For a specific Boolean function $f$, the ideal coherence is $\rho^{(f)}_{y,y\oplus\alpha} = \frac{1}{2^{n_q}} c_y c_{y\oplus\alpha}$, which has an absolute value of exactly $\frac{1}{2^{n_q}}$. As established in Appendix~\ref{app:quantum_analytics}, the visibility is given by $V_Q(\alpha \mid f) = 2 \sum_{y'} |\Re[\tilde{\rho}^{(f)}_{y, y\oplus\alpha}]|$. Substituting our damped coherence $\tilde{\rho}^{(f)}_{y,y\oplus\alpha} = \gamma_{y,\alpha}\,\rho^{(f)}_{y,y\oplus\alpha} + \kappa^{(f)}_{y,\alpha}$ into this expression and taking the ensemble average over all uniformly random functions $f$, the uncorrelated feeding terms $\kappa^{(f)}_{y,\alpha}$ average to zero. The expected visibility therefore simplifies precisely to the average damping factor:
\begin{equation}
    \mathbb{E}f[V_Q(\alpha \mid f)] = 2 \sum{y' \in {0,1}^{n_q-1}} \frac{\gamma_{y,\alpha}}{2^{n_q}} = \frac{1}{2^{n_q-1}} \sum_{y'} \gamma_{y,\alpha} := \bar{\gamma}(\alpha).
\end{equation}

\begin{figure}[htb!]
    \centering
    \includegraphics[width=0.9\columnwidth]{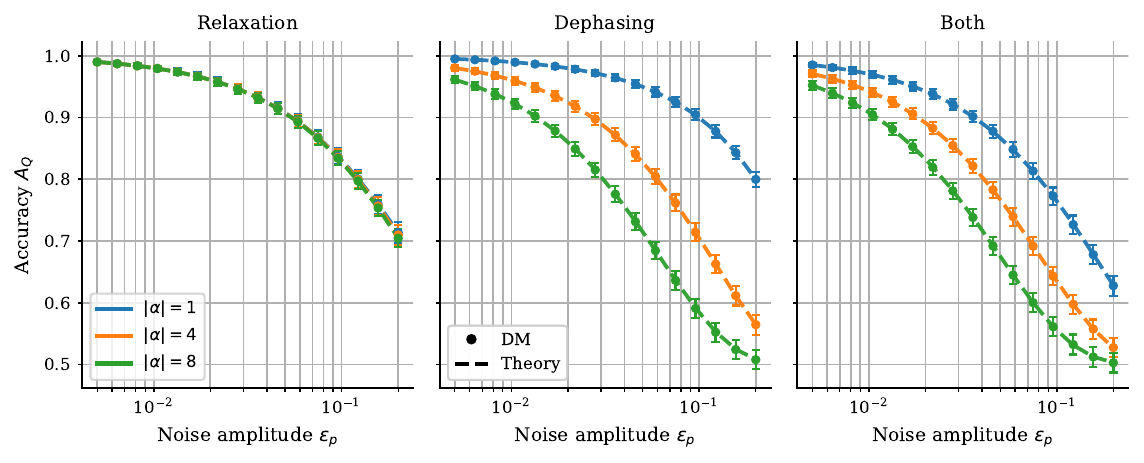}
    \caption{\textbf{Numerical validation of the analytical accuracy formulas in Table~\ref{tab:accuracy}}. The third panel illustrates the scenario where both relaxation and dephasing channels are applied to the initial state.}
    \label{fig:numerical_validation}
\end{figure}

For dephasing and depolarizing noise, \(\bar{\gamma}(\alpha)\) corresponds exactly to the $y$-independent factors derived above. For amplitude damping, since \(y\) is uniformly distributed, each passive qubit is equally likely to contribute a factor of \(1\) or \(1-\epsilon_p\). The average passive factor is \(\frac{1}{2}(1) + \frac{1}{2}(1-\epsilon_p) = 1-\epsilon_p/2\). Using the approximation \(1-\epsilon_p/2 \approx \sqrt{1-\epsilon_p}\), we find:
\begin{equation}
    \bar{\gamma}_{\mathrm{rel}}(\alpha) = (\sqrt{1-\epsilon_p})^{|\alpha|} \left(1-\frac{\epsilon_p}{2}\right)^{n_q-|\alpha|} \approx (\sqrt{1-\epsilon_p})^{n_q}.
\end{equation}
These analytical predictions are summarized in Table~\ref{tab:accuracy} and validated in Fig.~\ref{fig:numerical_validation}.

\begin{table}[h!]
\caption{Analytical predictions for the average accuracy of the fully quantum protocol under local preparation noise.}
\label{tab:accuracy}
\centering
\renewcommand{\arraystretch}{1.8}
\begin{tabular}{lc}
\hline\hline
Channel & Expected accuracy \(A_Q(\alpha)\) \\ \hline
Dephasing & \(\frac{1}{2}\left[1+(1-2\epsilon_p)^{|\alpha|}\right]\) \\
Depolarizing & \(\frac{1}{2}\left[1+\left(1-\frac{4\epsilon_p}{3}\right)^{|\alpha|} \left(1-\frac{2\epsilon_p}{3}\right)^{n_q-|\alpha|}\right]\) \\
Relaxation & \(\frac{1}{2}\left[1+(\sqrt{1-\epsilon_p})^{|\alpha|} \left(1-\frac{\epsilon_p}{2}\right)^{n_q-|\alpha|}\right]\) \\ \hline\hline
\end{tabular}
\end{table}

\newpage
\section{Relation between models of measurement circuit and realistic quantum architectures}
\label{app:real_hardware}

The fully quantum protocol relies on the measurement circuit $U(\alpha)$, which consists of a layer of CNOT gates controlled by the final qubit and targeting a specific active subset $\alpha$. Translating this fan-out structure to physical hardware, for instance, adapting to the limited connectivity shown in Fig.~\ref{fig:transpilation}, introduces distinct compilation overheads that fundamentally shape our simulated noise parameters.

\subsection{Comparison of the model devices}
As outlined in the main text, our numerical models explore distinct philosophies of quantum hardware design. Rather than perfectly mimicking specific commercial processors, Devices A, B, and C act as archetypes representing fundamental trade-offs in connectivity, gate fidelity, and coherence. 

Device A captures the paradigm of highly connected hardware. We assign it a conservative two-qubit gate fidelity of $99\%$, highlighting how all-to-all connectivity can dominate performance outcomes for heavy Hamming-weight concepts and larger qubit numbers, entirely bypassing the need for SWAP routing. In particular, for the circuit of interest, one-to-all connectivity allows the measurement circuit $U(\alpha)$ to be implemented with a depth of $\mathcal{O}(|\alpha|)$. Device B represents the highly optimized, local-connectivity approach. Here, exceptional gate fidelities ($99.9\%$) must compensate for the extensive SWAP networks required to execute fan-out operations. 

Device C bridges the gap to current reality, serving as a proxy for standard, cloud-accessible hardware or emerging, unoptimized scalable architectures. By sharing the $99\%$ gate fidelity of Device A but inheriting the square-lattice routing constraints of Device B, Device C allows us to identify the critical $n_q$ threshold where accuracy transitions from being gate-limited to being dominated by idling errors during long SWAP networks. Furthermore, demonstrating a resolvable quantum advantage on Device C indicates that the fully quantum protocol remains viable even on early-stage or standard-tier architectures.

We additionally use Devices B and C to isolate the influence of coherence times and dominant noise mechanisms. By setting the relevant quality factor $Q = T_{\mathrm{idle}}/T_{2q}$ to $2 \times 10^3$ for Device B and $2 \times 10^2$ for Device C, we observe how protocol performance degrades with an order-of-magnitude decrease in qubit quality. Furthermore, we shift the dominant idling error from $T_1$ thermal relaxation in Device B to $T_2$ dephasing in Device C. This distinction demonstrates that the fully quantum protocol is substantially more sensitive to $T_2$-type errors, directly mirroring the detrimental role of dephasing in the quantum data itself. 

Below, we detail five physical hardware platforms and discuss how their parameters motivate the theoretical constraints of our device archetypes.

\subsection{Trapped Ions}
Trapped ion architectures \cite{MosesIonHardware2024} provide strong physical motivation for the high-connectivity paradigm represented by Device A. These platforms feature native all-to-all connectivity, achieved through shared motional bus modes or the coherent transport of qubits \cite{MosesIonHardware2024, ChenIon30Qubits2024}. This natively accommodates the one-to-many structure of $U(\alpha)$ without requiring intermediate SWAP routing. Furthermore, ions support native multi-qubit entangling operations (e.g., global M\o lmer--S\o rensen gates \cite{molmer1999multiparticle}), allowing the theoretical $\mathcal{O}(|\alpha|)$ gate depth to be significantly compressed. 

\paragraph{Parameters.} Achieving high-fidelity single-qubit gates is a standard benchmark in these systems, with fidelities routinely exceeding $99.9\%$ \cite{loschnauer2025scalable,srinivas2021high}. Recently, $>99.99\%$ two-qubit gates have also been demonstrated \cite{hughes2025trappediontwoqubitgates9999,srinivas2021high}, making the $99\%$ fidelity of Device A a conservative estimate that accounts for potential degradation due to excessive shuttling or crosstalk \cite{MosesIonHardware2024} when implementing the fan-out structure. This allows us to explicitly demonstrate how two-qubit gate errors degrade the fully quantum protocol. Long coherence times ($T_2 > 1000$~s) \cite{wang2017single}, compared to typical gate times ($T_{2q} \sim 100~\mu$s) \cite{hou2024individually}, justify a massive quality factor above $10^6$ for Device A. Finally, the relatively slow cycle times of ion platforms ($\sim 0.1-1$~ms) \cite{leone2025resource} heavily penalize measure-first protocols, meaning the sample-efficient fully quantum protocol is particularly advantageous here.

\paragraph{State preparation and measurement circuit.} The combination of all-to-all connectivity and native multi-qubit operations based on the M\o lmer--S\o rensen interaction \cite{molmer1999multiparticle} paves the way for efficient Hamiltonian phase state preparation \cite{BostanciHPS2025}.

\subsection{Neutral and Rydberg Atoms}
Similar to trapped ions, neutral atom arrays \cite{ransford2025helios98qubittrappedionquantum} align closely with the Device A archetype. Highly flexible connectivity is achieved through dynamic optical tweezer rearrangement \cite{KimRydbergGraphs2023}, while multi-target Rydberg blockades \cite{ManovitzRobustEntangling2022} offer rapid entangling operations, including multi-qubit gates \cite{bai2026multipartitecontrollednotgatesusing}. 

\paragraph{Parameters.} While two-qubit gate fidelities in Rydberg systems are slightly lower than those of best-in-class trapped ions, recent demonstrations of $99.5\%$ fidelity on up to 60 atoms in parallel \cite{evered2023high} make the Device A baseline of $99\%$ another conservative estimate. Readout fidelity is also slightly reduced due to a reliance on fluorescence detection \cite{Adams_2019}. Coherence times ($T_2 > 10$~ms) \cite{picken2019entanglement} paired with fast Rydberg gate times ($T_{2q} \sim 0.2~\mu$s) \cite{bluvstein2024logical} yield a quality factor close to $10^6$ for Device A. Like trapped ions, total cycle times remain slow ($\sim 100-200$~ms) \cite{chiu2025continuous} due to the physical necessity of optical array reloading and cooling, heavily favoring algorithms with low sample complexity.

\paragraph{State preparation and measurement.} The combination of all-to-all connectivity and the Rydberg blockade mechanism \cite{evered2023high} in neutral atom systems similarly paves the way for random-phase state preparation using fast, native multi-qubit controlled-Z gates \cite{bai2026multipartitecontrollednotgatesusing}.

\subsection{Superconducting Qubits}
Solid-state superconducting architectures \cite{MarxerAbove99p9SC2025, kim2023evidence} realize the local-connectivity paradigm abstracted by Devices B and C. Constrained to planar nearest-neighbor layouts (e.g., square \cite{gambetta2017building} or heavy-hexagonal lattices \cite{hetenyi2024creating}), the fan-out structure of $U(\alpha)$ introduces a severe routing challenge: the control qubit must be sequentially moved adjacent to every target qubit via SWAP networks, scaling the effective circuit depth to $\mathcal{O}(|\alpha| n_q^{1.5})$ for the square lattice assumed in our models.

\paragraph{Parameters.} To offset this massive routing overhead, state-of-the-art processors leverage ultra-fast electrically driven gates with times below $100$~ns \cite{ding2023high, negirneac2021high} and fidelities exceeding $99.9\%$ \cite{MarxerAbove99p9SC2025}. Standard cloud-accessible platforms typically operate slightly above a more conservative $99\%$ fidelity \cite{last2020quantum,ibmquantum}, resonating with Device C. Because most superconducting qubits operate at the symmetric point to suppress flux noise, $T_1$ thermal relaxation becomes the dominant idling error \cite{DograMilliSecondCoherence2023}. Recently measured $T_1$ times span from $\sim 200~\mu$s \cite{place2021new} up to $\sim 1$~ms \cite{bland2025millisecond}, and remain around $T_1 \sim 50~\mu$s for cloud-accessible systems \cite{last2020quantum,ibmquantum}, allowing for quality factors above $5 \times 10^3$ for Device B and $5 \times 10^2$ for Device C. Crucially, their rapid $\sim 1~\mu$s cycle time \cite{google2025quantum} makes superconducting platforms highly amenable to the massive shot counts required by measure-first protocols.

\paragraph{State preparation and measurement.} Aside from attempts to realize long-range coupling \cite{marxer2023long} and analog-digital paradigms \cite{lamata2018digital}, these platforms can implement pseudorandom unitaries either through constant-depth Toffoli networks \cite{kim2022highfidelity} or via hardware-native transduction \cite{mirhosseini2020superconducting}.

\subsection{Spin Qubits}
Emerging silicon spin-qubit devices \cite{SteinackerSpinQubitCell2025} represent another highly scalable solid-state platform. Like superconducting qubits, they generally rely on planar nearest-neighbor connectivity via exchange interactions, incurring identical SWAP routing penalties \cite{sigillito2019coherent}. However, in the long term, scalable devices are expected to use coherent mobile qubits to realize long-range coupling \cite{vandersypen2017interfacing}.

\paragraph{Parameters.} While recent demonstrations on few-qubit spin devices have achieved impressive two-qubit fidelities ($99\%$ to $99.9\%$) \cite{noiri2022fast, tanttu2024assessment, matsumoto2025twoqubitlogicteleportationmobile, zhou2025high}, the difficulty of scaling these metrics across larger arrays justifies a conservative estimate of $99\%$ for Device C. Spin qubits are predominantly limited by $T_2^*$ dephasing originating from charge \cite{zajac2018resonantly,connors2022charge} or nuclear spin noise \cite{malinowski2017spectrum}. An echoed $T_2 \sim 20-100~\mu$s \cite{philips2022universal} alongside gate times of $T_{2q} \sim 50-200$~ns \cite{hendrickx2020fast, heinz2025fast} results in a quality factor typically above $2 \times 10^2$, as assumed in Device C. Like superconducting architectures, spin qubits feature extremely fast microsecond cycle times \cite{xue2022quantum,berritta2024real} and achieve above $99\%$ readout fidelity \cite{takeda2024rapid}, maintaining their viability for measure-first execution. 

\paragraph{State preparation.} The presence of electron spins with exchange interactions opens the possibility for native multi-qubit gates \cite{rodriguez2025threequbitentanglinggatessimultaneous,nguyen2025single}, and could potentially leverage native spin-spin interactions \cite{kandel2021adiabatic} for random-phase state preparation. For instance, analog Hamiltonian evolution under the Heisenberg interaction remains a promising avenue to explore.  

\paragraph{State measurement.} While Device C assumes square-lattice connectivity, spin qubits are increasingly being explored in $2 \times N$ geometries \cite{Micciche_good,siegel2024towards}, which can strongly benefit from recent developments in coherent qubit shuttling \cite{De_Smet_2025,matsumoto2025twoqubitlogicteleportationmobile,seidler2022conveyor}. In particular, the measurement circuit can be seen as a generalization of the stabilizer measurement circuit, which has already been realized using mobile ancillae at weight-4 \cite{undseth2026weight}. This would immediately unlock one-to-all connectivity for the measurement circuit, relaxing the severe SWAP routing overheads.

\subsection{Photonics}
Photonic quantum computing platforms offer a radically different hardware paradigm, primarily utilizing flying qubits rather than stationary matter qubits. Through programmable optical interferometer meshes and fast optical switches, photonic architectures can achieve highly reconfigurable, effectively all-to-all connectivity \cite{bourassa2021blueprint} matching the idealized Device A. However, the reliance on probabilistic gates makes effective two-qubit gate fidelity more difficult to define. 

\paragraph{State preparation and measurement.} While the dominant error mechanism in photonic systems is typically photon loss rather than gate infidelity or decoherence, an effective quality factor can still be defined as the ratio of the time required to implement the measurement circuit (dominated by optical switching and feedforward times) to the timescale of photon loss. With state-of-the-art integrated photonic platforms achieving sub-microsecond switching times \cite{bartolucci2023fusion} and loss rates corresponding to effective coherence times on the order of milliseconds \cite{bourassa2021blueprint}, photonic systems can achieve quality factors above Device A. Finally, due to the interference nature of the measurement circuit, this platform offers an interesting alternative to a gate-based implementation; however, analyzing the noise in such an approach remains outside the scope of our model. Nevertheless, if heralded state preparation is implemented \cite{zhong2020quantum}, this platform may be suitable for preparing random-phase states, as illustrated in recent experimental demonstrations up to $n_q=4$ \cite{huang2024demonstration}.

\begin{table}[h]
\centering
\caption{Physical hardware parameters informing the simulated device archetypes. Data are drawn from state-of-the-art benchmarks for trapped ions, neutral atoms, superconducting qubits, and emerging silicon spin architectures.}
\renewcommand{\arraystretch}{1.3}
\begin{tabular}{|l|c|c|c|c|c|}
\hline
\textbf{Metric} & \textbf{Trapped Ions} & \textbf{Neutral Atoms} & \textbf{Superconducting} & \textbf{Spin Qubits} & \textbf{Photonics} \\ \hline
1-Q Gate $F_{1q}$ & $\gtrsim 99.99\%$ \cite{loschnauer2025scalable,srinivas2021high} & $\gtrsim 99.99\%$ & $\gtrsim 99.99\%$ & $\gtrsim 99.9\%$ & N/A \\ \hline
2-Q Gate $F_{2q}$ & $\gtrsim 99\%$ \cite{hughes2025trappediontwoqubitgates9999,srinivas2021high} & $\gtrsim 99.5\%$ \cite{evered2023high,ransford2025helios98qubittrappedionquantum} & $\gtrsim 99.9\% / 99\%$ \cite{MarxerAbove99p9SC2025,ibmquantum} & $\gtrsim 99\%$ \cite{noiri2022fast,tanttu2024assessment,matsumoto2025twoqubitlogicteleportationmobile} & N/A \\ \hline
Idle Error & $T_2$  & $T_2$  & $T_1$ \cite{DograMilliSecondCoherence2023} & $T_2^*$ \cite{zajac2018resonantly,connors2022charge,malinowski2017spectrum} & Loss \\ \hline
Idling Time $T_{\mathrm{idle}}$ & $\gtrsim 1000$~s \cite{wang2017single} & $\gtrsim 40$~s \cite{picken2019entanglement} & $\gtrsim 200~\mu$s/$20~\mu$s \cite{place2021new,bland2025millisecond} & $\gtrsim 50~\mu$s \cite{philips2022universal} & N/A \\ \hline
Gate Time $T_{2q}$ & $\gtrsim 200~\mu$s \cite{hou2024individually} & $\gtrsim 0.2~\mu$s \cite{bluvstein2024logical} & $\gtrsim 100$~ns \cite{ding2023high,negirneac2021high} & $\gtrsim 200$~ns \cite{hendrickx2020fast,heinz2025fast} & $\gtrsim$ ps-ns \cite{bartolucci2023fusion} \\ \hline
Readout $1-\epsilon_r$ & $\gtrsim 99.9\%$ \cite{Myerson_2008} & $\gtrsim 99.5\%$ \cite{BarikRydbergTech2024} & $\gtrsim 99.5\% / 99.0\%$ \cite{google2025quantum} &$\gtrsim 99.0\%$ \cite{takeda2024rapid} & $\gtrsim 99\%$ \cite{bourassa2021blueprint} \\ \hline
Cycle Time $\Delta T$ & $\gtrsim 100$~ms \cite{leone2025resource} & $\gtrsim 200$~ms \cite{chiu2025continuous} & $\gtrsim 1~\mu$s \cite{google2025quantum} & $\gtrsim 1~\mu$s \cite{xue2022quantum,berritta2024real} & $\lesssim 1~\mu$s \cite{bartolucci2023fusion} \\ \hline
Connectivity & All-to-All \cite{MosesIonHardware2024,ChenIon30Qubits2024} & Reconfigurable \cite{KimRydbergGraphs2023}  & Planar (Square) \cite{gambetta2017building,hetenyi2024creating} & Planar/One-to-All \cite{sigillito2019coherent,vandersypen2017interfacing} & All-to-All \cite{bourassa2021blueprint} \\ \hline
Model & Device A & Device A & Devices B \& C & Device C & - \\ \hline
\end{tabular}
\end{table}
\newpage
\section{Shadow surrogate model}
\label{app:shadow_surrogate}

In this appendix, we derive the second-order fluctuation structure of density-matrix elements reconstructed by local-Clifford classical shadows and use it to construct a Gaussian surrogate model. The surrogate reproduces the finite-shot noise of the reconstructed density matrix while avoiding the cost of explicitly simulating a large number of randomized measurement circuits.

Our goal is not to state a generic worst-case shadow bound, but rather to characterize the reconstruction noise for the specific random-phase-state ensemble used throughout this work. We therefore work with \emph{task-averaged} second moments over the random Boolean functions \(f\), which can be evaluated exactly. This leads naturally to two closely related surrogate models: an \emph{exact second-order surrogate}, formulated in terms of the joint real-imaginary covariance, and a \emph{simplified proper-complex surrogate}, which neglects only an exponentially small pseudo-covariance term and is the model used in practice.

\subsection{Local-Clifford classical shadows}

Let \(\rho\) be an \(n_q\)-qubit state. In the local-Clifford shadow protocol, each measurement shot \(k\in\{1,\dots,n_c\}\) proceeds as follows. One samples a product unitary
\begin{equation}
    U_k=\bigotimes_{i=1}^{n_q} U_{k,i},
\end{equation}
where each \(U_{k,i}\) is drawn uniformly from the single-qubit Clifford group. The unitary \(U_k\) is applied to \(\rho\), the system is measured in the computational basis, and a classical bit string \(b_k\in\{0,1\}^{n_q}\) is obtained. Writing
\begin{equation}
    \ket{s_k}:=U_k^\dagger \ket{b_k},
\end{equation}
the corresponding single-shot shadow snapshot is
\begin{equation}
    \hat{\rho}_k
    =
    \bigotimes_{i=1}^{n_q}
    \left(
        3\ketbra{s_{k,i}}{s_{k,i}}-\mathbb{I}_2
    \right).
\end{equation}
The final estimator is the sample mean
\begin{equation}
    \hat{\rho}
    =
    \frac{1}{n_c}
    \sum_{k=1}^{n_c}
    \hat{\rho}_k.
\end{equation}
By construction,
\begin{equation}
    \mathbb{E}_{\mathrm{sh}}[\hat{\rho}\mid \rho]=\rho,
\end{equation}
where \(\mathbb{E}_{\mathrm{sh}}\) denotes the average over the random measurement protocol.

Throughout this appendix, basis labels \(n,m,p,q\in\{0,1\}^{n_q}\) are \(n_q\)-bit strings, and we write \(\hat{\rho}_{nm}:=\bra{n}\hat{\rho}\ket{m}\).

\subsection{Task-averaged second moments for random phase states}

We now specialize to the random phase states
\begin{equation}
    \rho^{(f)} = \ketbra{\psi_f}{\psi_f}, \qquad \ket{\psi_f} = \frac{1}{\sqrt{2^{n_q}}} \sum_{k\in\{0,1\}^{n_q}} (-1)^{f(k)} \ket{k}, \qquad c_k := (-1)^{f(k)} \in\{\pm1\},
\end{equation}
with \(f\) uniformly random. Their ensemble average is maximally mixed:
\begin{equation}
    \mathbb{E}_f[\rho^{(f)}] = \frac{\mathbb{I}}{2^{n_q}}.
\end{equation}
Because the distribution of a single shadow snapshot depends linearly on the input state, task-averaging over \(f\) is equivalent to evaluating snapshot second moments for the maximally mixed state. This yields exact task-averaged second moments.

For matrix-element observables \(O_{nm}:=\ketbra{m}{n}\), define the local-Clifford shadow sesquilinear form
\begin{equation}
    \langle O_{nm}, O_{pq}\rangle_{\mathrm{sh}}
    :=
    \sum_{P\in\{I,X,Y,Z\}^{\otimes n_q}}
    3^{\mathrm{wt}(P)}
    \Tr(O_{nm}P)\Tr(O_{pq}P)^*,
\end{equation}
and the corresponding bilinear form
\begin{equation}
    [ O_{nm}, O_{pq}]_{\mathrm{sh}}
    :=
    \sum_{P\in\{I,X,Y,Z\}^{\otimes n_q}}
    3^{\mathrm{wt}(P)}
    \Tr(O_{nm}P)\Tr(O_{pq}P).
\end{equation}
Then for a \emph{single} shadow snapshot,
\begin{align}
    \mathbb{E}_f\mathbb{E}_{\mathrm{sh}}\!\left[\hat{\rho}^{(1)}_{nm}\,\hat{\rho}^{(1)\,*}_{pq}\right] &= \frac{1}{4^{n_q}} \langle O_{nm}, O_{pq}\rangle_{\mathrm{sh}}, \\
    \mathbb{E}_f\mathbb{E}_{\mathrm{sh}}\!\left[\hat{\rho}^{(1)}_{nm}\,\hat{\rho}^{(1)}_{pq}\right] &= \frac{1}{4^{n_q}} [ O_{nm}, O_{pq}]_{\mathrm{sh}},
\end{align}
where \(\hat{\rho}^{(1)}\) denotes a single-shot snapshot. Since \(\hat{\rho}\) is the mean of \(n_c\) i.i.d. snapshots, all second-order cumulants of the sample mean are reduced by a factor \(1/n_c\).

\subsection{Single-element variance}

To characterize element-wise noise, we first compute the task-averaged variance scale of \(\hat{\rho}_{nm}\). For \(O_{nm}=\ketbra{m}{n}\), the shadow norm is
\begin{equation}
    \|O_{nm}\|_{\mathrm{sh}}^2
    :=
    \langle O_{nm},O_{nm}\rangle_{\mathrm{sh}}
    =
    \sum_{P\in\{I,X,Y,Z\}^{\otimes n_q}}
    3^{\mathrm{wt}(P)}
    |\Tr(O_{nm}P)|^2.
\end{equation}
Now \(\Tr(O_{nm}P)=\bra{n}P\ket{m}\) is nonzero only when
\begin{itemize}
    \item \(P_i\in\{X,Y\}\) on qubits where \(n_i\neq m_i\),
    \item \(P_i\in\{I,Z\}\) on qubits where \(n_i=m_i\).
\end{itemize}
Let \(w(n,m):=|n\oplus m|\) denote the Hamming distance between \(n\) and \(m\). Then each differing qubit contributes a factor \(6\), while each agreeing qubit contributes a factor \(4\), so
\begin{equation}
    \|O_{nm}\|_{\mathrm{sh}}^2 = 6^{w(n,m)}4^{n_q-w(n,m)}.
\end{equation}
Hence
\begin{equation}
    \frac{1}{4^{n_q}}\|O_{nm}\|_{\mathrm{sh}}^2 = \left(\frac{3}{2}\right)^{w(n,m)}.
\end{equation}

For the random phase states considered here, \(\rho^{(f)}_{nm} = \frac{1}{2^{n_q}}c_n c_m\), so \(|\rho^{(f)}_{nm}|^2 = 4^{-n_q}\) for all \(n,m\). Therefore the \emph{task-averaged} variance of the sample-mean estimator is
\begin{equation}
    \mathbb{E}_f\!\left[\Var_{\mathrm{sh}}(\hat{\rho}_{nm}\mid f)\right]
    =
    \frac{1}{n_c}
    \left[
        \left(\frac{3}{2}\right)^{w(n,m)}
        -
        4^{-n_q}
    \right].
\end{equation}

Two useful special cases are:
\begin{itemize}
    \item Diagonal elements (\(w=0\)). Since \(\hat{\rho}_{nn}\) is real,
    \begin{equation}
        \mathbb{E}_f\!\left[\Var_{\mathrm{sh}}(\hat{\rho}_{nn}\mid f)\right]
        =
        \frac{1}{n_c} \left(1-4^{-n_q}\right) \approx \frac{1}{n_c}.
    \end{equation}
    \item Off-diagonal elements (\(w\ge 1\)). The leading shadow-noise scale grows as \((3/2)^w/n_c\). Below we show that the corresponding covariance is block structured and that, after a canonical one-triangle parametrization, the remaining pseudo-covariance is exponentially small in \(n_q\).
\end{itemize}

\subsection{Ordinary covariance structure}

We now characterize the task-averaged ordinary covariance
\begin{equation}
    K_{(nm),(pq)} := \mathbb{E}_f\!\left[\Cov_{\mathrm{sh}}(\hat{\rho}_{nm},\hat{\rho}_{pq}\mid f)\right].
\end{equation}
Using the task-averaged second moment above, we obtain
\begin{equation}
    K_{(nm),(pq)}
    =
    \frac{1}{n_c}
    \left[
        \frac{1}{4^{n_q}} \langle O_{nm}, O_{pq}\rangle_{\mathrm{sh}}
        -
        \mathbb{E}_f[\rho^{(f)}_{nm}\rho^{(f)\,*}_{pq}]
    \right].
\end{equation}

A necessary condition for the shadow inner product to be nonzero is that the two matrix elements have the same bit-flip mask, \(n\oplus m = p\oplus q\). If the two masks differ, the corresponding Pauli support sets are disjoint, and the covariance vanishes.

\subsubsection{Diagonal sector}

For diagonal elements, all masks are zero, so all diagonal estimators are mutually correlated. Their covariance depends on the Hamming distance \(w(n,p)=|n\oplus p|\) between the basis states:
\begin{equation}
    K^{\mathrm{diag}}_{n,p}
    :=
    \mathbb{E}_f\!\left[\Cov_{\mathrm{sh}}(\hat{\rho}_{nn},\hat{\rho}_{pp}\mid f)\right]
    =
    \frac{1}{n_c} \left[ \left(-\frac{1}{2}\right)^{w(n,p)} - 4^{-n_q} \right].
\end{equation}
Thus the diagonal sector is dense but structured: the covariance decays with Hamming distance and alternates in sign.

\subsubsection{Off-diagonal sector}

For off-diagonal elements, the mask condition \(n\oplus m = p\oplus q =: \Delta \neq 0\) is necessary for a nonzero covariance. Let \(A(\Delta):=\{i:\Delta_i=1\}\) and \(P(\Delta):=\{i:\Delta_i=0\}\). Define \(r_\Delta(n,p) := \bigl|\{\, i\in P(\Delta): n_i\neq p_i \,\}\bigr|\), which is the Hamming distance between \(n\) and \(p\) restricted to the passive qubits. Combining the local factors yields
\begin{equation}
    \frac{1}{4^{n_q}} \langle O_{n,n\oplus\Delta}, O_{p,p\oplus\Delta}\rangle_{\mathrm{sh}}
    =
    \mathbf{1}_{\,n_{A(\Delta)}=p_{A(\Delta)}} \left(\frac{3}{2}\right)^{|\Delta|} \left(-\frac{1}{2}\right)^{r_\Delta(n,p)}.
\end{equation}

\subsection{Canonical one-triangle parametrization}

For each nonzero mask \(\Delta\), let \(h(\Delta)\) denote the most significant active qubit. Define the canonical representative set \(S_\Delta := \{\, n\in\{0,1\}^{n_q} : n_{h(\Delta)} = 0 \,\}\). Thus \(S_\Delta\) contains precisely one representative from each Hermitian-conjugate pair \(\rho_{n,n\oplus\Delta}\) and \(\rho_{n\oplus\Delta,n}\).

Restricting to \(n,p\in S_\Delta\), the second pairing \(p=n\oplus\Delta\) is impossible, and therefore \(\mathbb{E}_f\!\left[ \rho^{(f)}_{n,n\oplus\Delta} \rho^{(f)\,*}_{p,p\oplus\Delta} \right] = 4^{-n_q}\delta_{n,p}\). Hence, within the retained block,
\begin{equation}
    K^{(\Delta)}_{n,p}
    =
    \frac{1}{n_c}
    \left[
        \mathbf{1}_{\,n_{A(\Delta)}=p_{A(\Delta)}} \left(\frac{3}{2}\right)^{|\Delta|} \left(-\frac{1}{2}\right)^{r_\Delta(n,p)}
        -
        4^{-n_q}\delta_{n,p}
    \right].
\end{equation}
Thus the same-mask off-diagonal covariance is block structured: within a fixed mask \(\Delta\), two retained matrix elements are correlated only when they agree on all active qubits, and within such a block the covariance decays with the passive-bit Hamming distance.

\subsection{Relation matrix and residual impropriety}

For complex-valued off-diagonal elements, we also define the task-averaged relation matrix:
\begin{equation}
    R_{(nm),(pq)}
    =
    \frac{1}{n_c}
    \left[
        \frac{1}{4^{n_q}} [ O_{nm}, O_{pq}]_{\mathrm{sh}}
        -
        \mathbb{E}_f[\rho^{(f)}_{nm}\rho^{(f)}_{pq}]
    \right].
\end{equation}
After restricting to the canonical representative set \(S_\Delta\), the leading shadow-induced relation term vanishes identically. What remains is only the subtraction term from the mean, yielding:
\begin{equation}
    R^{(\Delta)}_{n,p} = -\frac{4^{-n_q}}{n_c}\delta_{n,p}.
\end{equation}
Thus the residual pseudo-covariance is diagonal and exponentially small in \(n_q\).

\subsection{Exact and simplified Gaussian surrogates}

Because \(\hat{\rho}\) is an average of \(n_c\) independent shadow snapshots, the multivariate central limit theorem implies that for sufficiently large \(n_c\), the reconstructed matrix elements are approximately jointly Gaussian. 

\paragraph{Diagonal sector.}
Since the diagonal entries are real, we sample \(d \sim \mathcal{N}(0,K^{\mathrm{diag}})\), with covariance matrix \(K^{\mathrm{diag}}\) given above.

\paragraph{Simplified proper-complex surrogate.}
Because the residual pseudo-covariance is exponentially small in \(n_q\), we use a simplified surrogate in which each retained off-diagonal block is sampled as a proper complex Gaussian with covariance \(K^{(\Delta)}\):
\begin{equation}
    z_\Delta \approx \mathcal{CN}\!\left(0,K^{(\Delta)}\right),
\end{equation}
thereby neglecting only the exponentially small relation matrix \(R^{(\Delta)}\). Finally, Hermiticity is imposed explicitly by setting \(\hat{\rho}_{n\oplus\Delta,n} = \hat{\rho}_{n,n\oplus\Delta}^*\) and assembling the blocks into the full Hermitian surrogate matrix \(\hat{\rho}_{\mathrm{surr}}\).

\subsection{Numerical validation}

To validate the surrogate model, we compare density matrices obtained from explicit classical-shadow simulations with those obtained by injecting the analytically derived Gaussian fluctuations into the ideal density matrix. We find that the surrogate accurately reproduces the scaling of the element-wise variances across the system sizes and shot counts relevant to this work.

\begin{figure}[htb!]
    \centering
    \includegraphics[width=1\linewidth]{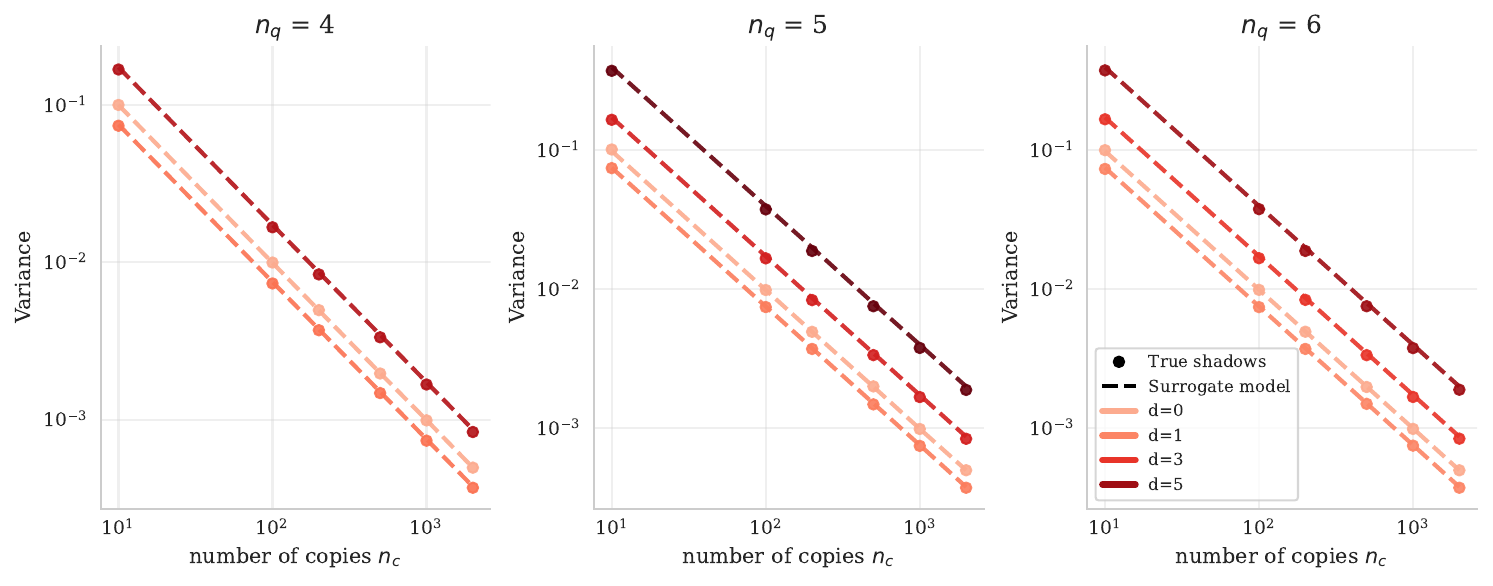}
    \caption{\textbf{Variance of the density matrix element}, grouped by the Hamming distance $d=w(n,m)$ in $\rho_{nm}$. Dots correspond to the true shadow tomography protocol, while dashed lines denote surrogate assumptions.}
    \label{fig:shadow_variance}
\end{figure}

\begin{figure}[htb!]
    \centering
    \includegraphics[width=0.8\linewidth]{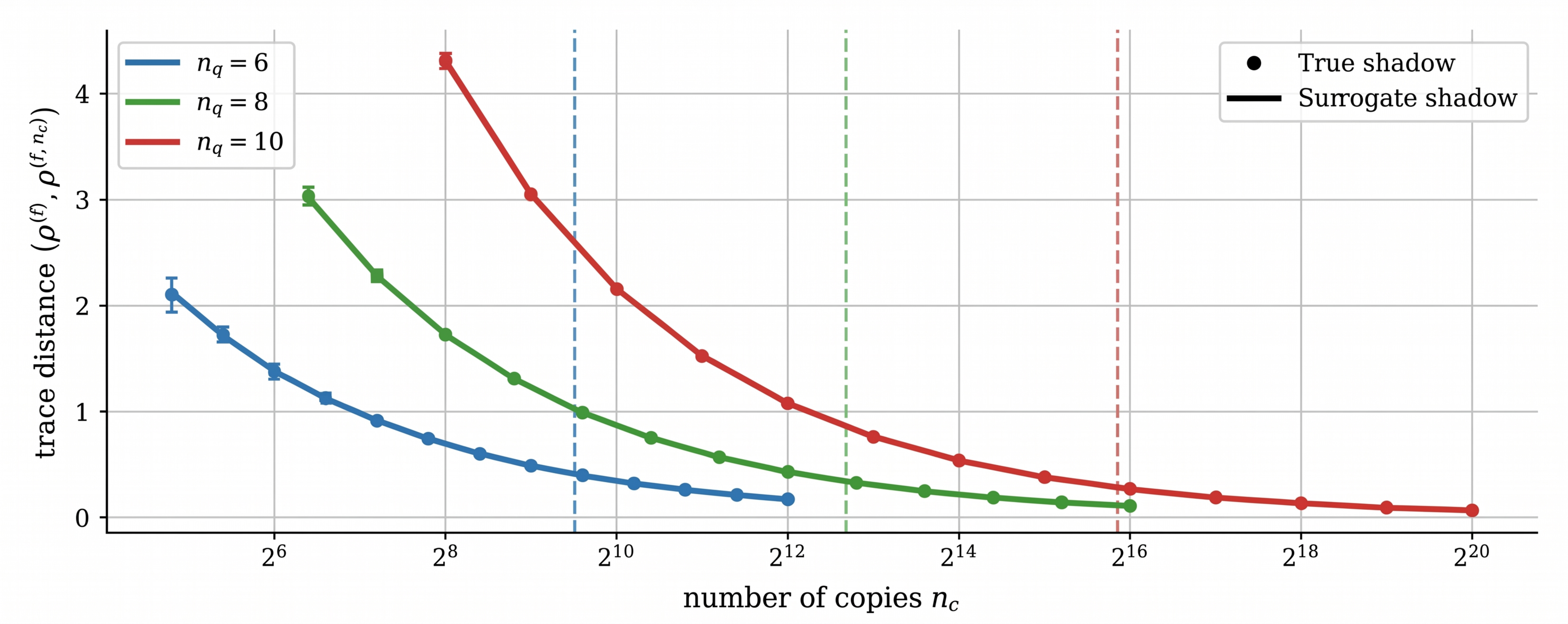}
    \caption{\textbf{Trace distance} between the true random binary phase state density matrix $\rho^{(f)}$ and the shadow-tomography reconstructed $\rho^{(f,n_c)}$ (points). Lines indicate the similar results obtained using the surrogate model. Different colors stand for different numbers of qubits.}
    \label{fig:trace_distance}
\end{figure}

\newpage
\, 
\newpage

\section{Sample Complexity of Measure-First Methods}
\label{app:measure_first_analytics}

This appendix provides a rigorous derivation of the sample complexity ($n_c$) required to infer properties of a Random Phase State using classical shadow tomography. We model the state under preparation noise, distinguishing between the survival of coherences on active qubits ($\bar{\gamma}_{\mathrm{act}}$) and passive qubits ($\bar{\gamma}_{\mathrm{pass}}$), as derived in Appendix~\ref{app:prep_noise}.

\subsection{Setup and Objective}
The target pure state is defined as:
\begin{equation}
    |\psi\rangle = \frac{1}{\sqrt{2^{n_q}}} \sum_{x \in \{0,1\}^{n_q}} s_x |x\rangle, \quad s_x \in \{+1, -1\}.
\end{equation}
We evaluate classical inference strategies by modeling the estimator for a target physical quantity as a normal random variable $\mathcal{N}(\mu, \sigma^2)$, where $\mu$ is the quantum signal and $\sigma^2$ is the shadow shot noise (variance). We define the Signal-to-Noise Ratio as $SNR = \frac{|\mu|}{\sigma}$. To reach a resolvable threshold ($SNR = 1$, yielding $\approx 84.1\%$ accuracy), the required number of snapshots $n_c$ scales as $n_c \propto SNR^{-2}$.

\subsection{Inference Protocols and Sample Complexity}

We analyze three distinct inference strategies, moving from local to global observable estimation.

\subsubsection{Protocol 1: Single Element Inference}
We first attempt to infer the sign of a specific off-diagonal density matrix element $\rho_{nm}$. 
\begin{itemize}
    \item Signal ($\mu$): The magnitude is exponentially small and degraded by noise: $\mu = 2^{-n_q} \bar{\gamma}_{\mathrm{act}}^{|\alpha|} \bar{\gamma}_{\mathrm{pass}}^{n_q-|\alpha|}$.
    \item Noise ($\sigma$): Based on Appendix~\ref{app:shadow_surrogate}, the standard deviation is $\sigma = \frac{1}{\sqrt{2n_c}} (1.5)^{|\alpha|/2}$.
\end{itemize}
The SNR decays exponentially with system size, yielding a sample complexity of:
\begin{equation}
    n_c \sim 4^{n_q} \left(\frac{1.5}{\bar{\gamma}_{\mathrm{act}}^2}\right)^{|\alpha|} \bar{\gamma}_{\mathrm{pass}}^{-2(n_q-|\alpha|)}.
\end{equation}

\subsubsection{Protocol 2: Cumulative Difference-Vector Strategies}
To boost the signal, we sum correlations over intermediate states $z$ to compute a cumulative estimator: $X = \sum_{z \in \Omega} \hat{\rho}_{y, y \oplus z} \hat{\rho}_{y \oplus \alpha, y \oplus z}$. The intermediate random phases cancel coherently. Using the combinatorial lemma derived in Section~\ref{app:lemma_summation}, we evaluate this over two domains:

Local Strategy ($\Omega = \text{Subspace}(\alpha)$): Summing only over the $2^{|\alpha|}$ states in the active subspace.
\begin{equation}
    SNR_{\text{local}} = \frac{2n_c}{4^{n_q}} \left(\frac{2\bar{\gamma}_{\mathrm{act}}}{\sqrt{3}}\right)^{|\alpha|} \bar{\gamma}_{\mathrm{pass}}^{n_q-|\alpha|} \implies n_c \sim 4^{n_q} \left(\frac{0.87}{\bar{\gamma}_{\mathrm{act}}}\right)^{|\alpha|} \bar{\gamma}_{\mathrm{pass}}^{-(n_q-|\alpha|)}
\end{equation}

Global Strategy ($\Omega = \{0,1\}^{n_q}$): Summing over the entire Hilbert space incorporates the passive "bulk". It is equivalent to taking an inner product of a $y$-row and $y\otimes \alpha$-column of an estimated density matrix.
\begin{equation}
    SNR_{\text{global}} = \frac{2n_c}{4^{n_q}} \left(\frac{2\bar{\gamma}_{\mathrm{act}}}{\sqrt{3}}\right)^{|\alpha|} \left(\frac{1+\bar{\gamma}_{\mathrm{pass}}^2}{\sqrt{3.25}}\right)^{n_q-|\alpha|} \implies 4^{n_q} \left(\frac{0.87}{\bar{\gamma}_{\mathrm{act}}}\right)^{|\alpha|} \left(\frac{1.80}{1+\bar{\gamma}_{\mathrm{pass}}^2}\right)^{n_q-|\alpha|} 
\end{equation}
While summing the bulk provides a slight mathematical advantage when $\bar{\gamma}_{\mathrm{pass}} \approx 1$, the $4^{n_q}$ pre-factor dominates in all cases. The exponential barrier remains insurmountable.
\subsubsection{Protocol 3: Principal eigenvector (Eigenshadow)}
\label{app:spectral_stability}

We derive the sample complexity required to extract the principal eigenvector of the shadow density matrix. The stability of the power iteration method depends on the spectral gap $\Delta \lambda$ between the principal eigenvalue and the noise floor. The principal eigenvalue $\lambda_1$ corresponds to the total coherent sum of the matrix elements. As shown in Section~\ref{app:lemma_summation}, summing these magnitudes over the hypercube simplifies cleanly via the binomial theorem. By defining the average effective coherence $\bar{\gamma}_\text{eff} = (\bar{\gamma}_{\mathrm{act}} + \bar{\gamma}_{\mathrm{pass}})/2$, the eigenvalue scales as:
\begin{equation}
    \lambda_1 \approx 2^{-n_q} (\bar{\gamma}_{\mathrm{act}} + \bar{\gamma}_{\mathrm{pass}})^{n_q} = 2^{-n_q} (2\bar{\gamma}_{\text{eff}})^{n_q} = \bar{\gamma}_{\text{eff}}^{n_q}.
\end{equation}
The bulk spectrum is dominated by the diagonal noise $\lambda_{\text{bulk}} \approx 2^{-n_q}$. Assuming we are operating in a regime where the signal is resolvable ($\bar{\gamma}_{\text{eff}}> 1/2$), the spectral gap is $\Delta \lambda \approx \lambda_1 = \bar{\gamma}^{n_q}_\text{eff}$. The shadow estimation error matrix introduces a noise variance $\mathcal{V}_{\text{noise}}$ that scales as $(2.5)^{n_q}/n_c$. For spectral stability, the noise perturbation must be small relative to the squared gap: $\mathcal{V}_{\text{noise}} \ll (\Delta \lambda)^2$.
\begin{equation}
    \frac{(2.5)^{n_q}}{n_c} \ll \bar{\gamma}_\text{eff}^{2n_q} \implies 
    n_c \gg \left( \frac{2.5}{\bar{\gamma}_\text{eff}^2} \right)^{n_q}.
\end{equation}

This formula captures two distinct regimes:
\begin{itemize}
    \item Strong Coherence Regime: When state preparation is high-quality ($\bar{\gamma}_\text{eff} \to 1$), we recover the absolute scaling of $n_c \sim (2.5)^{n_q}$.
    \item High Noise Limit: As noise destroys the phase coherences ($\bar{\gamma}_\text{eff} \to 0$), the denominator vanishes and the required sample complexity diverges to infinity ($n_c \to \infty$). This correctly reflects the physical reality that the eigenstate becomes fundamentally unresolvable from the completely mixed background.
\end{itemize}

\subsection{Summary of Scaling Laws}

\begin{figure}[htb!]
    \centering
    \includegraphics[width=0.99\linewidth]{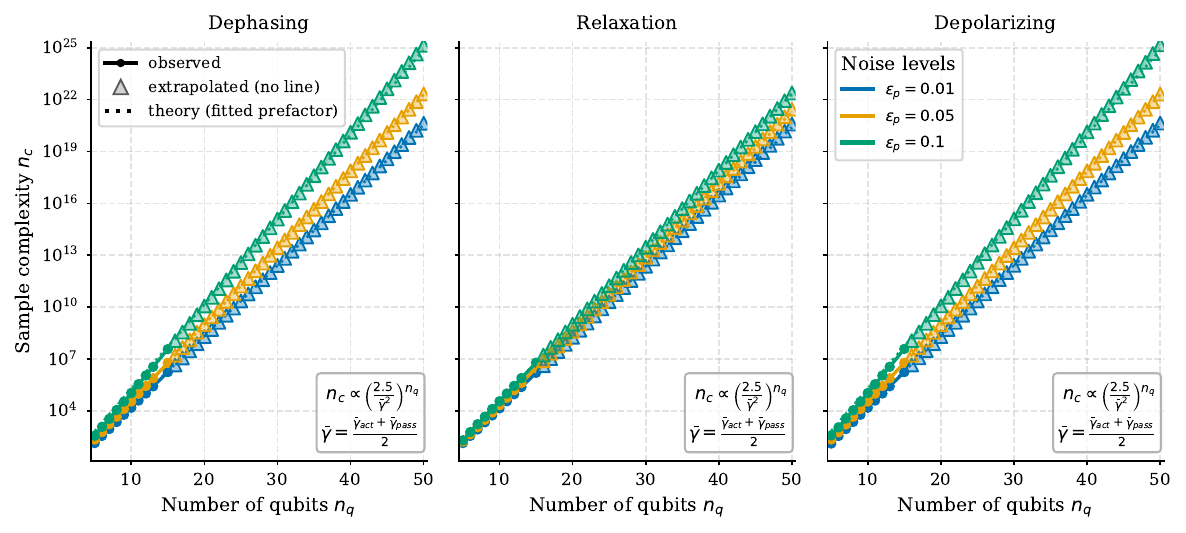}
    \caption{\textbf{Sample complexity scaling of the Eigenshadow method under various noise channels.} The required number of state copies $n_c$ is plotted as a function of the number of qubits $n_q$ for dephasing (left), thermal relaxation (middle), and depolarizing (right) noise. Solid lines with circles denote empirically observed data from exact simulations (up to $n_q=15$), while triangles indicate extrapolated requirements for larger system sizes. Dotted lines represent the corresponding theoretical scaling formula. Different colors correspond to distinct preparation noise amplitudes $\epsilon_p \in \{0.01, 0.05, 0.1\}$. In contrast to the Hypergraph method, the Eigenshadow method exhibits a consistent exponential scaling across all three noise channels, governed primarily by the averaged effective attenuation factor. Notably, it avoids the severe divergence under thermal relaxation, making it structurally more robust to population drift.}
    \label{fig:eigenshadow_scaling}
\end{figure}

Table \ref{tab:scaling_laws} summarizes the sample complexity required to achieve $SNR=1$ for each protocol. Fig.~\ref{fig:eigenshadow_scaling} numerically verifies the Eigenshadow scaling.

\begin{table}[h]
\centering
\renewcommand{\arraystretch}{1.5}
\begin{tabular}{l c c c}
\hline\hline
\textbf{Protocol} & \textbf{Target} & \textbf{Sample Complexity $n_c$} & \textbf{Limit ($\bar{\gamma}_{\mathrm{act}},\bar{\gamma}_{\mathrm{pass}} \to 1$)} \\
\hline
Single Element & $\tilde \rho^{(f, n_c)}_{nm}$ & $\sim 4^{n_q} \left(\frac{1.5}{\bar{\gamma}_{\mathrm{act}}^2}\right)^{|\alpha|} \bar{\gamma}_{\mathrm{pass}}^{-2(n_q-|\alpha|)}$ & $\sim 4^{n_q} (1.5)^{|\alpha|}$ \\
Local Sum & $\tilde \rho^{(f,n_c)}_{nm}$ & $\sim 4^{n_q} \left(\frac{0.87}{\bar{\gamma}_{\mathrm{act}}}\right)^{|\alpha|} \bar{\gamma}_{\mathrm{pass}}^{-(n_q-|\alpha|)}$ & $\sim 4^{n_q} (0.87)^{|\alpha|}$ \\
Global Sum &$\tilde \rho^{(f,n_c)}_{nm}$& $\sim 4^{n_q} \left(\frac{0.87}{\bar{\gamma}_{\mathrm{act}}}\right)^{|\alpha|} \left(\frac{1.80}{1+\bar{\gamma}_{\mathrm{pass}}^2}\right)^{n_q-|\alpha|}$ & $\sim 4^{n_q} (0.87)^{|\alpha|} (0.90)^{n_q-|\alpha|}$ \\
Eigenshadow &$\tilde \rho^{(f,n_c)}$& $\sim \left(\frac{10}{(\bar\gamma_{\text{act}}+\bar\gamma_{\text{pass})})^2}\right)^{n_q}$ & $\sim 2.5^{n_q}$ \\\hline\hline
\end{tabular}
\caption{Scaling comparison for Random Phase State inference. The Eigenvector protocol provides the most efficient classical scaling by avoiding element-wise inference.}
\label{tab:scaling_laws}
\end{table}

\subsection{Technical Lemma: Combinatorial Summation over the Hypercube}
\label{app:lemma_summation}

In deriving the signal and variance for the difference-vector protocols, we evaluate sums of the form:
\begin{equation}
    \mathcal{S} = \sum_{z \in \{0,1\}^{n_q}} \beta_{\mathrm{act}}^{w_{\mathrm{act}}(z, z \oplus \alpha)} \beta_{\mathrm{pass}}^{w_{\mathrm{pass}}(z, z \oplus \alpha)},
\end{equation}
where $w_{\mathrm{act}}$ and $w_{\mathrm{pass}}$ denote the Hamming distances restricted to the active and passive qubit sets respectively. We partition the qubit indices into the Active Set $\mathcal{A}$ (where $\alpha_k = 1$) and the Passive Set $\mathcal{P}$ (where $\alpha_k = 0$).

\paragraph{The Active Sum} For $k \in \mathcal{A}$, the bitwise distance $(z_k \oplus (z \oplus \alpha)_k)$ is strictly $1$ regardless of $z_k$. Summing over the $2^{|\alpha|}$ configurations of the active bits yields:
\begin{equation}
    \sum_{z|_{\mathcal{A}}} \beta_{\mathrm{act}}^{\sum_{k \in \mathcal{A}} 1} = 2^{|\alpha|} \beta_{\mathrm{act}}^{|\alpha|} = (2\beta_{\mathrm{act}})^{|\alpha|}.
\end{equation}

\paragraph{The Passive Sum} For $k \in \mathcal{P}$, $z_k = (z \oplus \alpha)_k$. The bitwise term is $0$ (if $z_k=0$) or $2$ (if $z_k=1$). Summing over the $n_q - |\alpha|$ passive bits is a binomial expansion:
\begin{equation}
    \sum_{z|_{\mathcal{P}}} \prod_{k \in \mathcal{P}} \beta_{\mathrm{pass}}^{2z_k} = (1 + \beta_{\mathrm{pass}}^2)^{n_q-|\alpha|}.
\end{equation}

Multiplying these independent contributions yields the master formula used in Section~\ref{app:measure_first_analytics}.2:
\begin{equation}
    \mathcal{S} = (2\beta_{\mathrm{act}})^{|\alpha|} (1 + \beta_{\mathrm{pass}}^2)^{n_q-|\alpha|}.
\end{equation}
For the signal sum, we substitute $\beta_{\mathrm{act}} = \bar{\gamma}_{\mathrm{act}}$ and $\beta_{\mathrm{pass}} = \bar{\gamma}_{\mathrm{pass}}$. For the variance sum, the base is determined purely by the shadow norm expansion, giving $\beta_{\mathrm{act}} = \beta_{\mathrm{pass}} = 1.5$.

\newpage
\section{Machine-learning method}
\label{app:ml_method}

In this appendix we describe the shadow-based machine-learning protocol used in the main text. The objective is to infer the class label
\begin{equation}
    \ell(f) := f(y)\oplus f(y\oplus\alpha)
\end{equation}
from a measure-first representation of a noisy random phase state. Rather than attempting to learn directly from an exponentially large shadow reconstruction of the full density matrix, we construct a low-dimensional summary tailored to the structure of the relational problem and train a regularized logistic classifier on it.

The resulting method has three ingredients. First, for fixed \(n_q\), \(y\), and \(\alpha\), we generate a balanced ensemble of random phase states labeled by \(\ell(f)\). Second, for each state, we use the shadow surrogate of Appendix~\ref{app:shadow_surrogate} to model the row, column, and diagonal sectors of the reconstructed density matrix that are relevant to the target coherence \(\rho_{y\oplus\alpha,y}\). Third, from these sectors, we construct shell-based features that summarize the physically relevant fluctuations and use them as input to a supervised classifier.

\subsection{Labeled random phase ensembles}

For each choice of qubit number \(n_q\), target bit string \(y\), and concept string \(\alpha\), we consider the ensemble of random phase states
\begin{equation}
    \ket{\psi_f}
    =
    \frac{1}{\sqrt{2^{n_q}}}
    \sum_{k\in\{0,1\}^{n_q}} (-1)^{f(k)}\ket{k},
\end{equation}
where \(f:\{0,1\}^{n_q}\to\{0,1\}\) is sampled uniformly. Each state is assigned the binary label
\begin{equation}
    \ell(f) := f(y)\oplus f(y\oplus\alpha).
\end{equation}
The datasets used in the numerical study are balanced; that is, they contain equal numbers of label-0 and label-1 examples. This balancing removes trivial class-imbalance effects and ensures that the reported validation accuracies reflect the discriminative content of the shadow-derived features rather than a biased label prior. For each triple \((n_q,y,\alpha)\), the same underlying ensemble is reused across preparation-noise channels and shadow-shot counts, so that different estimators are always evaluated on the same collection of random phase states.

In the machine-learning results reported in the paper, the target bit string is fixed to \(y=0^{n_q}\), while \(\alpha\) is varied over two representative Hamming-weight regimes, \(|\alpha|=n_q\) and \(|\alpha|=n_q/2\). The preparation-noise channels considered are dephasing, depolarizing, and zero-temperature amplitude damping, with noise strengths \(\epsilon_p\in\{0.01,0.05,0.1\}\). The shadow-shot budgets are chosen on an exponential grid $n_c = 2^{k n_q}$, with \(k\) scanned over a fixed range.

\subsection{Reduced shadow representation and shell-based feature design}

Our measure-first classifier is built on the shadow surrogate model introduced in Appendix~\ref{app:shadow_surrogate}, which serves as a proxy for performing classical-shadow tomography on the noisy state \(\rho'^{(f)}\) and reconstructing an estimator \(\hat{\rho}\) of the global density matrix. The classifier, however, does not use \(\hat{\rho}\) in full. Instead, it retains only the sectors directly connected to the relational signal carried by the target coherence \(\hat{\rho}_{y\oplus\alpha,y}\).

Concretely, we define the reduced shadow representation by
\begin{equation}
    r_t := \hat{\rho}_{\,y\oplus\alpha,t},
    \qquad
    c_t := \hat{\rho}_{\,t,y},
    \qquad
    d_t := \hat{\rho}_{\,t,t}.
\end{equation}
Thus each sample is represented by one row anchored at \(y\oplus\alpha\), one column anchored at \(y\), and the diagonal. This reduction is dictated by the structure of the learning task. All features used by the classifier are built from the target coherence \(\hat{\rho}_{y\oplus\alpha,y}=r_y=c_{y\oplus\alpha}\), the two anchor populations \(d_{y\oplus\alpha}\) and \(d_y\), and shell-averaged quadratic combinations of the retained row and column sectors. The resulting machine-learning protocol is therefore not a generic inference procedure on shadow-reconstructed states, but a task-specific measure-first estimator for the relational parity problem.

The reduced representation is also aligned with the fluctuation structure of local-Clifford shadows. As shown in Appendix~\ref{app:shadow_surrogate}, for the random-phase ensemble considered here, the \emph{task-averaged} shadow variance of an off-diagonal matrix element is
\begin{equation}
    \mathbb{E}_f\!\left[\Var_{\mathrm{sh}}(\Delta \hat{\rho}_{nm}\mid f)\right]
    =
    \frac{1}{n_c}
    \left[
        \left(\frac{3}{2}\right)^{w(n,m)} - 4^{-n_q}
    \right],
\end{equation}
where \(w(n,m)\) is the Hamming distance between the two basis states. The retained sectors therefore isolate precisely those matrix elements whose shadow fluctuations and coherence decay are most directly tied to the target parity observable. At the level of the surrogate, the diagonal sector is modeled by a correlated real Gaussian with exact trace preservation, while the retained off-diagonal entries inherit the corresponding task-averaged variance laws
\begin{equation}
    \mathbb{E}_f\!\left[\Var_{\mathrm{sh}}(\Delta r_t\mid f)\right]
    =
    \frac{1}{n_c}
    \left[
        \left(\frac{3}{2}\right)^{w(y\oplus\alpha,t)} - 4^{-n_q}
    \right],
\end{equation}
\begin{equation}
    \mathbb{E}_f\!\left[\Var_{\mathrm{sh}}(\Delta c_t\mid f)\right]
    =
    \frac{1}{n_c}
    \left[
        \left(\frac{3}{2}\right)^{w(t,y)} - 4^{-n_q}
    \right],
\end{equation}
together with an explicit identification of the shared central target coherence. These two expressions are written separately because, even in the gauge \(y=0^{n_q}\), one generally has \(w(y\oplus\alpha,t)\neq w(t,y)\); the row and column sectors therefore need not carry the same shadow-noise scale. For shell features, the intermediate index satisfies \(t\in\mathcal{S}_k\), so the anchor points \(t=y\) and \(t=y\oplus\alpha\) are excluded and treated separately through the anchor features \(d_y\), \(d_{y\oplus\alpha}\), and \(\mathrm{Re}[r_y]\).

For the feature families used in this work, no further off-diagonal covariance structure is required. The shadow-surrogate appendix shows that off-diagonal covariances are highly structured: they are zero unless two entries belong to the same XOR-mask sector, and even within a fixed sector they remain constrained by additional active- and passive-bit conditions. The shell features introduced below are built from the pair \(r_t=\hat{\rho}_{y\oplus\alpha,t}\) and \(c_t=\hat{\rho}_{t,y}\), whose masks are
\begin{equation}
    (y\oplus\alpha)\oplus t
    \qquad\text{and}\qquad
    t\oplus y,
\end{equation}
respectively. These differ by \(\alpha\), so for \(\alpha\neq 0\) they generically lie in different mask sectors, except at the special central target entry itself. Consequently, the leading second-order statistics relevant to the classifier are already captured by the correlated diagonal sector, the correct Hamming-distance variance of the retained off-diagonal entries, and the exact treatment of the shared target coherence.

This reduced representation leads naturally to shell-based summaries. Since the dominant shadow variance is organized by Hamming distance, shell averaging groups together matrix elements with the same leading fluctuation scale and therefore respects the natural geometry of the shadow noise. At the same time, quadratic combinations of the retained row and column sectors probe how preparation noise reshapes the local coherence environment around the target entry \(\hat{\rho}_{y\oplus\alpha,y}\). The selected feature families are designed precisely to capture this structure in a low-dimensional form.

\subsection{Hamming-shell feature families}

We now define the feature maps used by the classifier in terms of the reduced shadow representation \(\{r_t,c_t,d_t\}\). Since the numerical study is performed in the gauge \(y=0^{n_q}\), Hamming shells around the target location \(y\) are indexed simply by the Hamming weight of the intermediate bit string \(t\). For each \(k=1,\dots,n_q-1\), let
\begin{equation}
    \mathcal{S}_k
    :=
    \{\, t\in\{0,1\}^{n_q} : |t|=k \,\}
    \setminus
    \{\, y,\; y\oplus\alpha \,\}
\end{equation}
denote the shell of intermediate basis states of weight \(k\), excluding the two anchor points. The cardinality \(|\mathcal{S}_k|\) is therefore the number of retained intermediate states in shell \(k\) after excluding the anchor points \(y\) and \(y\oplus\alpha\). In particular, while the number of shell features is always \(n_q-1\), the shell sizes need not equal the naive binomial counts \(\binom{n_q}{k}\). In the special case \(|\alpha|=n_q\) with \(y=0^{n_q}\), the two anchor points are \(0^{n_q}\) and \(1^{n_q}\), which lie at Hamming weights \(0\) and \(n_q\), respectively. The exclusion in the definition of \(\mathcal{S}_k\) is then vacuous, but we retain it for uniformity across all \(\alpha\).

All feature families begin with the same three anchor quantities,
\begin{equation}
    d_{y\oplus\alpha},
    \qquad
    d_y,
    \qquad
    \mathrm{Re}[r_y],
\end{equation}
where \(r_y=\hat{\rho}_{y\oplus\alpha,y}\) is the central target coherence. These three terms capture the populations of the two relevant basis states together with the directly reconstructed relational signal. The different feature families then summarize the remaining row and column information across the Hamming shells.

\paragraph{\textbf{Baseline shell summary.}}
The most compact representation augments the three anchor terms by a single scalar for each shell,
\begin{equation}
    \phi_k^{\mathrm{base}}
    =
    \frac{1}{|\mathcal{S}_k|}
    \sum_{t\in\mathcal{S}_k}
    \mathrm{Re}[r_t]\,\mathrm{Re}[c_t].
\end{equation}
This yields a feature vector of dimension \(n_q+2\). The baseline summary indirectly probes target coherence through shell-averaged real row-column products, thereby capturing how coherence is distributed across intermediate pathways at different Hamming distances.

\paragraph{\textbf{Paired \((\mathrm{Re[\cdot]}\mathrm{Re[\cdot]},\mathrm{Im[\cdot]}\mathrm{Im[\cdot]})\) shell summary.}}
The paired feature family used in the main text keeps the same anchor terms and resolves the shell contribution into separate real and imaginary quadratic blocks:
\begin{align}
    \phi_k^{(\mathrm{rr})}
    &=
    \frac{1}{|\mathcal{S}_k|}
    \sum_{t\in\mathcal{S}_k}
    \mathrm{Re}[r_t]\,\mathrm{Re}[c_t],
    \\
    \phi_k^{(\mathrm{ii})}
    &=
    \frac{1}{|\mathcal{S}_k|}
    \sum_{t\in\mathcal{S}_k}
    \mathrm{Im}[r_t]\,\mathrm{Im}[c_t].
\end{align}
The resulting feature vector has dimension \(2n_q+1\). This representation allows the classifier to distinguish situations in which preparation noise affects the real and imaginary parts of the retained off-diagonal structure differently.

These feature maps provide two increasingly rich summaries of the same underlying object: the shell-resolved structure surrounding the target coherence \(r_y=\hat{\rho}_{y\oplus\alpha,y}\). The baseline summary retains only the coarsest quadratic information, whereas the paired summary decomposes it into components that respond differently to noise. In both cases, the feature dimension grows only linearly with \(n_q\), while the averaging over shells suppresses sample-specific fluctuations and preserves the physically relevant dependence on Hamming distance.

We also examined an alternative paired parametrization based on \(\mathrm{Re}[r_t c_t]\) and \(\mathrm{Re}[r_t c_t^{*}]\). This parametrization is related to the \((\mathrm{Re[\cdot]}\mathrm{Re[\cdot]},\mathrm{Im[\cdot]}\mathrm{Im[\cdot]})\) family by an invertible linear transformation on each shell and therefore does not introduce additional information in the unregularized linear setting. Any practical difference between the two parametrizations would arise only through regularization and feature normalization, rather than from genuine information content, so we do not discuss it separately.

\subsection{Classifier}

To convert the shell summaries into a binary prediction, we use a regularized logistic classifier acting on the feature vector \(x\in\mathbb{R}^d\). The model assigns to each sample a score
\begin{equation}
    s(x)= w^\top x + \beta_0,
\end{equation}
where \(w\) is the vector of feature weights and \(\beta_0\) is an intercept parameter. This score is then mapped to a probability for the two classes through the logistic function,
\begin{equation}
    p_\theta(\hat{\ell}=1\mid x)
    =
    \frac{1}{1+e^{-s(x)}}.
\end{equation}
Here \(\hat{\ell}\) denotes the predicted class label. We write the model parameters collectively as \(\theta=(w,\beta_0)\).

The parameters are obtained by minimizing the regularized binary cross-entropy
\begin{equation}
    \mathcal{L}(w,\beta_0)
    =
    -\frac{1}{N_{\mathrm{tr}}}
    \sum_{i=1}^{N_{\mathrm{tr}}}
    \bigl[
        \ell_i \log p_\theta(\hat{\ell}=1\mid x_i)
        +
        (1-\ell_i)\log p_\theta(\hat{\ell}=0\mid x_i)
    \bigr]
    +
    \frac{\lambda}{2}\|w\|_2^2,
\end{equation}
where \(\ell_i\in\{0,1\}\) is the ground-truth class associated with the relational task, and \(N_{\mathrm{tr}}\) denotes the number of samples in the training subset.
The \(\ell_2\) penalty suppresses large coefficients and stabilizes the fit, which is useful because nearby shells can carry partially redundant information.

We chose this classifier for two reasons. First, the shell representation is intentionally low-dimensional, scaling only linearly with \(n_q\), so a simple linear decision rule in feature space is sufficient to probe how much predictive information is already contained in the physically motivated summaries. Second, the resulting coefficients remain directly interpretable: they indicate which shells and which quadratic combinations contribute most strongly to the classification. In this sense, the model functions as a transparent readout layer on top of the shadow-derived observables, rather than as a high-capacity black-box learner.

As a minimal reference, we also consider a deterministic baseline classifier defined by thresholding the central target-coherence feature at zero,
\begin{equation}
    \hat{\ell}_{\mathrm{base}}
    :=
    \mathbb{1}\![\mathrm{Re}[r_y]>0].
\end{equation}
The comparison between this baseline and the trained logistic classifier isolates the information gained by incorporating the shell-resolved structure surrounding the target coherence, rather than relying only on a direct estimate of the single matrix element of interest.

\subsection{Training details}

Unless otherwise specified, all numerical experiments use the default training configuration. For each configuration \((n_q,y,\alpha,\epsilon_p,n_c)\), the available labeled dataset is divided into training and validation subsets with fractions \(0.8\) and \(0.2\), respectively, so that \(N_{\mathrm{tr}}\) denotes the size of the resulting training subset. The features are standardized using the mean and standard deviation of the training subset only, and the same affine normalization is then applied to the validation data. This preprocessing places all shell features on a common scale and prevents shells with larger raw variance from dominating the regularized objective.

The classifier is optimized with Adam using mini-batch size \(256\), learning rate \(10^{-2}\), \(\ell_2\) regularization strength \(10^{-3}\), and \(100\) training epochs. The model parameters are initialized at zero. The default experiments use a single training run per configuration.

\newpage
\section{Hypergraph method}
\label{app:hypergraph_method}

In this section, we provide a streamlined overview of the Hypergraph method. Full combinatorial derivations and extended numerical tests are available in Ref.~\cite{danaci_in_prep}.

\subsection{Algebraic Formulation and Pooled Reconstruction}
A Boolean function $f: \{0,1\}^{n_q} \to \{0,1\}$ can be uniquely represented in its Algebraic Normal Form (ANF) as a polynomial over $\mathbf{F}_2$:
\begin{equation}
    f(s) = \sum_{e \in \{0,1\}^n} w_e s^e = \boldsymbol{\phi}(s)^T \myvec{w}
\end{equation}
where $\myvec{w}$ is the vector of unknown binary coefficients, and $\boldsymbol{\phi}(s)$ is the feature vector of monomial evaluations $s^e = \prod s_j^{e_j}$. 

\paragraph{Example: 4-Qubit Hypergraph.} To make this formulation concrete, consider a 4-qubit system ($n_q=4$) with input bitstring $s = (s_0, s_1, s_2, s_3)$. The binary indicator vector $e$ defines which qubits interact, and the monomial $s^e$ represents that physical interaction:
\begin{itemize}
    \item Degree-1 (Vertices): $e=(1,0,0,0)$ yields $s_0$.
    \item Degree-2 (Edges): $e=(1,1,0,0)$ yields $s_0 s_1$, representing a standard two-qubit interaction (e.g., a $CZ$ gate).
    \item Degree-3 (Hyperedges):$e=(1,0,1,1)$ yields $s_0 s_2 s_3$, representing a multi-qubit interaction (e.g., a $CCZ$ gate).
\end{itemize}
The coefficient $w_e \in \{0,1\}$ indicates whether that specific interaction is present in the graph. For instance, a hypergraph containing the standard edges $(0,1)$ and $(1,2)$, alongside the hyperedge $(0,1,3)$, corresponds to the Boolean function:
\begin{equation}
    f(s) = s_0 s_1 \oplus s_1 s_2 \oplus s_0 s_1 s_3.
\end{equation}
In the vector formulation $\myvec{\phi}(s)^T \myvec{w}$, the feature vector $\myvec{\phi}(s)$ evaluates all $2^4 = 16$ possible monomials, and the sparse target vector $\myvec{w}$ contains $1$s exactly at the indices corresponding to $s_0 s_1$, $s_1 s_2$, and $s_0 s_1 s_3$, and $0$s elsewhere.

\vspace{1em} % Optional spacing to separate the example from the protocol

To learn these coefficients from an $n$-qubit binary-phase state, we employ separable measurements \cite{arunachalam2023optimalalgorithmslearningquantum}, where one qubit $l$ is measured in the $x$-basis and the remaining $n-1$ qubits are measured in the $z$-basis. In the noiseless case, valid measurement outcomes satisfy the phase-matching condition $b = f(y) \oplus f(\upsilon)$, where $y$ and $\upsilon$ differ only at the $l$-th bit.

Expanding this condition into its ANF representation yields the following linear equation over $\mathbf{F}_2$ for each sampled bitstring. 

\begin{equation}
  b^{(l)}_k = \Delta \boldsymbol{\phi}^{(l)}(s^{(l)}_k)^T \myvec{w}.
\end{equation}

Here $\bold{\Delta} \boldsymbol{\phi}^{(l)}(s^{(l)}_k)$ denotes the differential monomial feature vector. The vector element at the $\vb{e}$-th position is given by $\bold{\Delta} \boldsymbol{\phi}^{(l)}(s)_{\vb{e}} = \delta_{1, e_l} \prod_{j \neq l} s_j^{e_j}$. The formulation of the same framework by Arunachalam et al. involves the estimates of $n$ different Boolean derivatives in independent blocks \cite{arunachalam2023optimalalgorithmslearningquantum}. Rather than solving $n$ independent systems and using majority voting, we pool the sampled equations from all $n$ measurement configurations.
The pooled equations of the differential feature vector form the following single linear system.

\begin{equation}
    \mathbf{Y} \myvec{w} = \myvec{b}
\end{equation}

Here $\mathbf{Y}$ is the combined matrix of differential features, and $\myvec{b}$ is the vector of parity outcomes. The task reduces to solving this system over $\mathbf{F}_2$ to recover $\myvec{w}$.

\paragraph{Example: Building the Linear System.} To see how physical measurement outcomes populate this math, let us return to the 4-qubit example. Suppose we set our measurement configuration to $l=0$, meaning we measure Qubit 0 in the $x$-basis, and Qubits 1, 2, and 3 in the $z$-basis. 

A single shot on the quantum processor yields a 4-bit outcome—for instance, $(0, 1, 0, 1)$.
\begin{itemize}
    \item The Parity Vector ($\myvec{b}$): The $x$-basis outcome of the target qubit ($b=0$) goes directly into the right-hand side vector $\myvec{b}$.
    \item The Design Matrix ($\mathbf{Y}$): The $z$-basis outcomes of the spectator qubits ($s_1=1, s_2=0, s_3=1$) dictate the entries of the matrix $Y$. 
\end{itemize}
Because we measured configuration $l=0$, the differential feature vector $\Delta \myvec{\phi}^{(0)}$ only activates for monomials containing Qubit 0. To find the matrix entries, we factor out $s_0$ and substitute our $z$-basis outcomes into the remaining variables. For example, the column corresponding to the edge $s_0 s_1$ evaluates to $s_1 = 1$, meaning we place a $1$ in that column. The column for $s_0 s_2$ evaluates to $s_2 = 0$, placing a $0$ in that column.

Evaluating all 16 possible combinations for this single shot generates a binary sequence (e.g., $1, 1, 0, 1, \dots$). This sequence forms exactly one row in the design matrix $\textbf{Y}$. We repeat this process across $n_c$ shots and all $n$ measurement configurations, stacking the rows to build the complete monolithic system $\mathbf{Y} \myvec{w} = \myvec{b}$.

\subsection{Sorting and Critical Sample Needs per Noise Model}

In the presence of preparation noise, sampling yields invalid bitstrings that render the linear system $\mathbf{Y} \myvec{w} = \myvec{b}$ inconsistent. To recover the true coefficients, we sort the rows of the augmented matrix $[Y \mid \myvec{b}]$ in descending order based on empirical observation frequencies. A modified Gaussian elimination algorithm then forces pivots through the system, prioritizing the most frequently observed equations.

Let $\kappa$ denote the local sample budget (total shots) allocated to the $l$-th separable measurement instance. The empirical success probability of the sorting can be estimated in reference to the probabilities of (in)valid bitstrings sampled at $l$-th event given as $P_l( s_{(in)valid})$, which are multinomial distributions. 
The signal gap between (in)valid bitstrings is defined as the difference between the probabilities of (in)valid bitstrings.
\begin{equation}
    \Delta \lambda = P_l(s_{valid}) - P_l(s_{invalid})
\end{equation}

The validity condition imposed upon a bitstring $s$ in the context of the separable measurement $l$ is based on noiseless Boolean derivatives $D_l f(s_{\setminus l})$. A bitstring $s$ is valid if $s_l = D_l f(s_{\setminus l})$, and not valid otherwise. 
Here, $s_{\setminus l}$ denotes the $(n_q-1)$-bit spectator string formed by excluding the target $l$-th bit from the measurement outcome $s$. To evaluate the derivative at this configuration, let $y$ and $\nu$ denote the two $n$-bit strings that match $s$ on all spectator bits $s_{\setminus l}$, with target bits explicitly set to $y_l = 0$ and $\nu_l = 1$. The noiseless Boolean derivative evaluated at this spectator configuration is defined as $D_l f(s_{\setminus l}) = f(y) \oplus f(\nu)$.

Since the estimators of the empirical frequencies are unbiased, the mean empirical signal gap is just equal to the theoretical gap $\mathbb{E}[\Delta \lambda] = \Delta \lambda$. Similarly, the exact standard deviation, $\sigma_{\Delta \lambda}$ of the empirical signal gap can be also estimated from the multinomial distribution such that the Signal-to-Noise Ratio (SNR) for the sorting algorithm is given as follows.

\begin{equation}
    \text{SNR} = \frac{\mathbb{E}[\Delta \hat{\lambda}]}{\sigma_{\Delta \lambda}} = \frac{\Delta \lambda \sqrt{\kappa}}{\sqrt{P_l(s_{valid}) + P_l(s_{invalid}) - (\Delta \lambda)^2}}
\end{equation}

In the practically relevant quantum error regime, the noise floor heavily dominates the signal ($\Delta \lambda \ll P_l(s_{valid}) + P_l(s_{invalid})$), making the $(\Delta \lambda)^2$ term in the denominator negligible. The critical sample budget $\kappa_{crit}$ required to push a specific spectator configuration past the threshold of $\text{SNR} \ge 1$ is therefore inversely proportional to the square of the signal gap:
\begin{equation}
    \kappa_{crit} \approx \frac{P_l(s_{valid}) + P_l(s_{invalid})}{(\Delta \lambda)^2}
\end{equation}

The critical sample budget required to isolate a valid Boolean equation from the noise floor ($\text{SNR} = 1$) is derived by evaluating the ratio of the total probability mass to the squared signal gap for a given spectator string $s_{\setminus l}$. 

For $n_q$ independent dephasing channels, the total probability mass is exactly $1/2^{n_q-1}$, and the signal gap is uniform. The critical sample complexity is:
\begin{equation}
    n_c \sim \kappa_{crit}^{(deph)} = \frac{2^{n_q-1}}{(1 - \epsilon_p)^2}
\end{equation}
This defines a strict, uniform threshold. Every equation in the hypercube, regardless of its topological degree, breaches the noise floor simultaneously at this exact sample budget.

For $n_q$ independent depolarizing channels, the total probability mass remains $1/2^{n_q-1}$, but the signal gap is penalized by the filtered Boolean derivative. The critical sample complexity is:
\begin{equation}
    \kappa_{crit}^{(depol)}(s_{\setminus l}) = \frac{2^{n_q-1}}{c_{depol}^2 \left( \tilde{D}_l^{(depol)}(s_{\setminus l}) \right)^2}
\end{equation}

where $c_{depol} = 1 - \frac{4}{3}\epsilon_p$ is the coherence contraction factor for the single-qubit depolarizing channel. The noisy depolarizing Boolean derivative $\tilde{D}_l^{(depol)}(s_{\setminus l})$ is determined via the inverse Walsh-Hadamard transform, which acts as a low-pass filter on the Boolean Fourier spectrum: $\tilde{D}_l^{(depol)}(s_{\setminus l}) = \sum_{k \in \{0,1\}^{n_q-1}} \hat{D}_l(k) c_{depol}^{|k|} (-1)^{k \cdot s_{\setminus l}}$. Here, $\hat{D}_l(k) = \frac{1}{2^{n_q-1}} \sum_{x \in \{0,1\}^{n_q-1}} (-1)^{D_l f(x)} (-1)^{k \cdot x}$ represents the exact Walsh-Fourier coefficients of the noiseless derivative, and $|k|$ is the Walsh-Fourier degree (or Hamming weight of the parity interaction). The global minimum occurs when the underlying Boolean derivative $D_l f$ has the maximum possible topological complexity—specifically, when it is a pure parity function of maximal weight $d = n_q-1$. In this limit, its Walsh-Fourier spectrum contains mass only at the highest possible frequency $|k| = n_q-1$. The filtered derivative thus evaluates precisely to $\tilde{D}_l^{(depol)} = \pm c_{depol}^{n_q-1}$. Substituting this into the denominator yields the strict upper bound for depolarizing sample complexity:
\begin{equation}
\kappa_{max}^{(depol)} = \frac{2^{n_q-1}}{c_{depol}^{2n_q}} = \frac{2^{n_q-1}}{(1 - \frac{4}{3}\epsilon_p)^{2n_q}}
\end{equation}

For $n_q$ independent thermal relaxation  channels, both the total probability mass and the signal gap are heavily dependent on the spectator Hamming weight and the classical population bias. 

\begin{equation}
    \kappa_{crit}^{(relax)}(s_{\setminus l}) = \frac{2^{n_q-2} (2+\epsilon_p)(1+\epsilon_p)^{n_q-1-|s_{\setminus l}|}}{(1-\epsilon_p)^{|s_{\setminus l}|} \left[ \frac{\epsilon_p (-1)^{D_l f(s_{\setminus l})} (1+\epsilon_p)^{n_q-1-|s_{\setminus l}|}}{2} + c_{relax} \tilde{D}_l^{(relax)}(s_{\setminus l}) \right]^2}
\end{equation}

where $|s_{\setminus l}|$ is the Hamming weight of the spectator string, and $c_{relax} = \sqrt{1-\epsilon_p}$ is the coherence contraction factor for the thermal relaxation channel. The noisy thermal relaxation Boolean derivative is constrained by the asymmetric nature of the decay: $\tilde{D}_l^{(relax)}(s_{\setminus l}) = \sum_{k \le \overline{y \lor \nu}} (-1)^{f(y \oplus k) \oplus f(\nu \oplus k)} \epsilon_p^{|k|}$. The bitwise condition $k \le \overline{y \lor \nu}$ restricts the error string $k$ to be non-zero only at positions where the target bit is 0 and the spectator bits $s_{\setminus l}$ were measured as 0. 

Unlike unital channels, thermal relaxation introduces a highly asymmetric sampling complexity governed by the direct competition between non-unital population decay and surviving quantum interference. Averaged over randomly selected functions $f$, an effective flip probability of $\frac{1}{2}\epsilon_p$ yields a baseline scaling of $\mathcal{O}(2^{n_q}/[1-\epsilon_p]^{2n_q})$. However, this average-case bound underestimates the true bottleneck: the physical decay towards the $\ket{0\dots0}$ ground state heavily oversamples low-degree hyperedges. The critical sample budget diverges entirely when this classical population bias perfectly cancels the quantum signal. To isolate this strict upper bound, we assume an adversarial Boolean topology where all higher-order parity interactions ($k>0$) maximally oppose the base derivative. Letting $m = n_q - 1 - |s_{\setminus l}|$ denote the number of zeros in the spectator string, the signal gap minimization is driven by the specific integer dimension $m^*$ that forces the adversarial denominator closest to zero:
\begin{equation}
    m^* = \text{argmin}_{m \in \{0, \dots, n_q-1\}} \left| \frac{\epsilon_p (1+\epsilon_p)^m}{2} + c_{relax} \left( 2 - (1+\epsilon_p)^m \right) \right|
\end{equation}
Substituting this critical root yields the exact worst-case sample complexity limit:
\begin{equation}
    \kappa_{max}^{(relax)} = \frac{2^{n_q-2} (2+\epsilon_p)(1+\epsilon_p)^{m^*}}{(1-\epsilon_p)^{n_q-1-m^*} \left[ \frac{\epsilon_p (1+\epsilon_p)^{m^*}}{2} + c_{relax} \left( 2 - (1+\epsilon_p)^{m^*} \right) \right]^2}
\end{equation}
where $c_{relax} = \sqrt{1-\epsilon_p}$. At this combinatorial coordinate, the physical noise effectively erases the target gradient, creating statistical singularities that require exponentially massive or infinite macroscopic oversampling to resolve.

\newpage
\section{Details of extrapolation}
\label{app:extrapolation_details}

In this appendix we describe how the classical sample complexity \(n_c\) of the measure-first protocols is inferred from finite-size numerical data. We consider three classical competitors: Hypergraph, Eigenshadow, and the shell-based machine-learning classifier. For each protocol, noise channel, and noise strength, the goal is to determine the classical resource \(n_c\) required to match a target accuracy set by the fully quantum protocol. The relevant noise setting is specified by the preparation-noise strength \(\epsilon_p\) and the readout-error rate \(\epsilon_r\). Here, the label ``device'' refers to the hardware profile used to generate the quantum reference curve \(A_Q\), for example, the devices A, B, and C considered in the main text.

The extrapolation proceeds in five stages. First, each empirical accuracy curve is compressed into a threshold-crossing coordinate \(k_x\), where
\begin{equation}
    k = \frac{\log_2 n_c}{n_q},
    \qquad
    n_c = 2^{n_q k}.
\end{equation}
Second, bootstrap resampling is used to estimate the uncertainty of these threshold crossings and to construct the weights entering the finite-size fits. Third, for each threshold \(T\), the finite-size dependence of the crossing coordinate is modeled by
\begin{equation}
    k_x(T,n_q)=C_{\star}(T)+\frac{\beta_{\star}(T)}{n_q}.
\end{equation}
Fourth, this predictor is evaluated on the target curve
\begin{equation}
    T_\eta(n_q)=A_Q(n_q)-\eta,
\end{equation}
where \(\eta\ge 0\) is an allowed accuracy slack. Finally, uncertainty bands and extrapolation trust criteria are imposed so that only extrapolations supported by forward validation are reported. Throughout this appendix, CI68 and CI95 denote final uncertainty bands obtained with \(z_{\mathrm{band}}=1\) and \(z_{\mathrm{band}}\approx 1.96\), respectively.

\subsection{Threshold-crossing representation}

For each measure-first protocol, noise channel \(\mathrm{ch}\), preparation-noise strength \(\epsilon_p\), and system size \(n_q\), we begin from the empirical accuracy curve
\begin{equation}
    \widehat{A}_{\mathrm{MF}}(k;n_q,\mathrm{ch},\epsilon_p),
\end{equation}
viewed as a function of the normalized sampling coordinate \(k=\log_2(n_c)/n_q\). Rather than extrapolating the full accuracy curve directly, we compress it at fixed threshold \(T\in(1/2,1)\) to the crossing coordinate
\begin{equation}
    k_x(T,n_q;\mathrm{ch},\epsilon_p)
    :=
    \inf\left\{
        k:\,
        \widehat{A}_{\mathrm{MF}}(k;n_q,\mathrm{ch},\epsilon_p)\ge T
    \right\}.
    \label{eq:kx_def_appendix_nat_rewrite}
\end{equation}
This representation is natural because \(k_x\) maps immediately to classical sample complexity through
\begin{equation}
    \log_2 n_c(T,n_q)=n_q\,k_x(T,n_q),
    \qquad
    n_c(T,n_q)=2^{\,n_q k_x(T,n_q)},
    \label{eq:nc_from_kx_appendix_nat_rewrite}
\end{equation}
with the physical convention \(n_c\ge 1\).

Equation~\eqref{eq:kx_def_appendix_nat_rewrite} is the continuous definition. Numerically, the accuracy curve is available only on a discrete \(k\)-grid \(\{k_1,\ldots,k_m\}\), and the crossing is obtained by linear interpolation between neighboring grid points after monotonicization in \(k\) (described below). If
\begin{equation}
    \widehat{A}_{\mathrm{MF}}(k_j) < T \le \widehat{A}_{\mathrm{MF}}(k_{j+1}),
\end{equation}
the threshold is bracketed by \(k_j\) and \(k_{j+1}\), and the crossing is estimated as
\begin{equation}
    k_x
    =
    k_j
    +
    \frac{T-\widehat{A}_{\mathrm{MF}}(k_j)}
         {\widehat{A}_{\mathrm{MF}}(k_{j+1})-\widehat{A}_{\mathrm{MF}}(k_j)}
    \bigl(k_{j+1}-k_j\bigr).
\end{equation}
On the present datasets, this linear inversion was cross-checked against monotone cubic inversion based on PCHIP~\cite{fritsch1984method} interpolation and root finding; the discrepancy was small on the scale relevant to the subsequent fits, with median \(|\Delta k_x|\approx 4.0\times10^{-3}\) and \(95^{\mathrm{th}}\)-percentile \(\approx 1.4\times10^{-2}\).

Two censoring situations occur naturally. If \(\widehat{A}_{\mathrm{MF}}(k_1)\ge T\), the crossing is left-censored, meaning that the required \(k_x\) lies below the sampled range and the corresponding \(n_c\) is smaller than can be resolved on the grid. If \(\widehat{A}_{\mathrm{MF}}(k_m)<T\), the crossing is right-censored, meaning that the threshold is not reached on the sampled range and the required \(n_c\) is larger.

All fits are performed on the dense threshold grid
\begin{equation}
    \mathcal{T}=\{0.51,0.53,\ldots,0.97\}.
\end{equation}
The lower end of this grid probes the near-chance regime numerically. The upper end is set by the requirement that crossing support remain statistically usable: above \(T=0.97\), right-censoring becomes common and bootstrap crossing quantiles become unstable. A stricter reporting floor, \(T=0.52\), is imposed only at the stage where the fitted predictor is compared with the quantum target curve.

\subsection{Bootstrap estimation of the threshold surface}

For each fixed triple \((n_q,\mathrm{ch},\epsilon_p)\), we estimate the uncertainty of the threshold surface by bootstrap resampling. This yields, for each threshold \(T\),
\begin{equation}
    \left\{
        k_x^{(b)}(T,n_q)
    \right\}_{b=1}^{B},
\end{equation}
with \(B=1600\) in the production runs. Each bootstrap replicate is produced by resampling the raw data underlying the empirical accuracy curve and repeating the same threshold-inversion procedure described above.

The resampling unit is chosen to preserve the dependence structure intrinsic to data acquisition. For Hypergraph and Eigenshadow, the basic resampling object is a full empirical accuracy curve at fixed \((n_q,\mathrm{ch},\epsilon_p)\), resolved over the sampled \(k\)-grid. When the raw data are stored as aligned collections of such curves, entire curves are resampled with replacement, so that correlations along the \(k\)-direction are preserved. When an additional cluster structure is present, for example independent hypergraph instances with repeated draws, resampling is carried out at the cluster level, optionally together with resampling within cluster. For the machine-learning protocol, the bootstrap is performed over independent trial realizations of the empirical accuracy curve at fixed \((n_q,\mathrm{ch},\epsilon_p)\).

Before threshold inversion, every empirical accuracy curve is made non-decreasing in \(k\) by replacing it with its running maximum. This monotonicization reflects the physical requirement that increasing classical resources should not reduce the measured accuracy. It is applied both to the original empirical curve and to each bootstrap-resampled curve. This operation acts along the sampling coordinate \(k\) and should be distinguished from the optional monotonicity guard introduced later, which acts on the fitted profile \(T\mapsto k_x(T,n_q)\).

At each pair \((T,n_q)\), the bootstrap ensemble yields a median crossing location together with an empirical central interval
\begin{equation}
    [k_{x,\mathrm{lo}},k_{x,\mathrm{hi}}]
    =
    \left[
        Q_{\alpha/2}\!\left(k_x^{(b)}\right),\,
        Q_{1-\alpha/2}\!\left(k_x^{(b)}\right)
    \right],
\end{equation}
where \(Q_p(\cdot)\) denotes the empirical \(p\)-quantile. From this interval we define the effective uncertainty scale
\begin{equation}
    \sigma_k(T,n_q)
    =
    \frac{\mathrm{hw}_k(T,n_q)}{z_{\mathrm{boot}}},
    \qquad
    \mathrm{hw}_k(T,n_q)
    :=
    \frac{k_{x,\mathrm{hi}}-k_{x,\mathrm{lo}}}{2},
    \label{eq:sigma_k_appendix_nat_rewrite}
\end{equation}
with
\begin{equation}
    z_{\mathrm{boot}}=\Phi^{-1}(1-\alpha/2).
\end{equation}
In the production runs we use a \(90\%\) central bootstrap interval, so \(z_{\mathrm{boot}}\approx 1.645\).

The quantity \(\sigma_k\) is used only as an effective weighting scale in the weighted least-squares fits of the next subsection, through the inverse-variance weights \(w_i=\sigma_k^{-2}\). For sharply concentrated crossing distributions, this scale coincides naturally with a one-sigma interpretation; this is typically the case for the Hypergraph and Eigenshadow data, whose accuracy curves exhibit relatively sharp transitions in \(k\). More generally, \(\sigma_k\) should be interpreted as a robust proxy for the local uncertainty of the crossing rather than as an exact Gaussian standard deviation. Strongly censored slices, or slices with too few successful bootstrap inversions, are removed before fitting. Concretely, a bootstrap inversion is counted as successful only if the crossing is finite and non-censored within the sampled \(k\)-range, and any point with fewer than \(20\) successful bootstrap inversions is excluded from downstream fits. For the production choice \(B=1600\), this corresponds to a minimum successful-inversion fraction of \(1.25\%\). If the exclusions leave fewer than three valid \(n_q\) points within a threshold slice, that slice is not fitted.

The output of this stage is therefore a threshold surface \(k_x(T,n_q)\), represented by bootstrap medians, together with a corresponding uncertainty surface \(\sigma_k(T,n_q)\) that defines the weights of the finite-size regression.

\subsection{Finite-size extrapolation model}

At fixed threshold \(T\), we model the finite-size dependence of the crossing coordinate by
\begin{equation}
    k_x(T,n_q)
    =
    C_{\star}(T)+\frac{\beta_{\star}(T)}{n_q}.
    \label{eq:inv_n_model_appendix_nat_rewrite}
\end{equation}
Equivalently,
\begin{equation}
    \log_2 n_c(T,n_q)
    =
    n_q\,C_{\star}(T)+\beta_{\star}(T).
    \label{eq:inv_n_nc_appendix_nat_rewrite}
\end{equation}
Within this model, \(C_{\star}(T)\) is the asymptotic threshold cost per qubit,
\begin{equation}
    C_{\star}(T)=\lim_{n_q\to\infty}k_x(T,n_q),
\end{equation}
while \(\beta_{\star}(T)\) captures the leading finite-size correction over the range of system sizes accessible numerically.

Equation~\eqref{eq:inv_n_model_appendix_nat_rewrite} is used here as the minimal phenomenological extrapolation ansatz. Its adequacy was assessed empirically by forward-chaining validation in \(n_q\), performed on five interior threshold slices chosen by snapping the default percentile targets \(10\%,30\%,50\%,70\%,90\%\) to the available threshold grid. Across the \(27\) protocol--channel--noise predictor cases entering the CI68 and CI95 analyses, the median pooled root-mean-square residual in \(k\)-space was \(1.93\times 10^{-2}\), with interquartile range \((1.62\text{--}2.98)\times 10^{-2}\) and maximum \(4.78\times 10^{-2}\). The median signed pooled bias was \(+4.1\times10^{-3}\), and the median absolute pooled bias was \(5.6\times10^{-3}\). Within the observed range of \(n_q\), the retained cases therefore do not exhibit systematic deviations beyond this scale that would require a higher-order correction.

For each threshold \(T\), the coefficients \(C_{\star}(T)\) and \(\beta_{\star}(T)\) are estimated by weighted least squares from the non-censored points, using the weights \(w_i=\sigma_k(T,n_{q,i})^{-2}\). Threshold slices with fewer than three valid \(n_q\) points are dropped. We denote the fitted coefficients by \(\hat C(T)\) and \(\hat\beta(T)\), reserving the starred symbols for the model parameters themselves. The fitted coefficient functions are then interpolated over threshold using the piecewise-cubic Hermite interpolating polynomial (PCHIP), with linear interpolation as fallback when PCHIP support is unavailable. This yields the continuous predictor
\begin{equation}
    \hat{k}_x(T,n_q)
    =
    \hat{C}(T)+\frac{\hat{\beta}(T)}{n_q}.
\end{equation}
The predictor is subsequently evaluated at the threshold set by the quantum target curve.

\subsection{Evaluation against the quantum target curve}

Let
\begin{equation}
    A_Q(n_q;\mathrm{device},\mathrm{ch},\epsilon_p,\epsilon_r)
\end{equation}
denote the fitted accuracy of the fully quantum protocol for a given device, noise channel \(\mathrm{ch}\), preparation-noise strength \(\epsilon_p\), and readout-error rate \(\epsilon_r\). The readout contribution \(\epsilon_r\) enters only through this quantum target, because the classical measure-first estimators themselves do not depend directly on readout noise. For a tolerance parameter \(\eta\ge 0\), we define the target threshold
\begin{equation}
    T_\eta(n_q)
    :=
    A_Q(n_q;\mathrm{device},\mathrm{ch},\epsilon_p,\epsilon_r)-\eta.
    \label{eq:target_threshold_appendix_nat_rewrite}
\end{equation}
The extrapolated classical cost is then obtained by evaluating the predictor at this target:
\begin{equation}
    \log_2 \hat n_c(n_q)
    =
    n_q\,\hat{k}_x\!\left(T_\eta(n_q),n_q\right).
\end{equation}

When \(T_\eta(n_q)\) falls below the reporting floor \(0.52\), the target is treated as effectively chance-level. In that regime, the point is not used for fitting, calibration, or uncertainty quantification, although in figures it may be displayed using the visual convention \(n_c=1\). If \(T_\eta(n_q)\) exceeds the fitted threshold support, corresponding here to the upper grid edge \(0.97\), the point is treated as upper-threshold censored and no finite in-range crossing estimate is reported.

\subsection{Uncertainty quantification and extrapolation trust criteria}

A physically consistent predictor must satisfy
\begin{equation}
    T_1<T_2
    \quad\Longrightarrow\quad
    k_x(T_1,n_q)\le k_x(T_2,n_q)
\end{equation}
for every fixed \(n_q\): demanding higher target accuracy cannot reduce the required classical resource. Because the model \eqref{eq:inv_n_model_appendix_nat_rewrite} is fit independently at each threshold, small violations of this property can appear after interpolation in \(T\). We therefore allow an optional post-fit monotonicity guard in the threshold direction, which replaces the profile \(T\mapsto \hat k_x(T,n_q)\) by its running maximum on a dense 256-point threshold grid over \([0.51,0.97]\). This guard is used only as a conservative stabilization device in diagnostics and is not part of the default pipeline. It should be distinguished from the compulsory monotonicization of the raw accuracy curves in \(k\) described above.

The final uncertainty on \(\log_2 n_c\) combines a validation-based discrepancy term with a parametric weighted-least-squares term. The validation-based component is obtained from forward-chaining validation in \(n_q\). If the observed system sizes for a given protocol--channel--noise case are \(n_q^{(1)}<\cdots<n_q^{(m)}\), we fit the inverse-size model on the first \(t\) sizes and predict the next \(h\) sizes, with \(h=1,2,3\) wherever available. Each held-out prediction yields a residual in \(k\)-space,
\begin{equation}
    e_k = k_{x,\mathrm{obs}}-k_{x,\mathrm{pred}}.
\end{equation}
Pooling over the selected threshold slices and forecast horizons gives the residual set \(\{e_{k,r}\}_{r=1}^{N_{\mathrm{hor}}}\), from which we define
\begin{equation}
    \mathrm{RMSE}_{k,\mathrm{pooled}}
    :=
    \sqrt{
        \frac{1}{N_{\mathrm{hor}}}
        \sum_{r=1}^{N_{\mathrm{hor}}} e_{k,r}^{\,2}
    }.
\end{equation}
In addition, for each threshold slice \(s\) we compute a slice-wise RMSE and define
\begin{equation}
    \mathrm{RMSE}_{k,\max}
    :=
    \max_s \left(\mathrm{RMSE}_{k,s}\right).
\end{equation}
The ratio
\begin{equation}
    \frac{\mathrm{RMSE}_{k,\max}}{\mathrm{RMSE}_{k,\mathrm{pooled}}}
\end{equation}
is used as a measure of cross-slice instability.

To build the validation-based uncertainty floor, we use the empirical \(68.27\%\) quantile of the absolute residual distribution,
\begin{equation}
    \sigma_{\mathrm{CV},k}
    :=
    Q_{0.6827}\!\left(\{|e_{k,r}|\}_{r=1}^{N_{\mathrm{hor}}}\right),
\end{equation}
and map it to \(\log_2 n_c\) units by
\begin{equation}
    \sigma_{\mathrm{CV}}^{(\log_2 n_c)}(n_q)
    =
    n_q\,\sigma_{\mathrm{CV},k}.
\end{equation}
We also define an observed-range variant based only on one-step forecasts,
\begin{equation}
    \sigma_{\mathrm{CV},k}^{\mathrm{obs}}
    :=
    Q_{0.6827}\!\left(\{|e_{k,r}|:h_r=1\}\right),
\end{equation}
and an extrapolation variant based on the pooled one-, two-, and three-step forecasts,
\begin{equation}
    \sigma_{\mathrm{CV},k}^{\mathrm{extrap}}
    :=
    Q_{0.6827}\!\left(\{|e_{k,r}|:h_r\in\{1,2,3\}\}\right).
\end{equation}
In the remainder of this appendix, \(\sigma_{\mathrm{CV},k}\) without a superscript refers to \(\sigma_{\mathrm{CV},k}^{\mathrm{extrap}}\). These quantile-based scales are used for uncertainty bands, whereas \(\mathrm{RMSE}_{k,\mathrm{pooled}}\) and \(\mathrm{RMSE}_{k,\max}\) enter the trust criteria below.

The parametric contribution is obtained from the covariance of the weighted least-squares fit. Writing
\begin{equation}
    \hat\theta(T):=
    \begin{pmatrix}
        \hat C(T)\\[3pt]
        \hat\beta(T)
    \end{pmatrix},
    \qquad
    g(n_q):=
    \begin{pmatrix}
        n_q\\[3pt]
        1
    \end{pmatrix},
\end{equation}
one has
\begin{equation}
    \log_2 \hat n_c(T,n_q)=g(n_q)^\top \hat\theta(T).
\end{equation}
If \(\Sigma_\theta(T)\) denotes the covariance matrix of \(\hat\theta(T)\), then
\begin{equation}
    \sigma_{\mathrm{WLS}}^{(\log_2 n_c)}(T,n_q)
    =
    \sqrt{
        g(n_q)^\top \Sigma_\theta(T)\, g(n_q)
    }.
\end{equation}
In practice, \(\Sigma_\theta(T)\) is computed as \((X^\top W X)^{-1}\), with \(W=\mathrm{diag}(\sigma_k^{-2})\), and each matrix element is interpolated threshold-wise using the same rule as for the fitted coefficients. This interpolation does not enforce positive semidefiniteness exactly at every off-grid threshold; small negative propagated variances arising from numerical interpolation are therefore clipped to zero.

The final reported uncertainty band in \(\log_2 n_c\) is obtained by combining the validation-based floor and the parametric fit uncertainty in quadrature:
\begin{equation}
    \sigma_{\mathrm{eff}}^{(\log_2 n_c)}(n_q)
    =
    \sqrt{
        \left[
            \sigma_{\mathrm{WLS}}^{(\log_2 n_c)}
            \bigl(T_\eta(n_q),n_q\bigr)
        \right]^2
        +
        \left[
            \sigma_{\mathrm{CV}}^{(\log_2 n_c)}(n_q)
        \right]^2
    }.
\end{equation}
The reported bands are then
\begin{equation}
    \log_2 \hat n_c(n_q)
    \pm
    z_{\mathrm{band}}\,
    \sigma_{\mathrm{eff}}^{(\log_2 n_c)}(n_q),
\end{equation}
with \(z_{\mathrm{band}}=1\) for CI68-style bands and \(z_{\mathrm{band}}\approx 1.96\) for CI95-style bands. These intervals should be interpreted as approximate uncertainty bands rather than strict finite-sample guarantees, especially in the extrapolation regime. Empirical coverage diagnostics are summarized in Tables~\ref{tab:forward_validation_summary}--\ref{tab:coverage_summary}.

\begin{figure}
    \centering
    \includegraphics[width=1\linewidth]{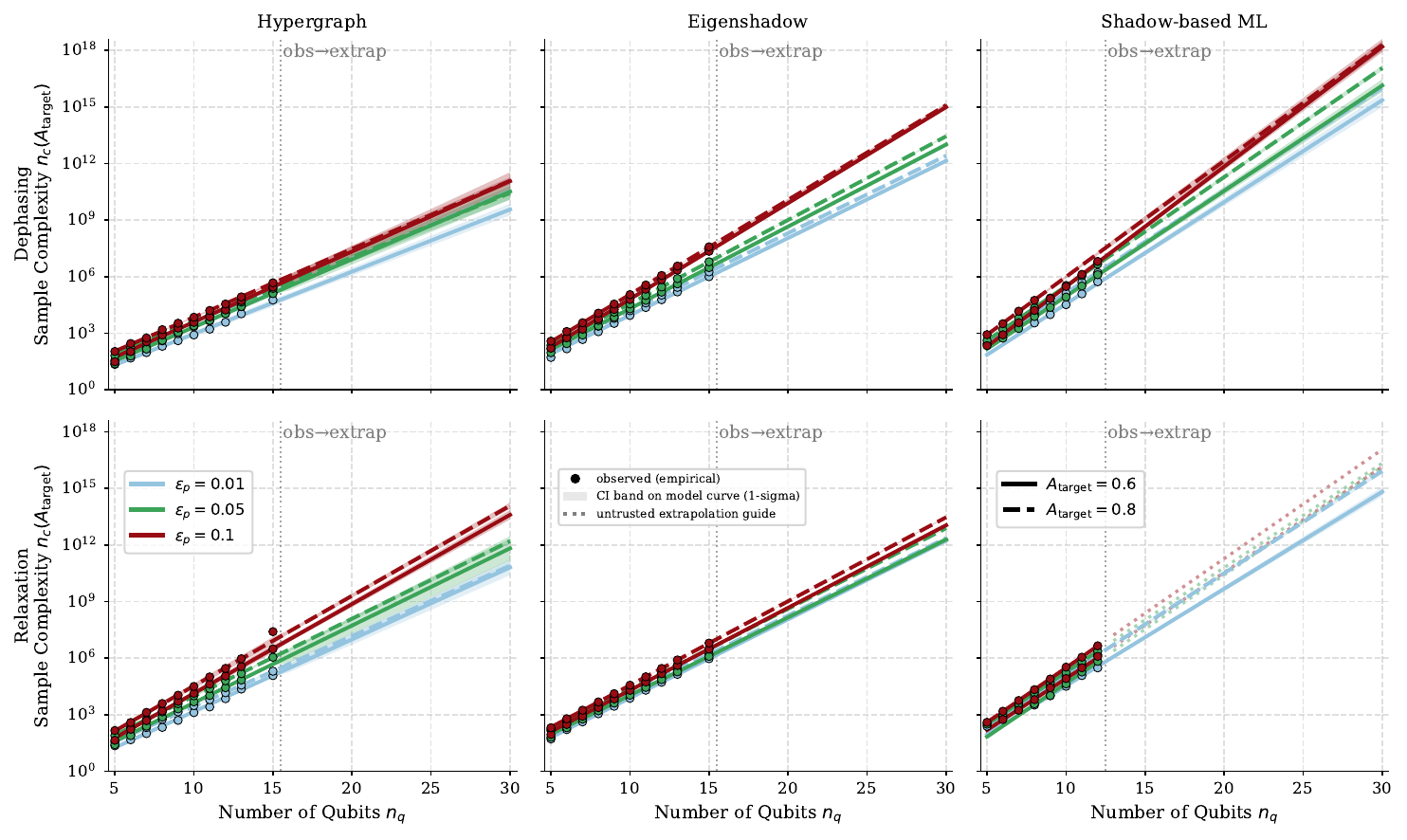}
\caption{\textbf{Validation of the extrapolation procedure.} 
The sample complexity $n_c(A_{\text{target}})$ is plotted against the number of qubits $n_q$ for three methods: Hypergraph, Eigenshadow, and Shadow-based ML. 
The top row demonstrates performance under a Dephasing noise model, while the bottom row shows a Relaxation noise model. 
Colors denote varying error probabilities $\epsilon_p$: $0.01$ (blue), $0.05$ (green), and $0.1$ (red). 
Line styles distinguish target accuracies, with solid lines representing $A_{\text{target}} = 0.6$ and dashed lines representing $A_{\text{target}} = 0.8$. 
Empirical observations are marked by circles, with shaded bands indicating the $1-
\sigma$ confidence interval on the model curve. 
A vertical dotted line designates the boundary between the observed data and the untrusted extrapolation guide.}
    \label{fig:placeholder}
\end{figure}

Finally, extrapolated points with \(n_q>n_{q,\max}^{\mathrm{obs}}\) are reported only when the extrapolation trust criterion is satisfied.
This requires, first, that forward-chaining validation be applicable (operationally: at least three threshold slices are available for validation, each usable slice contains at least three observed \(n_q\) points, and at least one held-out forecast horizon is produced); second, that the pooled number of validation horizons obey \(N_{\mathrm{hor}}\ge 12\); third, that the extrapolation-scale discrepancy at the observed-range boundary satisfy
\begin{equation}
    \sigma_{\mathrm{CV}}^{(\log_2 n_c)}\!\left(n_{q,\max}^{\mathrm{obs}}\right)\le 1.0;
\end{equation}
and fourth, that the instability ratio satisfy
\begin{equation}
    \frac{\mathrm{RMSE}_{k,\max}}{\mathrm{RMSE}_{k,\mathrm{pooled}}}\le 2.5.
\end{equation}
The bound \(1.0\) in \(\log_2 n_c\) units corresponds to at most a factor-of-two one-sigma discrepancy scale at the edge of the observed range. The ratio threshold \(2.5\) is a conservative stability cutoff chosen before final reporting and stress-tested against nearby alternatives. If any of these conditions fails, the extrapolated point is marked as untrusted and is omitted from the reported extrapolated curve. In the present study, all \(27\) protocol--channel--noise predictor cases satisfied these criteria in both the CI68 and CI95 analyses.

\begin{table}[h!]
\centering
\caption{Forward-validation and trust-gate diagnostics across all protocol--channel--noise predictor cases. These diagnostics are common to CI68 and CI95 because they are computed before the final reporting multiplier \(z_{\mathrm{band}}\) is applied.}
\label{tab:forward_validation_summary}
\begin{tabular}{lcc}
\hline
Statistic & Units & Value \\
\hline
Candidate predictor cases (Hypergraph/Eigenshadow/ML) & count & 27 (9/9/9) \\
Trusted extrapolation cases & count (\%) & 27/27 (100\%) \\
Median \(\mathrm{RMSE}_{k,\mathrm{pooled}}\) & \(k\) & \(1.93\times10^{-2}\) \\
IQR \(\mathrm{RMSE}_{k,\mathrm{pooled}}\) & \(k\) & \((1.62\text{--}2.98)\times10^{-2}\) \\
Maximum \(\mathrm{RMSE}_{k,\mathrm{pooled}}\) & \(k\) & \(4.78\times10^{-2}\) \\
Median signed pooled bias & \(k\) & \(+4.07\times10^{-3}\) \\
Median absolute pooled bias & \(k\) & \(5.60\times10^{-3}\) \\
\(N_{\mathrm{hor}}\) (min / median / max) & count & 21 / 60 / 60 \\
Median \(\mathrm{RMSE}_{k,h=1}\) & \(k\) & 0.0159 \\
Median \(\mathrm{RMSE}_{k,h=2}\) & \(k\) & 0.0198 \\
Median \(\mathrm{RMSE}_{k,h=3}\) & \(k\) & 0.0280 \\
Median \(\sigma_{\mathrm{CV},k}^{\mathrm{obs}}\) & \(k\) & 0.0145 \\
Median \(\sigma_{\mathrm{CV},k}^{\mathrm{extrap}}\) & \(k\) & 0.0167 \\
Median \(\sigma_{\mathrm{CV}}^{(\log_2 n_c)}(n_{q,\max}^{\mathrm{obs}})\) & \(\log_2 n_c\) & 0.171 \\
Maximum \(\sigma_{\mathrm{CV}}^{(\log_2 n_c)}(n_{q,\max}^{\mathrm{obs}})\) & \(\log_2 n_c\) & 0.581 \\
Median \(\mathrm{RMSE}_{k,\max}/\mathrm{RMSE}_{k,\mathrm{pooled}}\) & dimensionless & 1.63 \\
Maximum \(\mathrm{RMSE}_{k,\max}/\mathrm{RMSE}_{k,\mathrm{pooled}}\) & dimensionless & 2.17 \\
\hline
\end{tabular}
\end{table}

\begin{table}[h!]
\centering
\caption{Interval-width summaries in \(\log_2 n_c\) units. The extrapolated segment uses the larger validation floor \(\sigma_{\mathrm{CV},k}^{\mathrm{extrap}}\) relative to observed-range points. The CI95-to-CI68 half-width ratio is not uniformly equal to \(1.96\), because the tabulated values are medians over heterogeneous cases and system sizes rather than pointwise rescalings of a single underlying uncertainty.}
\label{tab:ci_width_summary}
\begin{tabular}{lcc}
\hline
Statistic & CI68 (\(z_{\mathrm{band}}=1\)) & CI95 (\(z_{\mathrm{band}}\approx 1.96\)) \\
\hline
Median half-width (all reported points) & 0.331 & 0.953 \\
Median half-width (observed \(n_q\) points) & 0.215 & 0.931 \\
Median half-width (extrapolated points) & 0.376 & 0.990 \\
\hline
\end{tabular}
\end{table}

\begin{table}[h!]
\centering
\caption{Empirical forward-validation coverage in \(k\)-space, aggregated as medians over all \(27\) predictor cases. Coverage at fixed \(n_q\) is identical in \(k\)-space and \(\log_2 n_c\)-space, since both residuals and reference scales are multiplied by the same factor \(n_q\). The aggregated median coverages reported here are therefore numerically very similar in the two representations.}
\label{tab:coverage_summary}
\begin{tabular}{lcc}
\hline
Nominal central coverage & Median observed coverage (\(h=1\)) & Median observed coverage (\(h\ge 2\)) \\
\hline
68.3\% (\(z=1\)) & 68.0\% & 60.0\% \\
90.0\% (\(z=1.64485\)) & 84.0\% & 80.0\% \\
95.0\% (\(z=1.95996\)) & 91.7\% & 82.9\% \\
\hline
\end{tabular}
\end{table}

\,
\newpage
\,
\newpage
\newpage
\,
\newpage
\section{Supplementary plots and data}
\label{app:extra_plots}

\begin{figure}[htb!]
    \centering
    \includegraphics[width=1\linewidth]{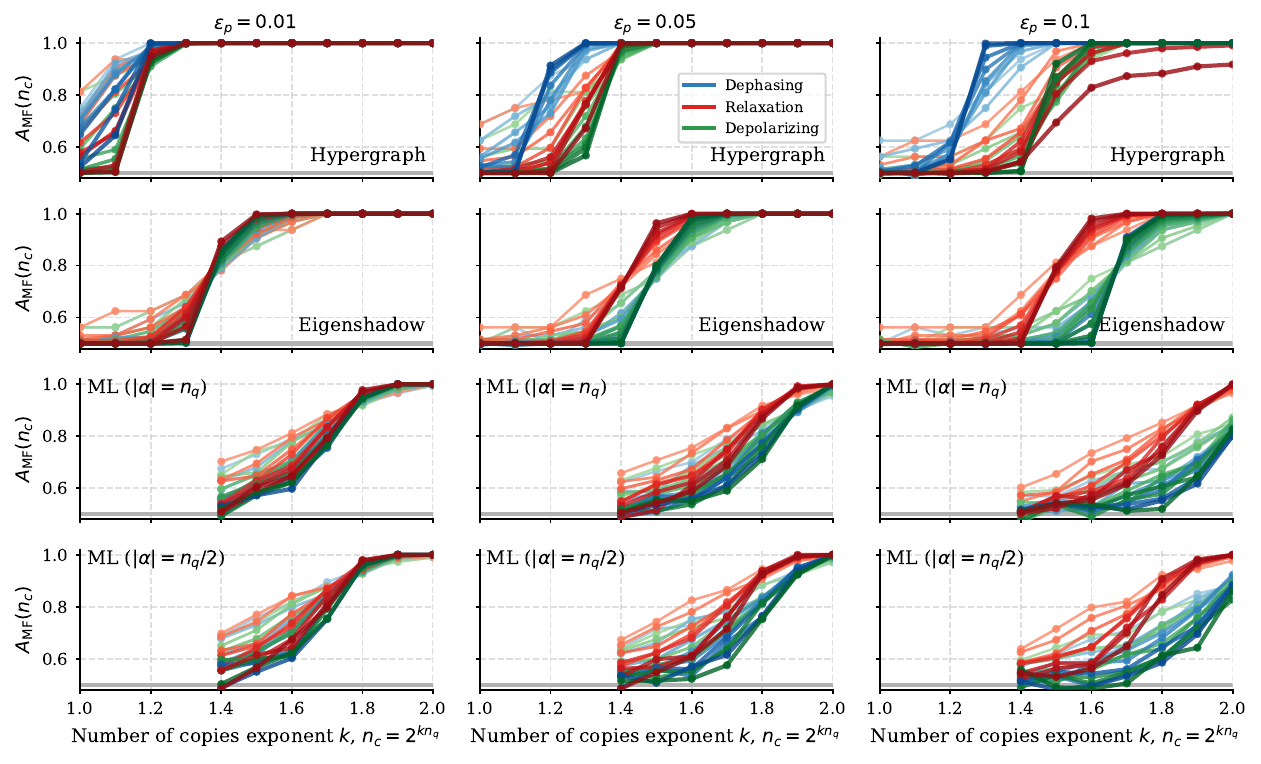}
    \caption{Supplementary version of Fig.~\ref{fig:eigenshadow_accuracies} with $\epsilon_p = 0.01,0.05,0.1$}
    \label{fig:all_mf}
\end{figure}

\begin{figure}[htb!]
    \centering
    \includegraphics[width=1\linewidth]{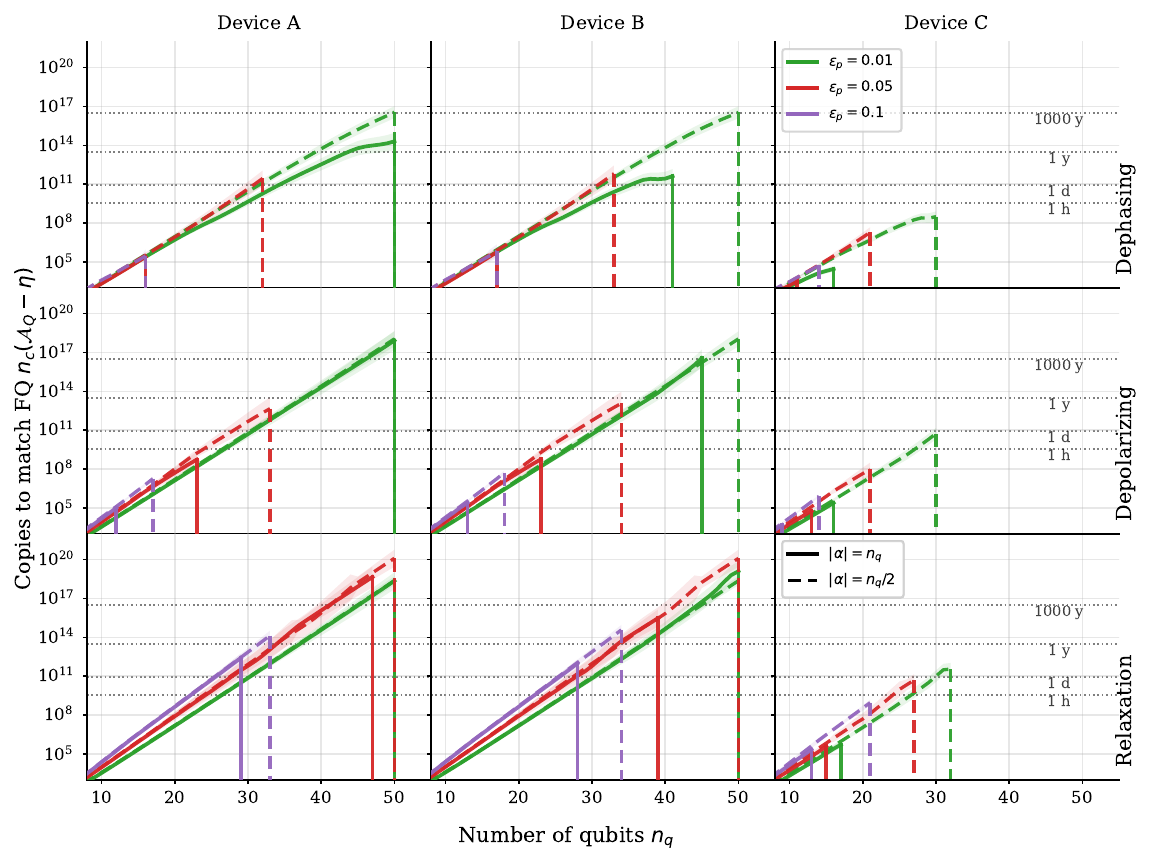}
    \caption{Supplementary version of Fig.~\ref{fig:scaling} with $\eta = 0.05$}
    \label{fig:eta5}
\end{figure}
\newpage

\end{document}